\documentclass[11pt]{article}
\pdfoutput=1

\usepackage{jheppub}
\usepackage[T1]{fontenc} 

\usepackage{amssymb}
\usepackage{amsfonts}
\usepackage{amsbsy}
\usepackage{amsmath}
\usepackage{amsthm}
\usepackage{graphicx}
\usepackage{epstopdf}
\usepackage{multirow}
\usepackage{latexsym} 
\usepackage{array}
\usepackage{color}
\input cyracc.def %sha
\newfont{\twelvecyr}{wncyr10 at 12pt}
\def\Z{\mathbb{Z}}
\def\F{\mathbb{F}}
\def\Q{\mathbb{Q}}
\def\C{\mathbb{C}}

\def\P{\mathbb{P}}

\def\n3a{t}

\def\ge{{\mathfrak{e}}}
\def\gso{{\mathfrak{so}}}
\def\gsu{{\mathfrak{su}}}

\def\gf{{\mathfrak{f}}}
\def\ggg{{\mathfrak{g}}}

\newcommand{\eq}[1]{(\ref{#1})}

\newcommand{\beq}{\begin{equation}}
\newcommand{\eeq}{\end{equation}}
\newcommand{\ba}{\begin{array}}
\newcommand{\ea}{\end{array}}
\newcommand{\bea}{\begin{eqnarray}}
\newcommand{\eea}{\end{eqnarray}}
\newcommand{\bean}{\begin{eqnarray*}}
\newcommand{\eean}{\end{eqnarray*}}
\newcommand{\eref}[1]{(\ref{#1})}
\newcommand{\sref}[1]{\S\ref{#1}}

\newcommand{\cO}{{\cal O}}
\newcommand{\comment}[1]{}

\title{\boldmath Geometric constraints in dual F-theory and heterotic string
  compactifications}
\author[1]{Lara B. Anderson}
\author[]{and}
\author[2]{Washington Taylor}

\affiliation[1]{Department of Physics\\
Robeson Hall, 0435\\
Virginia Tech \\
850 West Campus Drive \\
Blacksburg, VA 24061, USA}
\affiliation[2]{Center for Theoretical Physics\\
Department of Physics\\
Massachusetts Institute of Technology\\
77 Massachusetts Avenue\\
Cambridge, MA 02139, USA}

\emailAdd{{\tt lara.anderson} {\rm at} {\tt vt.edu}}
\emailAdd{{\tt wati} {\rm at} {\tt mit.edu}}

\preprint{MIT-CTP-4534}

\abstract{
We systematically analyze a broad class of dual heterotic and F-theory
models that give four-dimensional supergravity theories, and compare
the geometric constraints on the two sides of the duality.
Specifically, we give a complete classification of models where
the heterotic theory is compactified on a smooth Calabi-Yau threefold
that is elliptically fibered with a single section and carries smooth irreducible vector bundles, and the dual F-theory model has a
corresponding threefold base that has the form of a $\P^1$ bundle.  We
formulate simple conditions for the geometry on the F-theory
side to support an elliptically fibered Calabi-Yau fourfold.  We match
these conditions with conditions for the existence of stable vector
bundles on the heterotic side, and show that F-theory gives new
insight into the conditions under which such bundles can be
constructed.  In particular, we find that many allowed
F-theory models correspond to vector bundles on the heterotic side
with exceptional structure groups, and determine a topological
condition that is only satisfied for bundles of this type.  We show
that in many cases the F-theory geometry imposes a constraint on the
extent to which the gauge group can be enhanced, corresponding to
limits on the way in which the heterotic bundle can decompose.  We
explicitly construct all (4962) F-theory threefold bases for
dual F-theory/heterotic constructions in the
subset of models where the common twofold base surface is
toric, and give both toric and non-toric examples of the general
results.}
  
\begin{document}
\maketitle

\flushbottom

%--------------------------------
\section{Introduction and overview}\label{sec:het_f_motivation}

\subsection{Introduction}

Since the early days of string theory it has been known that a wide
range of different physical theories in four and higher dimensions can
be realized by compactifying ten-dimensional string theories (and
their more recently discovered higher-dimensional relatives M-theory
and F-theory) on different geometric spaces \cite{GSW, Polchinski}.
Tremendous effort has been expended in exploring the range of theories
that can be realized through such compactification.  While for
theories in higher dimensions with extended supersymmetry, the range
of possible string models has a tractable scope, for theories in four
dimensions with minimal supersymmetry known constructions seem to give
rise to such a vast ``landscape'' \cite{landscape-Susskind,
  flux-review} of possibilities that it is difficult to systematically
study the set of allowed models and the constraints that they impose
on 4D physics.  It is suspected, in fact, that the known constructions
of 4D ${\cal N} = 1$ theories from string theory may represent only
the tip of a much larger iceberg composed of compactifications
described by more general mathematical objects including non-K\"ahler
and non-geometric compactifications.
 
Nonetheless, it may be possible by analyzing specific string
constructions to ascertain some global constraints and systematic
features of the theories that arise from compactification of string
theory.  Recent work on globally classifying 6D string/F-theory
compactifications and associated constraints on 6D supergravity
theories \cite{universality, KMT-II, mt-clusters} suggests that a systematic
analysis is possible in six dimensions and may provide tools for a
similar treatment of some aspects of the space of 4D compactifications
\cite{Grimm-Taylor}.  In this paper we analyze how geometric
constraints on two general classes of string compactifications to 4D
are related, as a step towards a more systematic understanding of the
space of 4D ${\cal N} = 1$ theories that can arise from string theory.

Compactifications of heterotic string theory and F-theory provide two
corners of the string landscape where $4$-dimensional ${\cal N}=1$
supersymmetric theories with chiral matter and exceptional gauge
symmetries arise naturally. 
There is a tremendous literature on heterotic string
compactifications; some recent work has sought to explore and
enumerate the possible effective theories that can be obtained from
compactification of the heterotic theory on a smooth Calabi-Yau
(see \cite{Donagi:2000zf,Donagi:2000zs,Anderson:2007nc,Anderson:2008uw,Anderson:2009mh,Anderson:2013xka,Anderson:2012yf,Anderson:2011ns,Gabella:2008id} for some recent systematic studies). The duality between smooth heterotic
compactifications and equivalent 4D F-theory constructions has also been broadly
explored; see {\it e.g.}
\cite{Friedman-mw,
Hayashi:2008ba}.  
The effective
low-dimensional theories arising from compactifications of both
heterotic string theory and F-theory are highly constrained by the
background geometry of the compact dimensions.  Indeed, it is an
attractive possibility that these constraints might be strong enough
to characterize which effective theories can arise (in any dimension)
from heterotic or F-theory compactifications, or in the case of
compactifications to $4$ dimensions, used to characterize which string
geometries could be relevant for string phenomenology and give rise to
the low-energy physics we see in nature.

A major obstacle in any systematic attempt to classify the possible
compactification geometries and effective theories for either the
heterotic string or F-theory is the current limitation on our
mathematical understanding of the relevant geometries.  It is not
known, for example, whether the number of distinct diffeomorphism
classes of Calabi-Yau threefolds and fourfolds is even finite, 
much less how to characterize all the properties of the manifolds that
determine the effective theories.  There is, however, at least one
class of backgrounds, involving dual heterotic and F-theory
compactifications on elliptically fibered Calabi-Yau
threefolds and fourfolds, where the number of topologically distinct
string geometries is finite, and some systematic
analysis is possible.

For those theories that have dual heterotic and F-theory
constructions, the compactification geometries take the form
\cite{Morrison-Vafa-I, Morrison-Vafa-II}
\begin{equation}
\label{theduals}
\text{Heterotic on}~X_n,~~\pi_{h}: X_n
\stackrel{\mathbb{E}}{\longrightarrow} B_{n-1}~~\Leftrightarrow
~~\text{F-theory on}~Y_{n + 1},~~ \pi_{f}: Y_{n+1} \stackrel{K3}{\longrightarrow}
B_{n-1} 
\end{equation}
where $X_n$ is elliptically fibered over $B_{n -1}$ and the K3-fibered manifold $Y_{n +1}$ admits a more detailed description as an
elliptically-fibered Calabi-Yau $(n +1)$-fold with section over a base
${\cal B}_n$ which is itself $\P^1$ fibered over $B_{n -1}$.  The
classification of such dual theories can be done at increasing levels
of complexity by including successively more information about the
geometry.  At the coarsest level, the theories can be classified by
the topological type of the base $B_{n-1}$.  In dimensions eight and
six there is a unique $B_{n -1}$ associated with smooth
heterotic/F-theory dual pairs (respectively a point and $\P^1$), but
in four dimensions there are many distinct possible bases $B_{n -1}$.
At the next level of detail, theories can be classified by the
geometry of the F-theory base ${\cal B}_n$.  For any given $B_{n -1}$
there are in general many distinct topological types of ${\cal B}_n$
that characterize allowed F-theory geometries.  Theories with
different bases ${\cal B}_n$ (including those with different $B_{n
  -1}$) are connected on the F-theory side by tensionless string
transitions and on the heterotic side by small instanton transitions
\cite{dmw, Seiberg-Witten, Morrison-Vafa-II}.  We
primarily focus in this paper on classifying theories at the level of
${\cal B}_n$.  The choice of ${\cal B}_n$ on the F-theory side fixes
some of the topology of the dual heterotic bundles, but not all
(specifically, it fixes some components of the second Chern classes of
the bundles).  For a given choice of ${\cal B}_n$, Higgsing/unHiggsing
transitions in the effective theory, which correspond to deforming
along/tuning moduli in the F-theory and heterotic bundle pictures, can
modify the gauge group of the low-energy effective theory, and correspondingly
modify
the bundle structure group on the heterotic side.  For ${\cal N} = 1$
theories in 4D, $G$-flux on the F-theory side lifts some moduli and
can give disjoint sets of string vacua associated with
compactifications on a given ${\cal B}_n$.  For the most part, in this
paper we concentrate on features that depend only on the geometry of
${\cal B}_n$ and are independent of the moduli lifting and other
issues associated with $G$-flux.
While a further understanding of the consequences of  $G$-flux is
clearly desirable,  a good understanding of the underlying
geometric structure that we focus on in this paper
seems to be an important first step in a systematic understanding of
general 4D F-theory models.

In this work we focus on $4$-dimensional effective theories arising
from heterotic string theory on a smooth elliptically fibered
Calabi-Yau threefold and F-theory on a dual $K3$-fibered Calabi-Yau
fourfold that admits a compatible elliptic fibration and has a smooth
resolution. For both the Calabi-Yau threefold and fourfold geometries,
we consider only geometries where the elliptic fibrations admit a
(single) section.  We assume that the gauge bundle in the heterotic
theory is smooth, and that there are no heterotic $5$-branes wrapping
curves in the base $B_2$, which would be associated with singular
small instanton configurations\footnote{In some cases, the dual
  geometries we consider may include heterotic $5$-branes wrapping the
  elliptic fiber of the CY threefold. See \cite{Friedman-mw} for a
  discussion of the different roles that heterotic $5$-branes can
  play.} .

Beginning on the heterotic side of the duality, it is known that the
number of topological types of smooth elliptically fibered Calabi-Yau
threefolds with section is finite \cite{gross} (see also
\cite{KMT-II}, for a more constructive argument in the context of
Weierstrass models). In a heterotic dimensional reduction, the
  $10$-dimensional gauge field and the vacuum gauge field
  configuration over the Calabi-Yau threefold must be taken into
  account. These are described 
in the $E_8 \times E_8$ heterotic theory
by adding to the Calabi-Yau geometry a
  pair of holomorphic vector bundles $V_i$ ($i=1,2$) on $X_3$ with
  structure groups $H_i \subseteq E_8$. 
In the $SO(32)$ heterotic theory, only a single vector bundle is used.
For fixed bundle topology
  (specified by ${\rm rank} (V)$ and a total Chern class, $c(V)$), it
  is known that the moduli space of bundles\footnote{More precisely,
    the moduli space of Mumford semi-stable sheaves on $X_3$.}
  compatible with ${\cal N}=1$ supersymmetry in 4D has only finitely
  many components\footnote{Finiteness of the number of heterotic
    geometries here is established in two steps. First, the results of
    \cite{maruyama, langer} guarantee that for stable, hermitian bundles with
    fixed first and second Chern classes $c_1,c_2$ there are a finite
    number of possible values for the third Chern class $c_3(V)$ (note that in the case exceptional structure groups, $c_3$ is no longer a topological invariant). To
    argue that the number of heterotic geometries is finite we must
    further observe that $c_1 \equiv 0$ mod $2$, and the second Chern
    class is bounded as $0\leq c_2(V) \leq c_2(TX_3)$ by heterotic
    anomaly cancelation (see eq.\eref{anom_canc}).} \cite{maruyama,
    langer}.  Although these proofs of general finiteness are at present not
  constructive, it seems possible to systematically construct at least one important class of dual models.

As we discuss further in Section \ref{sec:duality}, when the
geometries are smooth on both the heterotic and F-theory sides the
base surface $B_2$ is restricted to be a \emph{generalized del Pezzo}
surface \cite{777}, of which there are a finite number of topologically distinct
types.  Over these bases a rational ($\mathbb{P}^1$) fibered threefold
${\cal B}_3$ must be built and here we restrict our attention to the
case where ${\cal B}_3$ can be constructed as a $\P^1$ bundle over
${\cal B}_2$. We demonstrate here that there are a finite number of
topologically distinct $\P^1$ bundles over any generalized del Pezzo
surface such that ${\cal B}_3$ supports an elliptically fibered
Calabi-Yau fourfold. Moreover,
we show that this finite set of ${\cal B}_3$'s can be enumerated for
any $B_2$.  The number of distinct branches of the moduli space of
Weierstrass models over any specific ${\cal B}_3$ corresponding to
distinct gauge group and matter contents is finite by a similar
argument to that for base surfaces in (section 6.5 of) \cite{KMT-II}.

This class of dual heterotic/F-theory models thus represents a reasonable
starting point with which we can get a first foothold into the problem
of classifying and characterizing 4D ${\cal N}=1$ string vacua and
their effective theories, as well as understanding constraints on the
effective theories arising from string geometry.
\footnote{As this paper
  was being completed, the paper \cite{Cvetic:2014gia} appeared, in
which magnetized brane models were considered over smooth elliptically
fibered Calabi-Yau threefolds over del Pezzo bases, and the number of
models in this class was shown to be finite.}

The general structure just detailed is illustrated clearly in the
simple case of 4D models where the base $B_2$ is toric.  The powerful
mathematical toolkit of toric geometry allows for simple and direct
computations in this class of examples.  While there are hundreds of
generalized del Pezzo surfaces $B_2$ that can act as bases of smooth
dual heterotic/F-theory Calabi-Yau threefolds and fourfolds, only 16
of these $B_2$'s are toric.  The direct enumeration of all associated
F-theory bases ${\cal B}_3$ (built as $\P^1$ bundles) is a straightforward calculation, which we
carry out in this paper as an example of the general theoretical
framework.

\subsection{Overview of main results}

For the convenience of the reader, we summarize here some of the main
results of the paper that we believe have some novelty, and indicate
where in the paper these results are described in more
detail.
\vspace*{0.05in}

\subsubsection{Classification and enumeration of models}

\subsection*{Classification of $\P^1$-bundle bases ${\cal B}_3$ for F-theory models with
  smooth heterotic duals}

As described at the conclusion of the previous section, we show in
\S\ref{sec:geometry-constraints} that there are a finite number of
$\P^1$ bundles ${\cal B}_3$ over smooth bases $B_2$ for F-theory
models with smooth heterotic duals on elliptically fibered Calabi-Yau
threefolds with section.  This follows from the fact that the number
of generalized del Pezzo surfaces $B_2$ is finite, and the number of
possible ``twists'' of the $\P^1$ bundle ${\cal B}_3$ over any $B_2$
is finite.  We construct explicit bounds on the twist that reduce the
classification of ${\cal B}_3$'s to a finite enumeration problem in
\S\ref{sec:F-theory-twists}, and write a simple set of topological
conditions that characterize allowed ${\cal B}_3$'s in
\S\ref{sec:constraints-global} and
\S\ref{sec:constraints-curves}. 
These results are quite general, and do not depend on toric geometry
or any other specific conditions on the F-theory base geometry beyond
the $\P^1$ bundle structure.
\vspace*{0.05in}

\subsection*{Enumeration of  models with toric $B_2$ and smooth
  heterotic duals}

As a concrete example of the general classification results,
we explicitly construct all F-theory bases  ${\cal B}_3$
that can be built as $\P^1$-bundles over toric surfaces $B_2$
giving rise to smooth elliptically fibered fourfolds for
F-theory compactifications with smooth heterotic duals.  For the 16
toric $B_2$'s we find  $4962$
threefolds ${\cal B}_3$, and
classify the generic
associated effective theories.  These manifolds add to
the dataset of Calabi-Yau fourfolds that have been systematically
studied to date (see
\cite{Klemm:1996ts,Brunner:1996pk,Brunner:1996bu,Brunner:1997bf,Knapp:2011wk,Gray:2013mja,Gray:2014fla}). These
results are 
described in Section \ref{sec:enumeration}. 
\vspace*{0.05in}

\subsubsection{Topological constraints on symmetries and spectra}

\subsection*{Matching geometric F-theory
constraints and heterotic bundle constraints}

We show that there is a close correspondence between the geometric
constraints on F-theory models and conditions for the existence of
smooth, slope-stable bundles in heterotic theories.  This extends
earlier work of Rajesh \cite{Rajesh:1998ik} and Berglund and Mayr
\cite{Berglund-Mayr}.  The details of this correspondence are
elaborated in Section \ref{sec:equivalence}.  Some of the most
interesting aspects of this correspondence arise when a constraint is
better understood on one side of the duality than the other.  In
particular, the next two items describe constraints on the gauge group
and bundle structure that are currently understood most clearly from
the F-theory perspective, while the last item below describes aspects
of matter content that are clearest from the heterotic point of view.

\vspace*{0.05in}

%\subsection*{Systematic correlations between topology and symmetries}

\subsection*{Heterotic bundles and the
  base-point free condition}

\noindent One of the most general methods known for explicitly
constructing bundles suitable for heterotic compactification is the
\emph{spectral cover construction} \cite{Friedman-mw,Friedman:1997ih}.
This construction is used to build bundles with structure group
$SU(N)$ or $Sp(N)$.  Irreducible bundles can only be constructed via a
spectral cover when the second Chern class of the bundle satisfies a
condition of base-point freedom.  We find that for $SU(N)$ or $Sp(N)$ structure groups the base point freeness condition can be derived from the F-theory geometry {\it independent of the assumption of any particular method of bundle construction}. Thus, for these structure groups base-point-freeness of (a part) of the second Chern class is required for all bundles in the moduli space.

In addition, we find that many F-theory models that should have smooth
heterotic duals violate this base-point free condition.  We show that
these are all associated with bundles having exceptional or $SO(8)$
structure groups and thus do not violate the above constraints for
$SU(N)$ and $Sp(N)$ bundles.  More general methods such as the {\it
  cameral cover construction} \cite{Ron_Elliptic,donagi,Donagi:1998vw}
and other approaches to constructing general $G$-bundles described in
\cite{Friedman-mw} based on a theorem of Looijenga \cite{Looijenga}
can in principle provide constructions of bundles with these more
general structure groups.  Our analysis gives a general classification
of situations in which bundles with exceptional structure groups are
expected to exist -- though in many cases explicitly describing the
properties of such bundles is an open problem in geometry.  The
property of base-point-freedom and its violation also has important
consequences for the problem of vector bundle deformations and
symmetry group breaking/enhancement.  The base-point free condition is
described in \S\ref{sec:spectral-cover}.  The corresponding F-theory
condition and circumstances for its failure are described in
\S\ref{sec:bpf}.  Examples of cases where this condition is violated
in F-theory are described in Sections \ref{sec:examples},
\ref{sec:enumeration}.

\vspace*{0.05in}

\subsection*{Limitations on gauge enhancement}

\noindent Geometric constraints on the F-theory side not only provide
a minimal gauge group for the low-energy theory given a
compactification topology, but can also limit the extent to which
the gauge group can be enhanced over a given base geometry.  For
example, in many situations $SU(2)$ and $SU(3)$ gauge groups are
constrained by F-theory geometry so that they cannot be in a broken
phase of an $SU(5)$ gauge group.  In these cases the
restriction is associated with the structure of
a codimension one singularity in the
F-theory geometry.  In other cases, codimension two singularities
related to matter fields constrain enhancement -- so that, for
example, in some cases an $E_6$ cannot be enhanced to an $E_7$.
These limitations on gauge enhancement are described in 
\sref{sec:enhancement-constraints} and Section
\ref{sec:het_cons}
with examples given in  Section \ref{sec:examples}. 

\vspace*{0.05in}

\subsection*{Chiral matter}

The circumstances under which the low-energy theory has chiral matter are
better understood on the heterotic side.  We identify a class of
situations in which chiral matter must arise due to the heterotic
geometry, with implications for the dual F-theory model when G-flux is
incorporated. On the heterotic side these correspond to bundles built
via the spectral cover construction with structure group given by
$SU(2n+1)$, giving rise to 4D GUT theories with, for example, $E_6$ or
$SU(5)$ symmetry \cite{Curio:1998vu}. In particular, in the dual F-theory geometries
enumerated in this work, we find that many examples of theories
with generic $E_6$ symmetry contain chiral matter. Examples of
this type are described in \sref{chiralsec}.

\vspace*{0.05in}

\subsection{Outline}

This paper is organized as follows: We begin in Section
\ref{sec:higher-dimensions} with a brief review of the duality between
heterotic string theory and F-theory in dimensions eight and six.  We
focus on the nature of heterotic/F-theory duality, the classification
of models, and constraints on the effective theory, illustrating
features and tools that are helpful in analyzing four-dimensional
compactifications in the remainder of the paper.  In Section
\ref{sec:duality} we summarize heterotic/F-theory duality in four
dimensions and describe the range of constructions of interest.
Section \ref{sec:F-theory-constraints} gives a more detailed
description of the geometric constraints on the F-theory side, and
Section \ref{sec:heterotic-constraints} describes the constraints on
both the Calabi-Yau threefold and bundle geometries on the heterotic
side.  In Section \ref{sec:equivalence} we compare the constraints on
the two sides and show when they are equivalent and when one side of
the duality provides new information about the geometry of the other
side.  Section \ref{sec:examples} contains some examples.  In Section
\ref{sec:het_cons} we summarize the consequences of our study for
heterotic bundle moduli spaces, and in Section \ref{sec:enumeration} we
describe the results of the systematic enumeration of all smooth
F-theory geometries with toric base $B_2$ and a smooth heterotic dual
construction. Finally, a brief summary of this work and associated
open questions are given in Section \ref{sec:conclusions}. Some
technical details are relegated to Appendices.

%%%%%%%%%%%%%%%%%%%%%%%%%%%%%%%%
\section{Lessons from heterotic/F-theory duality in higher dimensions}
\label{sec:higher-dimensions}

\vspace{10pt}

\subsection{Heterotic/F-theory duality in eight dimensions}

Beginning with the initial formulation of F-theory in $8$ dimensions
\cite{Vafa-F-theory, Morrison-Vafa-I} (see \cite{Denef-F-theory} for a
review), the duality of F-theory with the heterotic string has provided an
important window through which both theories can be better
understood. In $8$ dimensions, F-theory compactified on an
elliptically fibered $K3$ surface $Y_2$, $\pi: Y_2 \to \mathbb{P}^1$,
is dual in certain (separate) limits of its parameter space to the
perturbative $E_8 \times E_8$ and $SO(32)$ heterotic string theories
on $T^2$. In the case of the 8D $E_8 \times E_8$ heterotic theory,
this duality can be understood most explicitly in the weak coupling
limit of the effective theory, which is realized by taking the volume
of $T^2$ to be large in the heterotic theory.  The heterotic $T^2$
volume modulus is mapped into a complex structure modulus of the $K3$
surface in F-theory.  Geometrically, the $E_8 \times E_8$ limit corresponds to
decomposing the $K3$ surface into a (singular) fiber product of two
elliptically fibered $dP_9$ surfaces, glued together along an elliptic
curve\footnote{For the limit which produces the $SO(32)$ heterotic
  theory, the $K3$ degenerates into a fiber-product of rational
  surfaces, see \cite{Clingher:2003ui} for details.} -- the so-called
``stable degeneration limit'' \cite{Morrison-Vafa-I,Morrison-Vafa-II, Friedman-mw}.

In the 8D stable degeneration limit, all the features of the two
theories, including
%their parameters and vacua
the moduli parameterizing the vacua, can be matched exactly
\cite{Vafa-F-theory, Morrison-Vafa-II, LopesCardoso:1996hq,Lerche:1998nx, Clingher:2003ui,McOrist:2010jw}. 
For example, the possible gauge groups arising from different
configurations
of the heterotic flat gauge bundles on $T^2$ ({\it i.e.} Wilson
lines) can be matched to the symmetries arising from ADE degenerations
of the elliptic fiber of $K3$ that produce different non-Abelian
symmetries over points in the $\mathbb{P}^1$ base (corresponding to
the positions of $7$-branes in the language of Type IIB); these
degenerations were classified mathematically by Kodaira
\cite{Kodaira}.  From the point of view of the classification of
models and constraints, the 8D story is quite simple.  In this case
the base manifold $B_0$ is a point, and the F-theory base ${\cal B}_1
= \P^1$ is the unique $\P^1$ bundle over this point.  Thus, there is a
single moduli space of 8D models connected by ``Higgsing'' type
transitions that reduce or increase the size of the gauge group by
de-tuning or tuning moduli to modify the singularity structure of the
elliptic fibration.  On both the F-theory and heterotic sides, the
only constraint is that the gauge algebra ${\cal G}$ must have a root
lattice that can be embedded into the unique signature (2, 18)
unimodular lattice $\Gamma^{2, 18}$ (\cite{Ganor-Morrison-Seiberg},
reviewed in \cite{Taylor:2011wt}).

In lower dimensions, heterotic/F-theory duality is understood by
fibering the 8D duality over a nontrivial shared base manifold $B_{n
  -1}$. As in \eref{theduals}, a heterotic theory on an elliptically
fibered Calabi-Yau $n$-fold $X_n$, $\pi_h: X_n \to B_{n-1}$ is dual to
F-theory on a $K3$-fibered $(n+1)$-fold $Y_{n +1}$ with the same base, $\pi_{f}:
Y_{n+1} \to B_{n-1}$, in which the $K3$ fiber is in turn elliptically
fibered as described above. The elliptic and $K3$ fibrations are taken
to be compatible, and
both are chosen to have sections.  This duality has been studied
primarily in the stable degeneration limit \cite{Friedman-mw}, though
in this paper we describe aspects of the duality that are true more
generally, independent of this limit.

\subsection{Heterotic/F-theory duality in six dimensions}

\subsubsection{Dual 6D geometries}

For dual heterotic/F-theory compactifications to six dimensions, the
perturbative heterotic compactification space is a K3 surface that is elliptically
fibered over the common base $B_1 = \P^1$, and the dual F-theory
geometry is a Calabi-Yau threefold $Y_3$ that is elliptically fibered
with section over a Hirzebruch surface, $\pi: Y_3 \to \mathbb{F}_n$,
where the Hirzebruch surface ${\cal B}_2 = \F_n$ is itself a $\P^1$
bundle over $B_1$. As in 8D, codimension one singularities in the
elliptic fibration encode a gauge group in the F-theory picture, which
in 6D
can include non-simply laced groups when monodromy is present
\cite{Bershadsky:1996nh}.
Codimension two singularities encode matter fields.

In principle, heterotic/F-theory duality can be extended beyond the
set of smooth dual geometries by incorporating non-perturbative
effects such as NS$5$-branes in the heterotic theory. In this case,
the dual F-theory geometry $Y_3$ is an elliptic fibration over a more
general $2$-dimensional base ${\cal B}_2$, which is a blow-up of a
Hirzebruch surface \cite{Morrison-Vafa-II,Morrison-Vafa-I},
corresponding to a more general $\P^1$ fibration over $B_1 = \P^1$.
The base ${\cal B}_2 = \P^2$ can also be realized on the F-theory
side, {\it e.g.}  after a tensionless string transition from ${\cal
  B}_2 = \F_1$ \cite{Seiberg-Witten, Morrison-Vafa-I}.  For each
choice of ${\cal B}_2$, there is a connected moduli space of
elliptically fibered Calabi-Yau threefolds describing a set of 6D
theories connected by Higgsing and unHiggsing transitions.  The global
space of 6D F-theory compactifications \cite{KMT-II}
consists of a finite family of
such moduli spaces connected by tensionless string type transitions.
The connectivity of the set of moduli spaces associated with distinct ${\cal
  B}_2$'s corresponds to the mathematical framework of \emph{minimal
  surface theory} \cite{bhpv, Reid-chapters}, in which curves of
self-intersection $-1$ are blown down until a minimal surface (in this
case a Hirzebruch surface $\F_n$ with $n \leq 12$, $\P^2$, or the Enriques surface \cite{Grassi}) is reached.  A systematic
classification of F-theory bases ${\cal B}_2$ according to the
intersection properties of effective divisors is given in
\cite{mt-clusters}.  A complete enumeration of allowed bases is
in principle possible and has been carried out explicitly for toric
${\cal B}_2$'s \cite{mt-toric}, and the more general ``semi-toric''
class of $B_2$'s that admit a single $\C^*$ action
\cite{Martini-WT}.  The  global description of the 
moduli space is much more complicated on the heterotic side,
where multiple coincident small instantons must be analyzed
systematically (\S\ref{6d_symm}).  We restrict attention in this paper to smooth
heterotic/F-theory dual geometries where no small instantons arise,
which in 6D limits us to ${\cal B}_2 =\F_n$.

\subsubsection{Geometric conditions on vacua}\label{conditions_on_vac}

Even
in $6D$, each side of the duality encodes some  nontrivial
information about the geometry of its dual theory. In  early
explorations of F-theory \cite{Morrison-Vafa-I, Morrison-Vafa-II, Bershadsky:1996nh}, the degrees of freedom and
effective theory of heterotic compactifications were used to
develop the ``dictionary'' of how the dual
Calabi-Yau threefold geometry
determines the gauge symmetries and matter spectra in the F-theory description.  
F-theory can in turn be
used to enumerate possible heterotic backgrounds $(K3, V_1,V_2)$ and
to make useful statements about their properties.

One of the most significant aspects of this duality is the way that
the dual theories realize the condition for ${\cal N}=1$ (minimal)
supersymmetry in $6$ dimensions. On the F-theory side this appears as
the condition that the total elliptically fibered compactification
space is a Calabi-Yau threefold (more precisely, the manifold can be singular, in which case the resolved geometry is a
Calabi-Yau threefold; the F-theory description can be thought of as a
singular M-theory limit, as reviewed for example in
\cite{Denef-F-theory}).  On the heterotic side,
this condition corresponds to the statement that the compactification
manifold is a $K3$ surface and that the gauge bundles $(V_1,V_2)$ with
structure groups embedded into each $E_8$ factor satisfy the Hermitian
Yang-Mills equations \cite{GSW}, that is, that they are slope-(poly)stable \cite{duy1,duy2}.  Furthermore, the first Chern class of the
principal bundles must vanish\footnote{More precisely, since the
  generators of $E_8$ are traceless, all principal bundles ${\cal V}$
  that are sub-bundles of an $E_8$ bundle (i.e. that have structure
  group $H \subseteq E_8$) must have vanishing first Chern
  class. However, if the associated vector bundles $V, \wedge^2 V
  \ldots$ (arising in the heterotic theory from the decomposition of
  the ${\bf 248}$-dimensional adjoint of $E_8$) are reducible, then
  their first Chern classes can be non-zero and satisfy $c_1 (V_i)\equiv 0$
  (mod $2$).}, with $c_1 (V_i)\equiv 0$ (mod $2$).

The choice of a smooth heterotic/F-theory dual pair in 6D is
determined by a single integer. In the heterotic theory this appears
as the choice of a fixed second Chern class for the vector bundles
$c_{2}(V_{1,2})=12 \pm n$.  On the F-theory side, this corresponds to
the choice of Hirzebruch surface $\mathbb{F}_n$ for the two-fold base
${\cal B}_2$.  The constraint $n \leq 12$ (originally described in
\cite{Morrison-Vafa-I}) gives a simple example of the type of
geometric constraint that we explore later in this paper for 4D
compactifications.  On the heterotic side this constraint follows from
the slope-stability of the vector bundles and the heterotic anomaly
cancellation condition which relates the second Chern class of the
holomorphic tangent bundle of K3 with those of the gauge bundles
$V_{1, 2}$. On the F-theory side this constraint
follows from the fact that for
$n > 12$ the existence of an effective divisor in ${\cal B}_2= \F_n$
with self-intersection $-n < -12$ yields a singularity in any elliptic
fibration over ${\cal B}_2$ that cannot be resolved to yield a total
space that is a Calabi-Yau manifold.  Thus, in this case rather
different geometric considerations on the two sides give the same
analytic constraint on the structure of the allowed theories.

\subsubsection{Moduli and the stable degeneration limit}

It is worth briefly reviewing the moduli of heterotic/F-theory
compactifications in $6D$ and the interpretation of these moduli in
the two dual pictures. 
As described in \cite{Morrison-Vafa-I, Morrison-Vafa-II} and easily confirmed
in the toric description (reviewed in \S\ref{sec:duality-toric}),  a
Calabi-Yau threefold $Y_3$ that is elliptically fibered over the
Hirzebruch surface $\F_n$
can be described in Weierstrass form as
\begin{equation}
y^2=x^3+f(z_1,z_2)x+g(z_1,z_2)
\end{equation}
where $(z_1,z_2)$ are coordinates on $\mathbb{F}_n$ and
\begin{align}\label{fgdefs}
&f(z_1,z_2)=\sum\limits_{i=0}^{I} z_{1}^{i}f_{8+n(4-i)}(z_2) &I \leq
  8~\text{such that}~8+n(4-I) \geq 0\\ 
&g(z_1,z_2)=\sum_{j=0}\limits^{J} z_{1}^{j}g_{12+n(6-j)}(z_2) & J\leq
  12~\text{such that} ~12+n(6-J) \geq 0 
\label{eq:gs}
\end{align}
More abstractly, $f, g$ are sections of the line bundles ${\cal O} (-4K), 
{\cal O} (-6K)$,
where $K$ is the canonical class of
the base ${\cal B}_2$; these explicit expressions give a local
coordinate description of generic sections of these line bundles.

As is now well understood \cite{Morrison-Vafa-I, Morrison-Vafa-II, Friedman-mw}, the heterotic/F-theory
dictionary in 6D indicates that the ``middle'' polynomials (the
coefficients of $z_{1}^{4}$ and $z_{1}^{6}$ in $f$ and $g$,
respectively in \eref{fgdefs}, (\ref{eq:gs})) correspond to the moduli of the
heterotic $K3$ surface, while polynomials of low degree (coefficients
of $z_1^i$ with $i<4$ in $f$ and coefficients of $z_1^j$ with $j<6$ in
    $g$) parameterize one of the heterotic bundles $V_1$, and
    polynomials of high degree ($i>4$, $j>6$) parameterize the other
bundle $V_2$.  The bundle $V_1$ has structure group $H_1$, which is
embedded in $E_8$, and the resulting gauge group is the commutant
$G_1$ of $H_1 \subseteq E_8$.  On the F-theory side this corresponds to
a Calabi-Yau threefold with $7$-branes wrapping the $\mathbb{P}^1$ base of the $K3$-fibration, giving rise to symmetry $G_1$ encoded in the singularity
structure of the elliptic fibration at the point $z_1 = 0$.
Similarly, the second heterotic bundle $V_2$ has structure group $H_2$
with commutant $G_2$, associated with the singularity structure of the
F-theory elliptic fibration at $z_1= \infty$.  This correspondence can
be made precise in the stable degeneration limit, in which $Y_3 \to
Y_1 \cup_{K3} Y_2$ where $Y_{1,2}$ are (non-CY) $dP_9$-fibered
threefolds. In this limit, the infinitesimal deformation space, ${\rm
  Def}(Y_i)$, of $Y_{i}$ can be matched exactly to that of the bundles
${\rm Def}(V_i)$ and the $K3$ surface, ${\rm Def}(K3)$ (see \cite{Morrison-Vafa-I, Morrison-Vafa-II, Bershadsky:1996nh,Aspinwall:1998bw}
and \cite{Donagi:2012ts} for a modern treatment of this result in terms of
limiting mixed Hodge structures). That is,
\begin{equation}
h^{2,1}(Y_i)=h^1(K3,{\rm End}_0(V_i))+20
\end{equation}
The correspondence between the F-theory moduli in the Weierstrass
model  and moduli of the dual heterotic bundles is
particularly transparent in the spectral cover construction
(\ref{sec:spectral-cover}), where the polynomials
$f_k, g_k$ play a dual role in parameterizing the spectral cover
divisor on the heterotic side.

\subsubsection{Constraints on bundles and gauge symmetry}\label{6d_symm}

For a fixed topology of ${\cal B}_2$ on the F-theory side, the moduli
encoded in the functions $f, g$
(\ref{fgdefs}, \ref{eq:gs}) parameterize Weierstrass models for  all
elliptically (and $K3$) fibered threefolds $Y_3$ over the base
${\cal B}_2$.  Parts of this moduli space in principle give a complete
encoding of each dual heterotic
moduli space of sheaves with fixed
total Chern class, denoted ${\cal M}_{\omega}(\text{rank},c_1,c_2)$,
that are stable with respect to a chosen K\"ahler form\footnote{It is
  important to note that while heterotic/F-theory duality is believed to hold for the full moduli space of the the two theories, the explicit ``dictionary'' between degrees of freedom is only well understood for suitably weakly coupled regions of parameter space. On the heterotic side, this
  corresponds to moduli spaces of stable sheaves that are stable for
  the appropriate ``adiabatic'' choice \cite{Friedman-mw} of K\"ahler
  form on $K3$.}
 $\omega$ on $K3$. 
In this context,
nontrivial features of the heterotic and F-theory
geometries can be
exactly matched. 
Considering only F-theory on the  elliptically
fibered threefolds with the 13 Hirzebruch bases $\mathbb{F}_{n}$, $n=0, \ldots
12$, it is possible to deduce a number of  facts about the dual moduli
spaces of sheaves
${\cal M}_{\omega}(\text{rank},c_1,c_2)$. The first of these is that
for $c_1=0$, fixed $c_2$, and the structure group $H_i$ of $V_i$
fixed subject to $H_i \subseteq E_8$, ${\cal M_\omega}$ has only one component --
corresponding to a connected
deformation 
space of a dual
Calabi-Yau threefold
$Y_3$ as described above.\footnote{To avoid confusion it should be noted here that the rank appearing in the definition of the moduli space ${\cal M}_{\omega}(rank, c_1,c_2)$ refers to a fixed {\it fiber dimension} for a given vector bundle appearing in a representation of $H_i$ determined by the decomposition of the adjoint of $E_8$, {\it not} the rank of $H_i$ as a Lie group. For example, although $F_4$ has rank $4$ as a Lie group, one relevant moduli space for $F_4$ bundles appearing in heterotic theories would consist of rank $26$ vector bundles. See \S\ref{sec:hetmatter} for a general discussion of the relevant bundles, ranks and group representations.}

\begin{figure}
\begin{center}
\includegraphics[width=10cm]{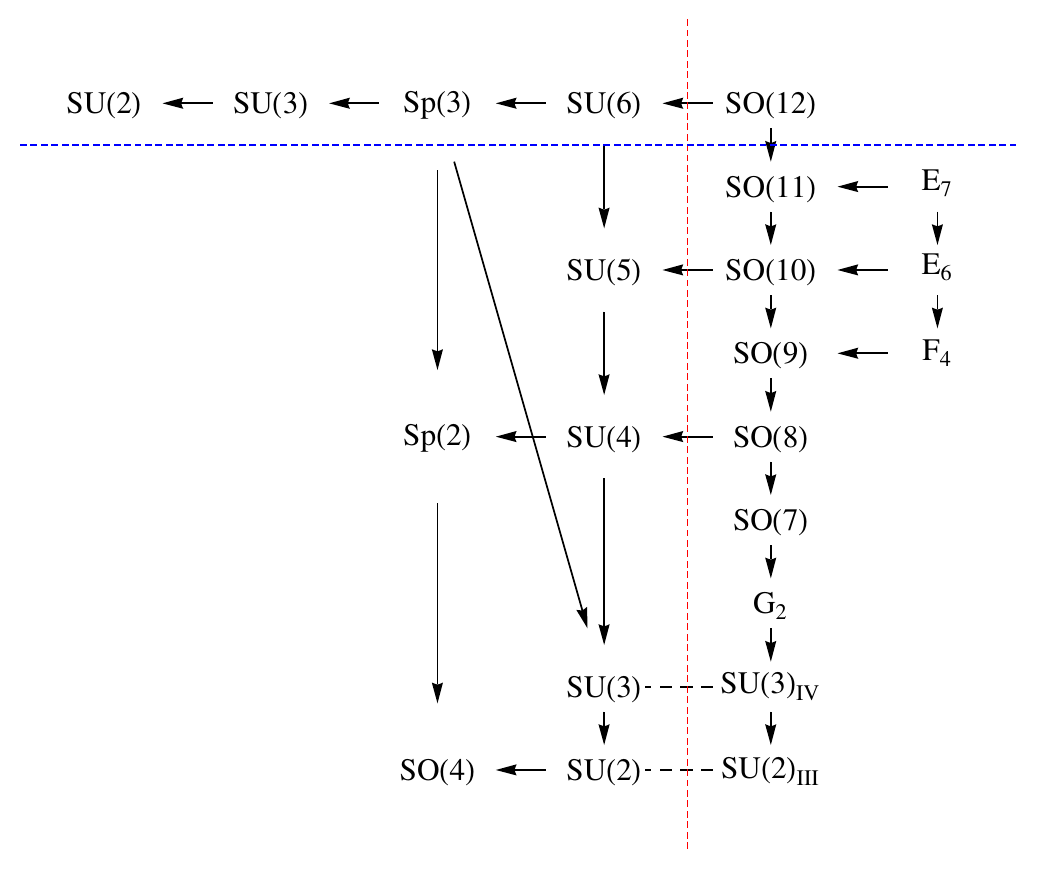}
\end{center}
\caption{\emph{The possible Higgsing/Enhancement chains for smooth
    heterotic/F-theory dual pairs; modified from
    \cite{Bershadsky:1996nh}.  Figure depicts Higgsing possibilities
    based on heterotic bundles with structure group $H \subset E_8$,
    which match with dual F-theory models.  F-theory gauge groups from
    Kodaira singularities with $f, g$ having nonzero degrees of
    vanishing lie to the right of the vertical red dashed line, such
    gauge groups can be forced from the geometry (geometrically
    ``non-Higgsable'') and cannot be unHiggsed to anything left of the
    line. 
The $SU(3)$'s and $SU(2)$'s connected near the bottom by horizontal
    dashed lines correspond to transitions between different Kodaira
    types in F-theory from type $IV, III$ to type $I_3, I_2$.  
The
top
   row above the horizontal blue dashed line corresponds to an
alternative
    Higgsing sequence from $E_8$ to $SU(3), SU(2)$ with non-standard
    commutants ({\it e.g.} $H =SU(3) \times G_2$ for upper $SU(3)$),
    generically associated with matter in the adjoint representation,
    which on the F-theory side involves wrapping on higher genus
    curves for 6D models.  Note that in F-theory models that do not
    have heterotic duals, further unHiggsing ({\it e.g.} to $SU(N >
    6)$) can occur.  Note also that in the heterotic theory some
    Higgsing chains lead to product gauge groups, as discussed further
    in text.  } \label{fig:chain}}
\end{figure}

The moduli space structure on the F-theory side is matched
non-trivially not only to the irreducibility of ${\cal
  M}_{\omega}(\text{rank},c_1,c_2)$ for {\it fixed rank}, but also
provides information about deformations of $V_i$ that change the
rank. A change in the rank (and hence structure group $H_i$) of the
bundles $V_i$ corresponds to changing the gauge symmetry of the $6D$
effective theory ($G_i \subseteq E_8$). In the F-theory geometry, this gauge symmetry
can be varied by changing the complex structure of $Y_3$.  By tuning
the complex structure to special loci in moduli space, it is possible
to augment the Kodaira singularity types of the elliptic fibration
over divisors in the base, which enhances the gauge group symmetry of
the 6D theory; this corresponds on the heterotic side to specializing
$V_i$ to bundles with smaller structure groups.  In the reverse
process, in cases where complex structure deformations exist to
break a symmetry, there is a simple realization in the effective
theory as a direct ``Higgsing'' of $G$ by charged
hypermultiplets. Complete ``chains'' of these breaking/enhancement
patterns have been determined for the dual $6D$ theories and matched
exactly to complex structure deformations of the corresponding
threefolds $Y_3$ \cite{Bershadsky:1996nh} (see also Figure \ref{fig:chain}).

While the gauge symmetry of a 6D theory can often be made smaller or
larger by Higgsing or un-Higgsing through de-tuning and tuning moduli
in the F-theory picture, there are also constraints on both sides that
can restrict the extent to which a gauge group can be broken or
expanded.  For $n\geq 3$, the threefolds $\pi: Y_3 \to \mathbb{F}_n$
are generically singular (though they admit a smooth resolution).
This means that in these cases there is a $6$-dimensional gauge symmetry
that cannot be Higgsed away by giving vevs to the hypermultiplets
associated to the complex structure moduli.  In the F-theory geometry,
this corresponds to the presence of a divisor of self-intersection
$-n$ in $\F_n$, over which the elliptic fibration must become singular
and has a Kodaira type associated with a nontrivial gauge group
factor.  On the heterotic side, this implies that for certain values
of the second Chern class $c_2$ there is a maximum structure group $H$
possible {\it for any bundle with that topology}, since $H \subseteq
E_8$ encodes the unbroken symmetry $G$ in the $6$-dimensional
heterotic theory. For example, if $c_2(V)=4$, the maximal structure
group of $V$ is $SU(2)$, for any such bundle on $K3$. It follows then
that the moduli space of stable sheaves, ${\cal M}_{\omega}(r, 0, 4)$
contains no locally free sheaves (i.e. smooth bundles) for $r>2$ (we
will sometimes  for brevity refer to such a moduli space as
``empty''). This corresponds in F-theory to the fact that $\pi: Y_3
\to \mathbb{F}_{8}$ is singular with a generic, non-Higgsable $E_7$
symmetry \cite{Morrison-Vafa-II}.

A fact that is perhaps not generally well appreciated is that there
are also cases in 6D where there are nontrivial constraints on the
ways in which a gauge group can be enhanced by ``unHiggsing'' the
generic model in a given component of the moduli space (i.e. possible up/rightward paths towards $E_7$ or $SO(12)$ in Figure \ref{fig:chain}).  In
particular, in any 6D F-theory construction where the low-energy
theory has a generic (non-Higgsable) gauge group $SU(3)$ 
arising from a Kodaira type $IV$ singularity
associated
with a curve $C$ of self-intersection $-3$, the group cannot be
enhanced to any larger $SU(N)$.  In F-theory this follows from the
fact that the Weierstrass coefficients $f, g$ must vanish to degrees
2, 2 on $C$, which is incompatible with an $A_n$ type singularity for
$n > 2$.  This condition corresponds on the heterotic side to a
constraint on the extent to which the $E_6$ bundle over K3 with
instanton number 9 can be deformed to a bundle with reduced structure
group. In the case at hand, the enhancement $SU(3) \to G_2$ ({\it
  i.e.,} the reduction of the $E_6$ bundle $V_{27} \to V_{26} \oplus
{\cal O}$ in\footnote{Here ${\cal O}$ denotes the trivial line bundle over $X_3$.} $F_4$) is possible, while the enhancement $SU(3) \to
SU(4)$ via $V_{27} \to V_{10}+V_{16} \oplus {\cal O}$ in $SO(10)$ is
not. Phrased differently, in terms of the Higgsing chains given in
Figure \ref{fig:chain}, if we begin with the lower right $SU(3)_{IV}$ for this
case, it is possible to move upwards (un-Higgsing) and to the right along
the right-hand path $G_2 \leftarrow SO(7) \leftarrow \ldots$, but
the group cannot be unHiggsed  to the path $SU(4) \leftarrow SU(5) \leftarrow \ldots$.
While in 6D, those F-theory models with smooth heterotic duals  
have gauge groups that can always be enhanced to $E_7$ (or $SO(12)$)
along {\it some} path in Figure \ref{fig:chain}, the obstacles to
gauge group enhancement can be stronger for more complicated 6D
F-theory models (without smooth heterotic duals), and in four
dimensions there are a number of constraints of this type.  We explore
the 4D constraints to gauge group enhancement in more detail in
\S\ref{sec:enhancement-constraints}.

The set of possible Higgsing/unHiggsing chains in dual pairs of 6D
heterotic/F-theory models contains a number of other interesting
features.  As depicted in Figure~\ref{fig:chain}, a heterotic $E_8$
symmetry can be broken to $SU(3)$ in several distinct ways, depending on
whether the commutant is $H = E_6$ or $H = G_2 \times SU(3)$, for example.  In the
latter case, explicit computation of the branching rules generically
gives matter in the adjoint representation for the smaller groups
along the Higgsing chain.  In 6D F-theory models, $SU(N)$ matter in
the adjoint representation is only possible when the gauge group
arises on a curve of higher genus in the F-theory base $B_2$.  This is
also true for $SU(N)$ models containing any representation other than
the fundamental and $k$-fold antisymmetric tensor representation
\cite{Sadov, kpt, mt-singularities}.  For F-theory models on $\F_m$
dual to smooth heterotic models, the gauge group factors live on
divisors of self-intersection $\pm m$.  The divisors of
self-intersection $-m$ with $m > 0$ are rigid and cannot support a higher
genus curve.  The divisors of self-intersection $+ m$, however, can be
taken with higher multiplicity, giving a higher genus Riemann surface.
For example, in $\F_2$ there are irreducible curves of genus one that
have twice the divisor class of the irreducible curve $\tilde{S}$ of
self-intersection $+ 2$.  An $SU(3)$ gauge group with an adjoint
results, corresponding to an $SU(3)$ factor in the dual heterotic
model with the non-standard commutant.  Note that for $m \neq 0$ this
can only happen on one side, so only one gauge group can have adjoint
matter representations and lie on the top line of
Figure~\ref{fig:chain}, corresponding on the heterotic side to the
fact that only bundles with $c_2 (V)> 12$ can have a structure group
such as
$G_2 \times SU(3)$.  Another interesting feature that can arise in
this context is the appearance of product gauge groups as one of the
factors $G_i \subset E_8$ in a heterotic model.  On the F-theory side
this corresponds again to a gauge group on a multiple of the divisor
class with positive self-intersection, now given as a sum of two
irreducible parts.

Constraints on heterotic bundles dual to F-theory models on ${\cal
  B}_2 = \F_n$ are particularly strong for $n=9,\ldots 12$, in which
cases the generic symmetry of $Y_3$ is $E_8$ (located on the patch
containing $z_1=0$ in $\mathbb{F}_n$). This corresponds on the
heterotic side to no structure group at all. That is, $H$ is trivial
and full $E_8$ symmetry is unbroken. 
%More precisely, it can be concluded that
In these cases, no smooth vector
bundles exist on $K3$ with $c_2=1,2,3$.  On the F-theory side this
corresponds to the fact that the $-9, -10$ or $-11$ curve in $\F_m$
must contain a point where the elliptic fibration is so singular that
the point in the base must be blown up for the total space to have the
structure of a Calabi-Yau threefold \cite{mt-clusters}.  On the
heterotic side this blowing up
corresponds to the shrinking of an instanton to a
point; in these cases there are {\it sheaves} (not locally free) with the desired
topology.  These cases go outside the smooth heterotic/F-theory dual
paradigm that we focus on here; we encounter 4D analogues of these
situations later but do not study them in detail.
%, however, there can be {\it sheaves} with this
%topology. 
F-theory geometry and heterotic/F-theory duality thus lead to the inclusion of small
instantons and sheafy degenerations in the heterotic picture, making
it clear that the structure of the physical theory is in agreement
with the mathematical notion of bundle/sheaf moduli spaces.
Mathematically, any attempt to construct a moduli space of bundles
alone results in a non-compact space.  It is only by including sheaves
({\it i.e.}, degenerations in the vector bundle) that a compact moduli space
${\cal M}_{\omega}$ arises. Physically, in the heterotic theory these
sheaves\footnote{More precisely, skypscraper sheaves supported over
  points in the $\mathbb{P}^1$ base of $K3$
  \cite{Aspinwall:1997ye,Aspinwall:1998he}.}  correspond to point-like
instantons (NS/M5 branes) on the K3 \cite{Morrison-Vafa-I,
  Morrison-Vafa-II}, in the cases above with ``instanton number''
$c_2=1,2,3$. Thus for $0 < c_2 < 4$ we note that the moduli space
${\cal M}_{\omega}$ is non-empty but contains only sheaves and no
smooth bundles.  In principle, this line of development could be used
to develop a full description of the 6D moduli space of non-perturbative vacua from the
heterotic side of the duality which could be matched to the geometric
F-theory description, though we do not pursue this further here.

%In this context, the heterotic ``small instanton transition'' has a
%beautiful realization via geometric transitions between the F-theory
%Calabi-Yau threefolds. In a heterotic small instanton transition, a
%smooth vector bundle on $K3$ with $c_2=k$ can be ``dissolved'' into $k$
%point like instantons\footnote{See \cite{XXX} for a description in
%  terms of the ADHM construction \cite{XXX} of sheaves.} in the
%language of heterotic M-theory \cite{XXX} then, such an instanton can
%be pulled off of one $E_8$ fixed planes into the $11$-dimensional
%$S^1/{\mathbb{Z}_2}$ interval where it becomes an $M5$-brane (visible
%as a tensor multiplet in the $6d$ theory). If this $M5$-brane is then
%dissolved onto the other $E_8$ fixed plane, we reach a new
%perturbative heterotic background with bundles $(V'_1, V'_2)$ with
%$c_2(V'_1)=k-1$ and $c_2(V'_2)=24-k+1$. In F-theory, this same
%transition is realized by a topology changing (conifold \cite{XXX})
%transition in the $3$-fold geometry taking $Y_3 \to Y'_3$. Dissolving
%a bundle into a $5$-brane corresponds to blowning up a point in the
%Hirzebruch base of $Y_3$. In many cases it is possible to blow-down
%these points again to obtain a different Hirzebruch base
%$\mathbb{F}_{n'}$ (equivalent to a perturbative heterotic theory
%associated to the new bundles $(V'_1, V'_2)$). An important
%observation here is that for a bundle with a given $c_2 \geq 4$, it is
%always possible to ``dissolve'' the bundle into small instantons and to
%perform a small instanton transition of the type described above.''

\subsubsection{Summary of 6D duality and relevance for 4D}

To summarize this review of the 6D story, in $6$ dimensions
heterotic/F-theory duality encodes a deep and non-trivial
correspondence between the moduli space of elliptically fibered
Calabi-Yau threefolds (and their stable degeneration limits) and the
moduli space of stable sheaves over $K3$.  In six dimensions,
essentially all of the information that can be inferred from the
F-theory geometry about the heterotic bundle moduli space, including
a) the irreducibility of ${\cal M}_{\omega}(r,c_1,c_2)$ and b) the
existence of a ``maximal'' rank/structure group for a given $c_2$, can
be independently determined using known mathematics to study the
heterotic geometry. Both of the facts, a) and b) were previously known
in the mathematics literature in the study of moduli spaces of stable
sheaves on $K3$ and Donaldson-Thomas invariants on $K3$ (see for example \cite{huybrechts_lehn}).

%with the geometry shown in Figure \ref{fig:geom}. 
In this paper, we ask many of the questions described above in the
context of 4D heterotic/F-theory duality, which relates the moduli
space of vector bundles $(V_1,V_2)$ on an elliptically fibered
Calabi-Yau threefold to the moduli of a $K3$ fibered Calabi-Yau
four-fold.  
%\begin{figure}
%\centering
%\includegraphics[width=3in, angle=90]{geom2.pdf}
%\caption{\emph{Add better figure here.}} \label{fig:geom}
%\end{figure}
%\lara{Do we need a figure like Figure 1? Or shall we just omit it entirely?}
In this context, however, the information obtained is made more
significant by the fact that far fewer mathematical techniques are
known for determining the moduli space of stable sheaves on Calabi-Yau
threefolds. Indeed, aside from a handful of examples with special
topology (see e.g. \cite{ellingsrud,Qin,Li_Qin,Li_Qin2,Friedman_Qin,Anderson:2010ty}), no systematic tools exist for constructing the
moduli spaces ${\cal M}_{\omega}({\rm rank}, c_1, c_2, c_3)$ or the
corresponding Donaldson-Thomas Invariants \cite{dt_original}. 

In the following sections we use heterotic/F-theory duality to
develop analogous statements to those made above for bundles on $K3$,
and explore a number of new features, unique to the 4D theory. These
include deriving strong upper and {\it lower} bounds (based on
$c_2(V)$)  on the bundle structure group $H$, as well as
constraints tying the matter spectrum of the theories to
topology.

\section{Heterotic/F-theory duality in four dimensions}
\label{sec:duality}

We now briefly review the duality between heterotic string and F-theory
compactifications for four-dimensional ${\cal N} = 1$ supergravity
theories \cite{Friedman-mw, Curio:1998bva}. We begin with a general abstract
formulation of the duality
in \S\ref{sec:duality-geometry}, 
and then characterize the possible compactification geometries that
are smooth on both sides of the duality
in \S\ref{sec:duality-bases}.
In \S\ref{sec:duality-toric}, we give a more detailed description of
constructions that involve a toric base surface $B_2$.

\subsection{Geometry of heterotic/F-theory duality}
\label{sec:duality-geometry}

We focus on the best understood class of dualities, in which the heterotic
compactification is on a smooth Calabi-Yau threefold $X_3$ that is
elliptically fibered with a single section over a base $B_2$, the
heterotic bundles are smooth and irreducible and there are no additional 5-branes wrapping curves in the base. In the dual F-theory
compactification, we consider a threefold ${\cal B}_3$ that is a $\P^1$ bundle
over $B_2$.  The F-theory compactification space ${\cal B}_3$ in turn acts as
a base for an elliptically-fibered Calabi-Yau fourfold $Y_4$.
Following \cite{Friedman-mw}, we
can construct the $\P^1$ bundle ${\cal B}_3$ 
as a
projectivization of a sum of two line bundles
\begin{equation}
{\cal B}_3 = \P ({\cal O} \oplus{\cal L}) \,,
\label{eq:m}
\end{equation}
where ${\cal L}$ is a general line bundle  on the base $B_2$.
On ${\cal B}_3$ we have the classes $R = c_1 ({\cal O} (1)), T = c_1 ({\cal
  L})$, where ${\cal O} (1)$ is a bundle that restricts to the usual
${\cal O} (1)$ on each $\P^1$ fiber.  There are sections $\Sigma_-$
and $\Sigma_+ = \Sigma_-+ T$ of  ${\cal B}_3$ that
satisfy $\Sigma_- \cdot \Sigma_+ = 0$, corresponding to the relation $R
(R + T) = 0$ in cohomology.

An F-theory model on ${\cal B}_3$ is dual to a heterotic model on  $X_3$.
For the $E_8 \times E_8$ heterotic theory, the bundle
decomposes as $V_1 \oplus V_2$,  and the curvatures split as (see Appendix \ref{sec:app_identities})
\begin{equation}
\frac{1}{30}  {\rm Tr}\; F_i^2 = \eta_i \wedge \omega_0 + \zeta_i \,,
\;\;\;\;\; i = 1, 2
\label{eq:f2-decomposition}
\end{equation}
where $\eta_i, \zeta_i$ are (pullbacks of) 2-forms and 4-forms on
$B_2$ and $\omega_0$ is Poincar\'{e} dual to the section.  The Bianchi
identity gives $\eta_1 + \eta_2 = 12c_1 (B_2)$.  Heterotic/F-theory
duality is possible when
\begin{equation}
\eta_{1, 2} = 6c_1 (B_2)\pm T \,, \;\;\;\;\;
(E_8 \times E_8) \,.
\label{eq:eta}
\end{equation}
This correspondence between $\eta_i$
and $T$ was identified by Friedman, Morgan, and Witten for
bundles in
the stable degeneration limit in \cite{Friedman-mw}.  It was shown
more generally in
\cite{Grimm-Taylor} that this correspondence follows directly from the
structure of axion-curvature squared terms in the dimensionally
reduced supergravity action, independent of the stable degeneration
limit or type of bundle construction.
For the $SO(32)$ heterotic string, the analysis \cite{Grimm-Taylor}
of the axion-curvature
squared terms in the 4D supergravity action constrains the twisting
$T$ of the bundle on the F-theory side to satisfy
\begin{equation}
 T = 2c_1 (B_2) \,,\;\;\;\;\;
(SO(32))
\label{eq:n}
\end{equation}
for a dual $SO(32)$ heterotic compactification to exist.
This generalizes the corresponding  6D case where the $SO(32)$
heterotic theory
is dual to F-theory on $\F_4 \; (n = 4
\Rightarrow T = 4H = 2c_1 (B_1 = \P^1))$.

Note that we assume that the elliptic fibration on the heterotic
side has precisely one section.  It is possible that fibrations with
more than one independent section
({\it i.e.}, with nontrivial
Mordell-Weil group)
-- or with multisections -- may admit some more general kind of F-theory
dual.  We leave this interesting question for further work.
On the F-theory side, elliptic fibrations without a global section
were explored in
\cite{Braun:2014oya}, and included into the moduli space of 
Weierstrass models in \cite{Morrison:2014era}.
As described in \cite{Morrison:2014era}, such models can be understood from
Higgsing abelian $U(1)$ symmetries, which can in turn be understood
from Higgsing nonabelian symmetries, all of which should have a clear
parallel between the F-theory and heterotic descriptions, though we do
not pursue this here.

For elliptic fibrations $X_3$ with one section and smooth total space,
not all topological features of the heterotic bundle are
determined by the dual F-theory geometry.  Knowing the base $B_2$ and
the twist $T$ of the dual F-theory $\P^1$ bundle
({\it i.e.}, knowing ${\cal B}_3$) allows for the
identification of all components of $c_2 (V_i)$ except for $\zeta_i$
in (\ref{eq:f2-decomposition}); these components  together satisfy $\zeta_1 +
\zeta_2 = 11c_1 (B_2)^2 + c_2 (B_2)$ (Appendix
\ref{sec:app_identities}, \cite{Friedman-mw}).  The F-theory
fourfold geometry also does not fix $c_3(V_i)$; 
for this we must consider in addition $G$-flux on $Y_4$.  
These features each correspond to one
undetermined parameter on the heterotic side.  It is interesting that
many of the consequences that can be derived from F-theory for the structure
of heterotic bundles -- discussed later in this paper --
are largely independent of any possible freedom in these two parameters on the
heterotic side.

A central piece of information on which we focus  in our analysis is
the generic gauge group in the low-energy 4D supergravity theory
corresponding to a given heterotic/F-theory dual pair.  On the
F-theory side, the threefold base ${\cal B}_3$ defined by $B_2$ and
$T$ supports an elliptic fibration that may have singularities along
certain divisors.  The Kodaira classification of such singularities
indicates the presence of nonabelian gauge group factors in the 4D
supergravity theory \cite{Morrison-Vafa-I, Morrison-Vafa-II}.  There
can only be a smooth heterotic dual when the only nonabelian gauge
group factors are associated with the divisors $\Sigma_-, \Sigma_+$.
For any given base ${\cal B}_3$, there is therefore a minimal gauge
group $G = G_1 \times G_2$, corresponding to singularity structures
present over $\Sigma_-, \Sigma_+$ in a completely generic elliptic
fibration over ${\cal B}_3$.  When such a gauge group is present, it
implies that the largest possible structure group for a bundle on the
heterotic side over $B_2$ with the topological data $\eta_{1, 2}$
fixed by \eq{eq:eta} is $H = H_1 \times H_2$ where the commutant of
$H_i$ in $E_8$ is $G_i$.  The singularity structure of the F-theory
elliptic fibration is determined in terms of a Weierstrass model $Y^2
= X^3 + f X + g $ over ${\cal B}_3$, parameterized by $f$ and $g$,
which are as before
sections of line bundles ${\cal O} (-4K), {\cal O} (-6K)$ where $K$ is the
canonical class of ${\cal B}_3$.  The minimal
group factors $G_i$ are
determined by the minimal degrees of vanishing of $f, g$ as listed in
Table~\ref{t:Kodaira}, and in some cases distinguished by monodromy
around the singular divisor \cite{Morrison-Vafa-II, Bershadsky:1996nh,
  kmss}.  Note that the Kodaira singularity dictates the physical
gauge algebra only; different theories may have gauge groups that
differ by a discrete factor that does not affect the algebra.  In much
of this paper we are somewhat cavalier about the distinction between
gauge algebra and gauge group, but the reader should keep in mind that
in most cases the only structure fixed by the local singularity
structure of the F-theory geometry is the gauge algebra.
  
The only gauge factors listed in Table~\ref{t:Kodaira} are those that
can be forced to arise from the structure of the F-theory base
threefold, independent of the choice of elliptic fibration, in cases
with a smooth heterotic dual.  This is analogous to the generic gauge
groups for 6D F-theory models over Hirzebruch surfaces $\F_m$
\cite{Morrison-Vafa-I}, with for example a generic gauge group of
$E_6$ over the curve of self-intersection -6 in $\F_6$.  For more
general F-theory models ({\it i.e.}, those without smooth heterotic
duals) there can be more complicated minimal gauge groups, some
involving multiple gauge group factors -- in analogy to the the
general set of structures arising in maximally Higgsed 6D F-theory
constructions, which can contain ``non-Higgsable'' matter
\cite{mt-clusters}.  In general, the gauge group in a particular model
may be larger than the minimum group dictated by the structure of the
base.  For example, over some F-theory base threefolds it is possible
to tune the Weierstrass coefficients $f, g$ to have an $A_4$
singularity corresponding to a gauge factor $SU(5)$ over certain
divisors, though this group does not arise as an automatic consequence
of the geometry of any base corresponding to a singularity that arises
in the generic elliptic fibration over that base.  While a wide
variety of models with different gauge groups can be tuned over each
base, we are focused here on the minimal gauge group for each base.

\begin{table}
\begin{center}
\begin{tabular}{|cc |c |c |}
\hline
deg $f$ & deg $g$ &  ${\cal G}_i$ & ${\cal H}_i $\\
\hline
1 & 2 &  $\gsu_2  $&$ \ge_7  $\\
2 & 2 &  $\gsu_3,\gsu_2 $ &$ \ge_6,\ge_7  $\\
2 & 3 &  $\gso_8,\gso_7,\ggg_2 $ &$ \gso_8,\gso_9,\gf_4 $\\
3 & 4 &  $\ge_6,\gf_4 $ &$ \gsu_3,\ggg_2 $\\
3 & 5 &  $\ge_7 $ &$ \gsu_2 $\\
4 & 5 & $ \ge_8 $ & {\rm trivial}\\
\hline
\end{tabular}
\end{center}
\caption[x]{\footnotesize  
The gauge algebra summands associated with
group factors $G_i$ arising
in 4D supergravity theory from divisors on which Weierstrass
parameters $f, g$ vanish to various degrees, and the
associated structure
group factors for dual heterotic bundles.}
\label{t:Kodaira}
\end{table}

In a number of places we will need to know the precise gauge algebra
associated with given degrees of vanishing of $f, g$ over the divisors
$\Sigma_\pm$, including the effects of monodromy.  As described in
\cite{Bershadsky:1996nh, kmss}, this can be determined by performing
an 
expansion $f = f_0 + f_1 z + f_2z^2+ \cdots, g = g_0 + g_1 z + g_2z^2
+ \cdots$ around the divisor $D$ of interest, where $z$ is an
algebraic coordinate that vanishes on $D$.  
%Because our results are
%different than those appearing elsewhere in the literature, we list
%them here: 
Since the precise conditions that determine the monodromy are
expressed differently in various places in the literature, we collect
here a succinct summary of the possible situations.
When ${\rm deg}~f = 2, {\rm deg}~g = 2$, the gauge algebra
is $\gsu_3$ when $g_2$ is a perfect square, and $\gsu_2$ otherwise.
Similarly, when ${\rm deg}~f = 3, {\rm deg}~g = 4$ the algebra is
$\ge_6$ when $g_4$ is a perfect square, and $\gf_4$ otherwise.  The
case ${\rm deg}~f = 2, {\rm deg}~g = 3$ is somewhat more complicated;
in this case, the algebra depends on the factorization properties of
the cubic $X^3 + f_2X + g_3$.  If this cubic can be algebraically
factorized into a product of three terms to the form $ (X-a) (X-b) (X
+ (a + b))$ then the gauge algebra is $\gso_8$, if it factorizes into
the form $(X-a) (X^2 + aX + b)$ the algebra is $\gso_7$, and if it
does not factorize algebraically, the algebra is $\ggg_2$.  We use
these conditions in the analysis in several places in the remainder of
the paper.

It is also worth noting that in the cases $T = 2c_1 (B_2)$ that admit
an $SO(32)$ heterotic dual, the twist $T$ -- and hence the base ${\cal
  B}_3$ on the F-theory side -- is fixed uniquely for any $B_2$.  The
corresponding minimal gauge group is always $\gso_8$, matching with the
expectation from the heterotic side.
This is shown in \S\ref{sec:constraint.32}

\subsection{Possible base surfaces for smooth heterotic/F-theory duals}
\label{sec:duality-bases}

There are only a limited class of bases $B_2$ over which the elliptic fibration
geometry
is a smooth Calabi-Yau on the heterotic side.  The set of
complex base surfaces over which an elliptically fibered Calabi-Yau
threefold exists can be classified according to the intersection
structure of effective divisors on the base \cite{mt-clusters}.  When
the base contains a curve of self-intersection $-3$ or below, the
total space of the elliptic fibration becomes singular, and in general
the heterotic theory acquires an enhanced gauge group.  
Note that there are some
special cases where an F-theory construction on a $\P^1$ bundle
over a base surface with $-3$ curves can apparently exist without an
extra nonabelian gauge group, though there is still no smooth
heterotic dual in such cases as the dual Calabi-Yau geometry would be
singular; an example of such a model is given in
\S\ref{sec:examples.3-no-group}.
Our
restriction to models with smooth heterotic duals means that we limit
our analysis here to bases $B_2$ that only have effective curves of
self-intersection $-2$ or above.  

The set of bases $B_2$ that contain no curves of self-intersection
$-3$ or below consist of the \emph{del
  Pezzo} surfaces $dP_n$, given by $\P^2$ blown up at $n \leq 9$
points,  the bases
$\F_0 =\P^1 \times \P^1$ and $\F_2$ (which is a limit of $\F_0$ with a
$-2$ curve),
and the broader class of \emph{generalized del Pezzo} surfaces, which are in
general described as limits of del Pezzo surfaces containing curves of
self-intersection $-2$ \cite{777}.
For each $n,1 <n < 9$, there are generalized del Pezzo surfaces
corresponding to limits of $dP_n$ with a set of $-2$ curves having an
intersection structure corresponding to any proper subgraph of the
extended Dynkin diagram $\hat{E}_{n}$ \cite{pinkham}.  For $n = 9$, the
classification is slightly more complicated.  There are 279 rational
elliptic surfaces with different combinations of $-2$ curves,
corresponding to generalized del Pezzo surfaces with $n = 9$; these
surfaces are classified in \cite{Persson, Miranda-Persson}.  Over each
of the possible base surfaces $B_2$ there are a
wide range of possible twists $T$ giving different geometries on the
F-theory side.  Each such geometry will correspond to a different
class of bundles on the heterotic side on  the Calabi-Yau describing an
elliptic fibration over $B_2$.  In principle, all possible F-theory
bases ${\cal B}_3$ with a smooth heterotic dual can be classified by
determining all allowed twists $T$ for each del Pezzo and generalized
del Pezzo (and for $\F_0 = \P^1 \times \P^1$).

\subsection{Heterotic/F-theory duals with toric base surfaces}
\label{sec:duality-toric}

A particularly simple class of bases $B_2$ can be described using
toric geometry.  In \cite{mt-toric}, the complete set of toric
bases for elliptic threefold fibrations was enumerated.  Here we are only interested in those
cases where all effective curves have self-intersection $-2$ or above,
which restricts us to only 16 possible bases: the del Pezzo surfaces
$dP_n$ with $0 \leq n \leq 3$, the surfaces $\F_0$ and $\F_2$, and 11
other toric generalized del Pezzo surfaces with various combinations
of curves of self-intersection $-2$.  These 16 bases are listed in
Table~\ref{t:table}.

Following standard methods in toric geometry \cite{Fulton,
  Knapp-Kreuzer}, we characterize the base $B_2$ by the toric fan,
consisting of vectors $v_0, \ldots, v_{k +1} \in N_2 =\Z^2$.  
This describes a base with $h_{1, 1} (B_2) = k$.
We can
choose a basis in which $v_0 = (1, 0)$ and $v_{k +1} = (0,1)$.
The coordinates of the remaining vectors
defining $B_2$ can be written as
$v_i = (x_i, y_i)$ in this basis.
Any toric $\P^1$ bundle over $B_2$ can then be described  in terms
of a 3D toric fan
\begin{eqnarray}
w_0 & = &  (1, 0, 0)\label{eq:w0}\\
w_i & = &  (x_i, y_i, t_i), \; 1 \leq i \leq k
\label{eq:wi}\\
w_{k +1} & = &  (0, 1, 0)\\
w_{k + 2, k + 3} =
s_\pm & = &  (0, 0, \mp 1) \label{eq:ws}\,,
\end{eqnarray}
where $s_\pm$ correspond to the divisors $\Sigma_\pm$, and $w_a, 0
\leq
a
\leq k +1$, are 3D rays that project to $v_a$, with the third component
$t_i$ parameterizing the twist $T$ that defines the $\P^1$ bundle.  
The vanishing of the third component of $w_0, w_{k +1}$ is a
coordinate choice on $N_3 =\Z^3$ used to eliminate two redundant
degrees of freedom in the twist $T$.
For convenience we will use indices $i\in\{1, \ldots, k\}, a \in\{0,
\ldots, k +1\},$ and $\alpha \in\{0, \ldots, k + 3\}$, and write $r =
k + 4$ for the total number of rays generating the 3D toric fan.

The toric language gives a simple description of the monomials
available in the Weierstrass description of the F-theory model.
If the set of 1D rays
describing the toric threefold ${\cal B}_3$ are $w_\alpha, \alpha = 0,
\ldots, r-1$, then the monomials in $f$ are in one-to-one
correspondence with the elements $m = (a, b, c)\in M = N^*$ of the dual lattice
whose inner product with all of the $w_\alpha$ is not less than -4,
\begin{equation}
{\cal F} =\{ m \in N_3^*:
\langle m, w_\alpha \rangle \geq -4 , \forall \alpha\}\,.
\label{eq:b}
\end{equation}
Similarly, the monomials in $g$ are associated with
\begin{equation}
{\cal  G} =\{ m \in N_3^*:
\langle m, w_\alpha \rangle \geq -6 , \forall \alpha\}\,.
\label{eq:c}
\end{equation}
As a simple example, for the case ${\cal B}_3 =\P^1 \times\P^1
\times\P^1$, which is the trivial $\P^1$ bundle over $\F_0 =\P^1
\times \P^1$, the rays $w_\alpha$ are the basis vectors $(\pm 1, 0,
0), (0, \pm 1, 0), (0, 0, \pm 1)$, and the monomials in $f, g$ are the
triplets $(a, b, c) \in \Z^3$ with $| a |, | b |, | c | \leq 4, 6$.

For a toric F-theory base, we can compute the
(anti)canonical class of $B_2$ directly from the toric description.
There are two equivalence relations on the set of divisors
$D_a$ associated with the rays $v_a$, giving
\begin{eqnarray}
D_0 & \sim &  \sum_{i = 1}^{k}-x_i D_i \\
D_{k +1} & = &  \sum_{i = 1}^{k}  -y_i D_i \,.
\end{eqnarray}
We have then
\begin{equation}
-K_2 = c_1 (B_2) = \sum_{a = 0}^{k +1}  D_a
= \sum_{i}(1-x_i-y_i) D_i \,.
\label{an_equation}
\end{equation}

Similarly, we have
\begin{equation}
-K_3 = 2 \Sigma_- -K_2 + T \,,
\label{eq:g}
\end{equation}
where $T = \sum_{}^{}t_iD_i $.
As discussed below for more general F-theory geometries with a
heterotic dual, the formula (\ref{eq:g}) also follows straightforwardly from the
definition of ${\cal B}_3$ given in \eref{eq:m} and the adjunction
formula \cite{Friedman-mw}.

\section{F-theory constraints}
\label{sec:F-theory-constraints}

In this section we describe the geometric constraints on $\P^1$-bundle
threefold bases in the class of models with smooth heterotic duals.
In Section \ref{sec:geometry-constraints} we describe constraints on
the threefold geometry, and in Section
\ref{sec:enhancement-constraints} we describe further constraints on
the gauge group of the corresponding 4D supergravity theory and the
extent to which it can be enhanced through ``unHiggsing'' by moving on
the Calabi-Yau moduli space.

\subsection{Constraints on threefold  base geometry}
\label{sec:geometry-constraints}

The basic conditions on an F-theory threefold base geometry ${\cal
  B}_3$ are that there are no codimension one or codimension two loci
with singularities  worse than the $\ge_8$ singularity in the Kodaira classification.  These
conditions can be described in terms of constraints on the base
geometry $B_2$ and twist $T$ describing ${\cal B}_3$ as a $\P^1$
bundle over $B_2$.  We begin (\S\ref{sec:constraints-general})
with a brief overview of the general F-theory constraints, which are
easy to make explicit in the toric context
(\S\ref{sec:constraints-toric}).  In \S\ref{sec:F-theory-twists}, we
use the general conditions to derive a  set of local
constraints on the twist $T$ associated with specific divisors in the
base.  The toric description of these constraints is given in
\S\ref{sec:twist-constraints}.    In
\S\ref{sec:constraints-global}, we derive a simple set of necessary
conditions associated with the divisors $\Sigma_\pm$.  Combined with
the constraints on $T$, this gives a set of
conditions that are
necessary, but not sufficient,  for the existence of a good
F-theory compactification geometry.
In \S\ref{sec:constraints-curves}, we include more general,
nonlocal conditions associated with curves in ${\cal B}_3$, which give
a set of sufficient conditions for an acceptable F-theory model,
subject to issues from codimension three singularities and G-flux that
we do not address here (see
\S\ref{chiralsec} for some relevant aspects of G-flux for 4D F-theory compactifications).

In the analysis in this section we repeatedly use a basic result from
algebraic geometry, which states that if an effective divisor $A$ on a
surface $S$ has a negative intersection $A \cdot D < 0$ with an
irreducible effective divisor $D$ having negative self-intersection $D
\cdot D < 0,$ then $A$ {\it contains $D$ as a component}, meaning that
$A = D + X$ with $X$ effective.  This means in particular that any section
$s\in{\cal O} (A)$ must vanish on $D$.  This result was used in
\cite{mt-clusters}  to identify the ``non-Higgsable clusters'' that
classify the intersection structure of base surfaces ${\cal B}_2$ for
6D F-theory compactifications.  More generally, an effective divisor 
may contain a number of rigid divisors $D_i$ with multiplicity
$\gamma_i$ by repeated applications of the preceding rule
\begin{equation}
A = \sum_{i}\gamma_iD_i + X, \;\;\;\;\; (X \; {\rm effective}) \,.
\label{eq:Zariski}
\end{equation}
When such a decomposition is carried out over the rational numbers
$\gamma_i \in\Q$, it is called the {\it Zariski decomposition} of
$A$.  While in higher dimensions the Zariski decomposition can be more
subtle, for surfaces the computation of the terms $\gamma_iD_i$, known
as the {\it base locus} of $A$ is straightforward.  For example, if
$D$ is a curve of self-intersection $D \cdot D = -2$, and $A \cdot D =
-4$, then $A = 2D + X$ with $X$ effective, $X \cdot D = 0$.

In a number of places in this section we focus on curves in ${\cal
  B}_3$ of the form $C = \Sigma_\pm\cap D$, where $D$ is a divisor on
${\cal B}_3$ pulled-back from a corresponding divisor in the base surface
$B_2$.  We will generally use $C$ for the curve in ${\cal B}_3$, while
$D$ can refer either to the divisor in ${\cal B}_3$ or in $B_2$,
depending on context.

\subsubsection{General constraints from F-theory geometry}
\label{sec:constraints-general}

We begin with a general statement of the F-theory constraints that
hold for any geometry.
For a good F-theory model to exist on a base ${\cal B}_3$ there must be a
Calabi-Yau fourfold that is elliptically fibered over ${\cal B}_3$
\cite{Vafa-F-theory, Morrison-Vafa-I, Morrison-Vafa-II}
%\footnote{More
%  generally, F-theory models can exist for genus one fibrations
%  without section \cite{Braun:2014oya}, but we restrict here to
%  elliptic fibrations.}.  
% covered earlier.
As described in \S\ref{sec:duality-geometry},
when the Weierstrass coefficients $f, g$ and the discriminant 
$\Delta = 4f^3 + 27g^2$ vanish  on a divisor in ${\cal B}_3$, the 
corresponding
4D supergravity theory
gets a nonabelian gauge group contribution depending upon the Kodaira
type of the corresponding singularity in the elliptic fibration.
When the vanishing degrees of $(f, g, \Delta)$ reach or exceed (4, 6,
12) on a divisor, the fibration becomes too singular to admit a Calabi-Yau
resolution.  Thus, a constraint on ${\cal B}_3$ is that $-nK_3$ must admit a
section of vanishing degree $< n$ for $n = 4$ or 6 on any
irreducible effective divisor
$D$.  Similarly, $f, g$ cannot vanish to orders $4, 6$ on any curve,
or the curve would need to be blown up, giving a different base
structure, for a Calabi-Yau resolution of the singular elliptic
fibration to exist.  This provides a
strong set of constraints on bases ${\cal B}_3$
that admit good F-theory models.  The constraint on codimension three
loci (points) on the base is less clear; if the degrees of $f, g$
reach $8, 12$ on a point then the point must be blown up for a good
Calabi-Yau resolution.  On the other hand,  if the degrees of
vanishing reach $4, 6$ on a codimension three locus but do not exceed
$8, 12$ then the model may be problematic yet cannot be blown up
directly \cite{Morrison-codimension-3}.  We focus in this paper on the
constraints associated with codimension one and two loci, associated
with gauge groups and matter content in the low-energy theory.
We include therefore in our analysis models with codimension 3
singularities, leaving the resolution of the status of these models to
future work.
Codimension three singularities are discussed further in
\S\ref{chiralsec}.

\subsubsection{Constraints for toric bases}
\label{sec:constraints-toric}

The F-theory constraints described above are particularly simple to
describe for toric F-theory bases ${\cal B}_3$
using the explicit description of the Weierstrass monomials as
elements of the dual lattice, as described in the previous section.
\footnote{Related constraints in the toric language of ``tops'' were
  described in \cite{Braun:2013nqa}.}
The degrees of vanishing of $f, g$ on the divisor
$D_\alpha$ associated with the ray $w_\alpha$
are given by
\begin{equation}
{\rm deg}_{D_\alpha} f = {\rm min}_{m \in{\cal F}} \langle m, w_\alpha
\rangle + 4, \;\;\;\;\;
{\rm deg}_{D_\alpha}  g = {\rm min}_{m \in{\cal  G}} \langle m, w_\alpha
\rangle+ 6\,.
\label{eq:d}
\end{equation}
These are easily computed for any given base and divisor.
For a good F-theory base, these degrees cannot both reach or exceed
$4, 6$ on any divisor $D_\alpha$.  
When the degrees are nonzero, they indicate the presence of a generic
gauge algebra factor on $D_\alpha$, according to Table~\ref{t:Kodaira}.
The degrees of vanishing of $f, g$ on the toric curve $D_\alpha \cap
D_\beta$ are given by
\begin{equation}
{\rm deg}_{{D_\alpha} \cap D_\beta} f = {\rm min}_{m \in{\cal F}}
\langle m, w_\alpha + w_\beta \rangle + 8, \;\;\;\;\; {\rm
  deg}_{{D_\alpha} \cap D_\beta} g = {\rm min}_{m \in{\cal G}} \langle
m, w_\alpha + w_\beta \rangle+ 12\,.
\label{eq:e}
\end{equation}
Again, these degrees cannot both reach or exceed $4, 6$ on any curve
or the F-theory base must be blown up along that curve to give a new base.
When these degrees reach or exceed 1, 2 along a given curve they generally
indicate the presence of ${\cal N} = 2$ matter transforming under
gauge groups carried on the divisors $D_\alpha, D_\beta$, in analogy
to the 6D situation.  In some toric 4D
cases, however, there are no such nonabelian gauge
groups.
An example of such a situation is described explicitly
in \S\ref{sec:examples-fm}, with ${\cal B}_3$ 
a $P^1$ bundle over $B_2 = \F_1$.  Such
codimension two singularities may simply represent cusps where the
discriminant becomes singular, as occurs in 6D compactifications
(see {\it e.g.} \cite{Grassi-Morrison}), or
may in some cases represent matter charged under $U(1)$ gauge factors.

\subsubsection{F-theory bounds on twists}
\label{sec:F-theory-twists}

We now specialize to the class of F-theory geometries that have smooth
heterotic duals as described in Section \ref{sec:duality}.  In this case,
the relation (\ref{eq:g}) derived above in the toric context holds
more generally from the adjunction formula applied to ${\cal B}_3$ defined in \eref{eq:m}
\begin{equation}
-K_3 = 2 \Sigma_- -K_2 + T \,.
% \label{eq:}
\end{equation}
Writing the Weierstrass functions $f,
g$ locally in a region around the locus $\Sigma_-$ defined by the
coordinate $z = 0$, where $\Sigma_+$ is at $z = \infty$,
we have
\begin{eqnarray}
f & = &  f_0 + f_1 z + f_2z^2 + \cdots f_8z^8 \label{fdef1}\\
g & = &  g_0 + g_1 z + g_2z^2 + \cdots g_{12} z^{12} \label{gdef1}\,.
\end{eqnarray}
The term $f_kz^k$ vanishes to order $k$ on $\Sigma_-$ and to order
$8-k$ on $\Sigma_+$, and $f_k$ is a section of ${\cal O} (-4K_3- k
\Sigma_- -(8-k) \Sigma_+)$.  Similarly, $g_kz^k$ vanishes to order $k$
on $\Sigma_-$ and to order $12-k$ on $\Sigma_+$, and $g_k$ is a
section of ${\cal O} (-6K_3- k \Sigma_- -(12-k) \Sigma_+)$.
Thus, $f_k$ and $g_k$ are sections of
${\cal O} (-4K_2-(4-k)T)$
and
${\cal O} (-6K_2-(6-k)T)$ respectively.  We can use this fact to
determine constraints on the possible twists $T$ compatible with any
particular base $B_2$.

A set of necessary conditions on $T$  can be determined by
imposing the condition that $f, g$
should not vanish to orders 4, 6 on any curve $C$ in ${\cal B}_3$ that
is of the form $\Sigma_-\cap D$, where $D$ is associated with an
irreducible effective divisor on $B_2$.  We consider the various
possibilities depending upon the self-intersection of $D$ in $B_2$,
using the fact that $(K_2 + D) \cdot D = -2$ when $D$ is a rational
curve.
We focus on divisors $D$ 
with non-positive self-intersection
$D \cdot D \leq 0$, since determining conditions on these divisors is
sufficient to bound the total number of twists $T$ over any base $B_2$.
Note that, as shown in \cite{mt-clusters}, any higher genus divisor of
negative self-intersection in the base $B_2$ gives a singular elliptic
fibration that cannot be resolved to a Calabi-Yau, so it is sufficient
to restrict attention to rational curves $D$.

We begin with the case $D \cdot D = -2$, where $K_2 \cdot D = 0$.  Consider then the
intersection
\begin{equation}
(-nK_2-(n-k)T) \cdot D = -(n-k) T \cdot D, \; \; k < n\,.
\label{eq:fg-intersection}
\end{equation}
When $T \cdot D > 0,$ this intersection is negative,
and $(-nK_2-(n-k)T)$ contains $D$ as
a component for $n = 4, 6$ when $ 0 \leq k < n$.
This means that the sections $f_k, g_k$ must vanish on $D$.
If $T \cdot D \geq 2$, then  $f_k, g_k$ must vanish on $D$ 
at least
to order
$4-k, 6-k$ respectively, which would mean that $f, g$ vanish to
degrees 4, 6 on $C$.  Thus, the twist must satisfy $T \cdot D < 2$ for
  any rational curve $D$ in the base having self-intersection $D
  \cdot D =-2$.  
A
  similar argument for $\Sigma_+$ shows that
$T \cdot D > -2$, so $|T \cdot D | \leq 1$ for any $-2$ curve in
$B_2$.

Now, consider the case $D \cdot D = -1$.  In this case, $-K_2 \cdot D
= 1$, so we have $ (-nK_2-(n-k)T) \cdot D = n -(n-k) T \cdot D$.  This
is less than or equal to $-(n-k)$ for all $k <n = 4, 6$ when $T \cdot
D > 6$, which would force $f, g$ to vanish on $D \cap \Sigma_-$ to
degrees 4, 6, with a similar constraint with the opposite sign for
$\Sigma_+$, so we have a bound in this case of $| T \cdot D | \leq 6$.
(For example, if $T \cdot D = 6$, then $(-6K_2-T) \cdot D = 0$, so
$g_5$ need not vanish on $D$, though all other $f_k, g_k$ vanish on
$D$ to degree $n-k$.)

Similar reasoning shows that  analogous constraints
hold for curves of self-intersection 0 and for curves of
more negative self-intersection; the complete set of
constraints for the twist over any rational curve of
(non-positive) self-intersection $-n$
is
\begin{eqnarray}
n = 0: & \hspace*{0.1in} &  |T \cdot D | \leq 12 \label{eq:t-0}\\
n = 1: & \hspace*{0.1in} &  |T \cdot D | \leq 6\label{eq:t-1}\\
6 \geq
n  \geq 2: & \hspace*{0.1in} &  |T \cdot D | \leq 1 \,.\label{eq:t-2}
\\
n  \geq  7: & \hspace*{0.1in} &  T \cdot D  = 0 \,.\label{eq:t-big}
\end{eqnarray}
We only need the results for $n \leq 2$ in this paper.

These bounds provide strong constraints on the twists that are allowed
for a $\P^1$ bundle over any base $B_2$.  As promised in
\S\ref{sec:het_f_motivation},
these reduce the problem
of identifying all smooth F-theory bases ${\cal B}_3$ with smooth
heterotic dual geometries to a finite enumeration problem, since the
curves of negative or 0 self-intersection in the base $B_2$ form a
connected set.  We summarize the results of a complete enumeration of
all twists over toric bases with smooth heterotic duals
in \S\ref{sec:enumeration}.

\subsubsection{Toric bounds on twists}
\label{sec:twist-constraints}

\begin{figure}
\begin{center}
\includegraphics[width=10cm]{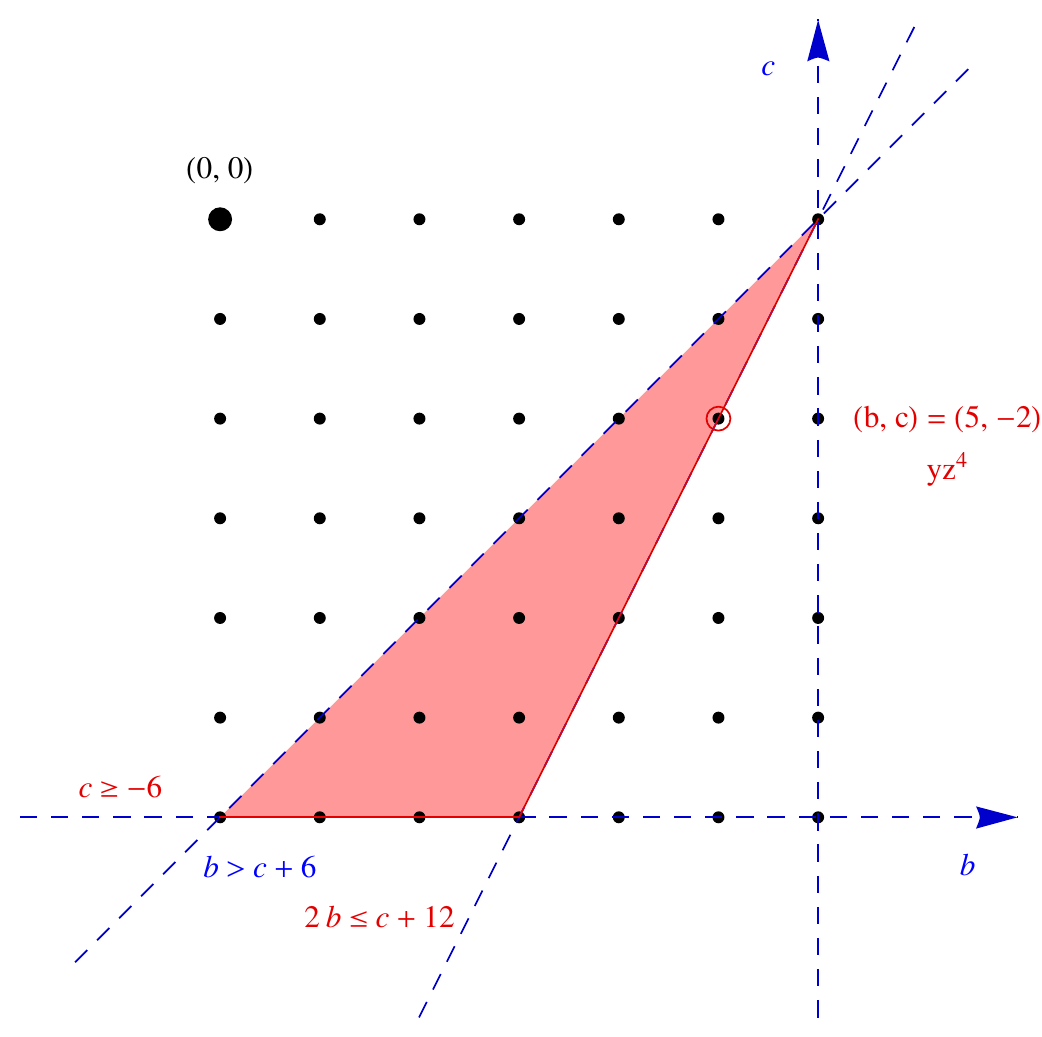}
\end{center}
\caption[x]{\footnotesize  Constraints imposed on dual monomials in
  the Weierstrass function $g$
  parameterized by $a,
  b, c$ in the toric description of a twist $\tilde{t}_i = 1$ over a
  curve of self-intersection $-n = -2$; the depicted constraints
on $b, c$ correspond to
  weakest conditions, which hold at $a = -B= -6$.  For a smooth
  F-theory geometry, at least one monomial in the shaded region (not
  including the boundary at $b > c + 6$, or the corresponding region
  for $f (B = 4)$ must be nonzero.  The circled   point
  $(5, -2)$ corresponds to the monomial $yz^4$ in coordinates where $z
  = 0$ corresponds to $\Sigma_-$ and $y = 0$ corresponds to
$D_i$.  This point is relevant in ruling out gauge algebra factors
  $\ge_7,\ge_8$ on $\Sigma_-$ under these conditions (\S\ref{sec:bpf}).)}
\label{f:twist-constraint}
\end{figure}

The bounds on twists can be seen explicitly in the toric context.
We can identify the bounds on the twist $t_i$ over the divisor
$D_i$ associated
with a given base ray $v_i$ by
considering the local geometry of the $\P^1$ bundle over the
sequence of rays $v_{i -1}, v_i, v_{i +1}$.  (If $i = 0$ or $i = k
+1$, we replace $i -1$ or $i +1$ with $k +1$ or $0$ respectively in
the obvious fashion to respect the cyclic ordering of rays).  We can
choose a basis for $N_2$ so that $v_{i -1} = (1, 0)$ and $v_{i} = (0,
-1)$ (note that this is a different choice of basis than that used in 
(\ref{eq:w0}-\ref{eq:ws})).  Associated with the three 2D rays
$v_{i -1}, v_i, v_{i +1}$ there are twists $t_{i -1}, t_i, t_{i +1}$,
associated with the extension of the corresponding 3D rays $w_{i-1},
w_i, w_{i +1}$
in the third
dimension (\ref{eq:wi}).  In the 3D toric lattice $N_3$ we can
perform a linear transformation taking
\begin{equation}
w_{i-1} =(1, 0, t_{i -1}) \rightarrow  \tilde{w}_{i-1} =(1, 0, 0), \;\;\;\;\;
w_i =(0, -1, t_i) \rightarrow \tilde{w}_i =(0, -1, 0) \,.
\label{eq:transformation}
\end{equation}
Since we are assuming that the base $B_2$ is smooth, the third ray has
the form $w_{i +1} = (-1, -n, t_{i +1})$, where  the integer $-n$ is the
self-intersection of the divisor $D_i$ \cite{Fulton}.  The linear
transformation (\ref{eq:transformation}) takes
\begin{equation}
w_{{i +1}} \rightarrow \tilde{w}_{i +1} =
(-1, -n, \tilde{t}_i), \; \; {\rm where} \;
\tilde{t}_i = t_{i-1} + t_{i +1} -nt_i  \,.
\label{eq:f}
\end{equation}
The parameter $\tilde{t}_i= T \cdot D_i$ determines the nontrivial
part of the twist around the ray $v_i$, and can be constrained
geometrically depending upon $n$, to reproduce the conditions
(\ref{eq:t-0}-\ref{eq:t-big}).
As in the general situation described in the previous subsection,
the strongest constraint on the twist component $\tilde{t}_i$
comes from the condition that
$f, g$ do not have degrees $4, 6$ on the curves associated with
$w_i+ s_\pm$.  Let us assume that $\tilde{t}_i \geq 0$.  Then we
have the following constraints on the monomials associated with
$m = (a, b, c)\in N_3^*$ for $f, g$ with $B = 4, 6$
\begin{eqnarray}
\langle m, \tilde{w}_{i -1} \rangle \geq -B & \; \; \; \rightarrow \;
\; \; & a \geq -B \label{eq:inequality-1}\\
\langle m, \tilde{w}_{i} \rangle \geq -B & \; \; \; \rightarrow \; \; \; & b  \leq B\label{eq:inequality-2}\\
\langle m, s_- \rangle \geq -B & \; \; \;
\rightarrow \; \; \; & c \geq -B \label{eq:inequality-3}\\
\exists \langle m, \tilde{w}_{i} + s_- \rangle   <  -B & \; \; \; \rightarrow \; \; \; & b
-c> B\label{eq:inequality-4}\\
\langle m, \tilde{w}_{i +1} \rangle \geq -B & \; \; \; \rightarrow \; \; \; &
-a-nb+\tilde{t}_i c \geq - B  \,. \label{eq:inequality-5}
\end{eqnarray}
All of these inequalities except \eq{eq:inequality-4} must be
satisfied for all monomials in ${\cal F},{\cal G}$.  Inequality
(\ref{eq:inequality-4}) on the other hand, need only be satisfied by
at least one point $m$, to avoid having a 4, 6 singularity on $w_i +
s_-$.  But this means that there must be at least one integral
point satisfying all of these inequalities.  The second through fourth
inequalities define a simple triangle in the $b$-$c$ plane within
which any solutions must lie (Figure~\ref{f:twist-constraint}).  The
final constraint imposes a condition that restricts the solutions
within this triangle.  This constraint is weakest when $a$ is
maximally negative, so if there are any solutions in $b, c$ for any
$a$ they will also be acceptable for $a = -B$.  Therefore, we need
only ask whether there can exist any solutions of the form $(-B, b,
c)$ to inequalities (\ref{eq:inequality-2}-\ref{eq:inequality-5}).
For $n = 0$, \eq{eq:inequality-5} becomes $\tilde{t}_ic \geq -2B$, and
since $c < 0$ from \eq{eq:inequality-4} and \eq{eq:inequality-2}, we
need $\tilde{t}_i \leq 2B$ for a solution of all the inequalities to
exist.  This is weakest for $B = 6$, so we have the constraint
$\tilde{t}_i \leq 12$ when $n = 0$.  For $n = 1$, \eq{eq:inequality-5}
becomes $b \leq \tilde{t}_i c + 2B$, which combined with
\eq{eq:inequality-4} becomes $c + B < \tilde{t}_i c + 2B$, so
$(\tilde{t}_i -1) (-c) < B$ and since $-c > 0$ we need $\tilde{t}_i
\leq 6$.  Finally, for $n = 2$, \eq{eq:inequality-5} becomes $2b \leq
\tilde{t}_i c + 2B$, which combined with \eq{eq:inequality-4} becomes
$2c < \tilde{t}_i c$, and since $c < 0$ this implies $\tilde{t}_i \leq
1$.  To summarize, we have reproduced the constraints
(\ref{eq:t-0}-\ref{eq:t-big}) in the toric context
\begin{eqnarray}
n = 0: & \hspace*{0.1in} &  |\tilde{t}_i | \leq 12 \label{eq:tt-0}\\
n = 1: & \hspace*{0.1in} &  |\tilde{t}_i | \leq 6\label{eq:tt-1}\\
n  \geq 2: & \hspace*{0.1in} &  |\tilde{t}_i | \leq 1 \,.\label{eq:tt-2}
\end{eqnarray}
The constraints on coefficients in $g$ $(B = 6)$ for $n = 2,
\tilde{t}_i = 1$ are shown in Figure~\ref{f:twist-constraint}.

\subsubsection{General constraints on geometry from $f, g$ on divisors}
\label{sec:constraints-global}

In \S\ref{sec:F-theory-twists} we used curves of the form $C
=\Sigma_\pm \cap D$ in the base ${\cal B}_3$ to determine bounds on
the individual components $T \cdot D$ of the twist $T$ parameterizing
the $\P^1$ bundle over $B_2$.  We now describe more general
constraints on $T$ associated directly with the divisors $\Sigma_\pm$.
Note that the divisors $D$ on the base cannot give further constraints
on $T$ since $f, g$ cannot vanish on $D$ to higher degree than they do
on the corresponding curve in $B_2$.  This follows from the fact that
$f_4, g_6$ are sections of $-4K_2, -6K_2$ respectively.  Thus, we need
only consider constraints associated with the degrees of vanishing of
$f, g$ on the divisors $\Sigma_\pm$.

As described in \S\ref{sec:F-theory-twists}, the components
$f_k, g_k$ of the discriminant locus that vanish to degree $k$
on $\Sigma_-$ are associated with sections of the line bundles
${\cal O} (-4K_2-(4-k)T)$
and
${\cal O} (-6K_2-(6-k)T)$ respectively.  In particular, $g_5$ is a
section of
${\cal O} (-6K_2-T)$.  If $-6K_2-T$ is effective,
then this line bundle admits a section, so generically $g_5$ is
nonzero and $g$ does not vanish to degree 6 on $\Sigma_-$.  On the
other hand, if $-6K_2-T$ is not effective then there are no sections
and $g_5 = 0$.  Furthermore, if $-6K_2-T $ is not effective
then $-6K_2-nt$ cannot be effective for $n \geq 2$, since
\begin{equation}
n (-6K_2-T) = -6K_2-nt + (n-1) (-6K_2)
% \label{eq:}
\end{equation}
and $-6K_2$ is effective, as is any positive combination of effective
divisors.  Similarly, if $-6K_2-T$ is not effective then $-4K_2-mt$
cannot be effective for $m \geq 1$ since
\begin{equation}
m (-6K_2-T) = -4K_2-mt + (6m-4) (-K_2) \,,
% \label{eq:}
\end{equation}
where again the last term is effective.
This shows that $-6K_2-T$ is effective if and only if $f, g$ do not
vanish to degrees 4, 6 on $\Sigma_-$.  A parallel argument shows that
$-6K_2+T$ is effective if and only if
 $f, g$ do not
vanish to degrees 4, 6 on $\Sigma_+$.

From these considerations we can distinguish several possible
configurations of allowed geometries on the F-theory side

\noindent {\bf A)} One possibility is that 
\begin{equation}
T = -6K_2\,.
% \label{eq:}
\end{equation}
  In this case
all coefficients of $f_k, g_k$ vanish up to but not including degrees
(4, 5).

\noindent {\bf B)} If $T \neq -6K_2$ then $g_5$ is not a constant and
must vanish on some curves $D$ in $B_2$.  To avoid having $f, g$ both
vanish to degrees 4, 6 on the associated curve $C = \Sigma_-\cap D$ in ${\cal B}_3$ at
least one other coefficient $f_{k \leq 3}, g_{k \leq 4}$ must be
nonvanishing.  But by a parallel argument to the above this means that
\begin{equation}
-4K_2-T \; \; {\rm is\ effective} \,.
% \label{eq:}
\end{equation}

\noindent {\bf C)} If there is any curve $D$ in the base with
self-intersection $D \cdot D = -2$, where the associated component of
the twist is $T
\cdot D= +1$, then by the analysis of Section
\ref{sec:F-theory-twists} we know that $f_3, g_5$ both vanish on
$D$.  This means that one of $f_{k \leq 2},g_{k \leq 4}$ must be
nonvanishing.
Again, a parallel argument to the above  means that in this case
\begin{equation}
\exists D:
D \cdot D = -2,
T \cdot D = +1 \;
\Rightarrow
\;
-3K_2-T \; \; {\rm is\ effective} \,.
% \label{eq:}
\end{equation}

Any good F-theory base geometry must satisfy these conditions and must
fit into one of the 3 categories (A-C).  A similar set of conditions
hold for $\Sigma_+$ where the sign is changed for $T$ on all equations.

\subsubsection{General constraints on geometry from $f, g$ on curves}
\label{sec:constraints-curves}

The conditions (A-C) derived in
\S\ref{sec:constraints-global}, along with the local twist conditions
(\ref{eq:t-0}-\ref{eq:t-big}),
give a set of {\it  necessary} conditions that must be satisfied for
any F-theory compactifications on a space ${\cal  B}_3$ that is a
$\P^1$ bundle over a base $B_2$ without divisors of self-intersection
below -2.  These conditions are not, however, sufficient.  While the
conditions derived in the preceding subsection  are both necessary and
sufficient for $f, g$ to be well-behaved on divisors, the local
constraints on twist components $T\cdot D$ are not sufficient to
guarantee that $f, g$ are well-behaved on all curves 
$C = \Sigma_- \cap D$ with $D$ a divisor in $B_2$.
In general, ``nonlocal'' effects from other divisors can limit the
range of allowed twists more stringently than the local conditions
(\ref{eq:t-0}-\ref{eq:t-big}).
Note, however, that there are no further constraints associated with
curves formed from the intersection of two divisors $D_i, D_j$ since
the vanishing on such curves cannot be greater than at the
corresponding points in $B_2$.

As an example of a further constraint following from the interaction
between twists on different divisors, consider a base $B_2$ that
contains two divisors  $D_1, D_2$ with $D_1 \cdot D_1 = D_2 \cdot D_2
= -2, D_1 \cdot D_2 = 1$.  While the bound (\ref{eq:t-2}) seems to
allow $T \cdot D_1 = T \cdot D_2 = 1$, if this were to hold then $f,
g$ would vanish to degrees 4, 6 on $D_1 \cap \Sigma_-, D_2 \cap
\Sigma_-$.  This can be seen by considering (as usual, for $n= 4, 6$
and $0 \leq k < n$)
$Q = (- nK_2 - (n- k) T)$, which satisfies
\begin{eqnarray}
Q \cdot D_1 & = &  - (n- k) T \cdot D_1 = - (n- k) <0 \\
Q \cdot D_2 & = &  - (n- k) T \cdot D_2 = - (n- k) <0 \,.
\end{eqnarray}
We then have a decomposition (\ref{eq:Zariski}) of $Q$ of the form
\begin{equation}
 Q =\gamma D_1 + \eta D_2 + X \,,
\label{eq:Zariski-q}
\end{equation}
with $X \cdot D_1\geq 0, X \cdot D_2 \geq 0,$ so
$(\eta -2  \gamma)\leq - (n- k), 
(\gamma -2 \eta) \leq - (n- k),$ from which it follows that
$\eta \geq n-
k, \gamma \geq n- k$.

A general statement of the sufficient conditions on $T$ for $f, g$ 
to be well-behaved on all curves $D \cap \Sigma_\pm$
(and all divisors $\Sigma_-$)
is that for each divisor $D$ in $B_2$, there is at least one $n, k$ 
($n = 4$ or 6, $k < n$)
so
that $Q_{n, k}= - nK_2 - (n- k) T$ has a decomposition of the form
(\ref{eq:Zariski}) with $\gamma_D < k$.  This condition is
automatically satisfied for any base $B_2$ without -2 curves, since
for a -1 curve $D$ as long as $-6K_2 - T$ is effective,
$Q_{6, 5}\cdot D = 6 - T \cdot D \geq 0$ for $T \cdot D \leq 6$,  so
$Q_{6, 5}$ is effective and has sections that do not vanish on any -1
curves $D$  in the base.

This shows that while the conditions described
in
\S\ref{sec:constraints-global}
and
(\ref{eq:t-0}-\ref{eq:t-big})
are sufficient for a model to have acceptable $f, g$  on all divisors
and curves when the base is del Pezzo (or $\P^2$ or $\F_0$) with no -2 curves,
when the base has -2 curves
the more general conditions stated above must be
included to give a set of sufficient conditions.

In formulating these conditions, note again that we have not
considered potential problems with codimension 3 singularities or
G-flux, which may make a geometry unsuitable for F-theory
compactification even when the sufficient conditions discussed here
are satisfied.  These other issues are deferred to future work.
In specific, however, note that any curve on which $f, g$ vanish to
degrees $4, 5$ but are not constants will generically have points
where $f, g$ vanish to degrees 4, 6, which are of codimension 3 in the
full base ${\cal B}_3$.

\subsection{Constraints on gauge enhancement}
\label{sec:enhancement-constraints}

The constraints described so far limit the possible geometries that
can be used for F-theory compactification, and for any given geometry
impose a minimum gauge group that cannot be broken without changing
the F-theory base $B_2$.
In some situations, the vanishing conditions on $f, g$ also impose
constraints that limit the extent to which gauge group factors in the
effective supergravity theory can be enhanced by ``unHiggsing'' matter
fields to form larger gauge groups.  These constraints give nontrivial
limitations on bundle structure in the dual heterotic picture.  We
consider two specific types of such constraints.  In the first type,
generic $SU(2)$ and $SU(3)$ gauge groups are constrained from being
enhanced to $SU(N)$ with $N > 3$; in the second type, constraints
are associated with codimension two loci in the F-theory picture.
In both cases, the restriction on enhancement is related to the
absence of sufficient matter to represent a Higgsed phase of a theory
with higher symmetry.

Note that while the bounds considered here on gauge enhancement are a
consequence of the geometry, it is possible that in some cases G-flux
may decrease the size of the gauge group, as discussed further in
later sections.  This would not affect the upper bounds described in
this section.

\subsubsection{Constraints on $SU(2)$ and $SU(3)$ enhancement}\label{su2su3_sec}

One unusual feature of the gauge groups $SU(2)$ and $SU(3)$ is that
they can be realized in two separate ways in F-theory, associated with
two different Kodaira singularities.  There is a standard $A_{N -1}$
realization of $SU(N)$ where $f$ and $g$ do not vanish on a given
divisor but $\Delta$ vanishes to order $N$ (an $I_{N}$ singularity in
the Kodaira classification).  Another realization of $SU(2)$ arises
when $f, g, \Delta$ vanish to degrees $1, 2, 3$ and
$SU(2)$ or $SU(3)$  can be
realized when $f, g, \Delta$ vanish to degrees $2, 2, 4$ (Type III and
IV singularities in the Kodaira classification).  
While type III and IV singularities are in one sense simply special
limit points on the loci of type $I_2$ and $I_3$ singularities, their
physical properties are rather different.
When
$\gsu_2$ or $\gsu_3$  gauge algebras are forced to exist on a
divisor $D$ by
type III or type IV singularities in
the F-theory
geometry, there is no way to tune the Weierstrass moduli to realize
any $SU(N)$ gauge group with $N > 3$ on $D$ since this can only happen
from a type $I_N$ singularity where $f, g$ do not vanish on $D$.
This
means that for many theories with generic $SU(2)$ or $SU(3)$ gauge
symmetries there is in principle no branch of the theory with enhanced
$SU(N)$ gauge symmetry (particularly no $SU(5)$ gauge symmetry).

As a concrete example of where this constraint is relevant, we begin
by considering the 6D case of F-theory compactification on $\F_3$.  In
this case, the base ${\cal B}_2 = \F_3$ contains a divisor $\Sigma$
with self-intersection $\Sigma \cdot \Sigma = -3$ on which $f, g,
\Delta$ necessarily vanish to degrees 2, 2, 4.  The resulting $SU(3)$
gauge group in the corresponding 6D supergravity theory \emph{cannot}
be enhanced to $SU(4)$ or any higher $SU(N)$ by tuning the Weierstrass
moduli to get an $A_{N -1}$ singularity.  This is clear in the
low-energy 6D theory, as there is no matter charged under the $SU(3)$,
such as would arise under Higgsing from a larger gauge group.  As we
discuss in Section \ref{sec:examples}, this corresponds in the heterotic dual picture
to a constraint on how the associated bundle over K3 can be
decomposed.

A similar constraint occurs for a wide range of 4D F-theory
compactifications.  In many 4D compactifications the 
structure of the base ${\cal B}_3$ is such that $f, g, \Delta$ are
forced to vanish to degree $1, 2, 3$ or $2, 2, 4$, giving rise to a
non-Higgsable $SU(2)$ or $SU(3)$ gauge group.  In such cases these
gauge groups cannot be enhanced to higher $SU(N)$ with $N > 3$
anywhere in the
moduli space, though the $SU(2)$ gauge factors can generally be enhanced to
$SU(3)$ by tuning moduli.
We describe some specific examples where these kinds of constraints
appear in
Section \ref{sec:examples}.

\subsubsection{Constraints from codimension two loci}

In another class of situations, the extent to which a gauge group
factor can be enhanced is limited by the degrees of vanishing of $f,
g, \Delta$ on a locus of codimension 2.  This occurs when the
enhancement of the factor $G$ on a divisor $D$ to a given Kodaira
singularity type automatically raises the degrees of vanishing of $f,
g, \Delta$ on a codimension two locus to $4, 6, 12$ or beyond.

In six dimensions, constraints of this type only appear for F-theory
constructions without smooth heterotic duals.  A simple example is
when the base ${\cal B}_2$ contains two effective irreducible divisors
$C_1, C_2$ with self and mutual intersections $C_1 \cdot C_1 = -2, C_2
\cdot C_2 = -3, C_1 \cdot C_2 = 1$. In such a situation, the 6D theory
has a non-Higgsable gauge group with Lie algebra summands $\gsu_2
\oplus\ggg_2$, and there is matter charged under both groups, as
described in \cite{mt-clusters}.  While naively the $G_2$ factor can
be tuned to an $E_6$ or $F_4$ by increasing the degrees of vanishing
of $f$ and $g$ on $C_2$ to $3, 4$, doing this raises the degrees of
vanishing of $f, g$ to $4, 6$ on the intersection point $C_1 \cdot
C_2$, leading to a point in the base that must be blown up.  Writing
the Weierstrass coefficients explicitly, in a coordinate system where
$z, w$ vanish on $C_1, C_2$ respectively we have $f = a z w^2 +{\cal
  O} (z, w)^4$, $g = bz^2 w^3 +{\cal O} (z, w)^6$, from which the
above conclusions follow directly.

While this kind of enhancement constraint only arises for 6D F-theory
models that do not have heterotic duals, in 4D the issue is much more
general.  One key class of examples, which we discuss further in
Section \ref{sec:equivalence}, are 4D compactifications of F-theory
that have heterotic duals violating the base-point free condition (see
\S\ref{sec:spectral-cover}).  In such situations an $F_4$ symmetry can
have an obstruction to enhancement to an $E_6, E_7$ or $E_8$ on the
F-theory side as such an enhancement would lead to a $4, 6$ vanishing
of $f, g$ on a codimension two locus.

\section{Heterotic constraints}
\label{sec:heterotic-constraints}

As noted in \S\ref{sec:het_f_motivation}, the compactification of the
$E_8 \times E_8$ heterotic string gives rise to a number of
consistency constraints, linking the topological data of the
Calabi-Yau threefold with that ({\it i.e.}, Chern classes) of the vector
bundles $V_1, V_2$. We will be interested here in these bounds on
topology, as well as in the conditions for supersymmetric ${\cal
  N}=1$ heterotic vacua.

We consider a pair $(V_1,V_2)$ of vector bundles on a Calabi-Yau
threefold with structure groups $H_i \subseteq E_8$, $i=1,2$, which
break each $E_8$ factor to the commutant $G_i$ of $H_i$
in $E_8$. The bundles must satisfy the topological constraints
\begin{align}
&c_1(V_1)\equiv c_1(V_2)\equiv 0~(\text{mod}~2) \\
&{\rm ch}_2(TX_3)-{\rm ch}_2(V_1)-
{\rm ch}_2(V_2)+[W]_{\rm eff}=0\label{anom_canc}
\end{align}
The first of these conditions is equivalent to the vanishing of the
second Steifel-Whitney class of the bundles $V_i$, a necessary
condition for the existence of spinors; in the case of irreducible
principal bundles, this  reduces to the condition $c_1=0$ (see the
discussion in Section \ref{conditions_on_vac}).
Henceforth in this paper we focus attention on irreducible bundles and
take $c_1 = 0$. The second constraint
on the second Chern characters of the bundles is the familiar 10D
anomaly cancellation condition we have already encountered in
\S\ref{sec:duality-geometry} in the context of heterotic/F-theory
duality. The last term in \eref{anom_canc} is a non-perturbative
contribution arising from NS5-branes (equivalently M5 branes, in
heterotic M-theory), where $[W]_{\rm eff}$ denotes the total class of 
effective curves wrapped by $5$-branes. In this work, we will not
include $5$-branes wrapping curves in the base $B_2$
({\it i.e.}, degenerations of the bundle corresponding to sheaves supported
over curves) and as a result, any possible term $[W]_{\rm eff} \neq 0$
will not affect \eref{eq:eta}, the definition of $\eta$ given in
Section \S\ref{sec:duality-geometry}. In some cases, non-perturbative
effects in the form of $5$-branes wrapping the elliptic fibers may be
present (see \cite{Friedman-mw} for a discussion of such heterotic
$5$-branes and G-flux).

For a supersymmetric vacuum, the vanishing of the $10$-dimensional
gaugino variation 
requires each bundle $V_i$ to satisfy the well-known
``Hermitian-Yang-Mills'' equations \cite{GSW}
\begin{equation} \label{hym}
F_{ab}=F_{{\bar a}{\bar b}}=0~~~~,~~~~g^{a{\bar b}} F_{a{\bar b}}=0
\end{equation}
The first half of these conditions, namely the vanishing of
$F^{2,0}=F^{0,2} = 0$, is by definition the condition that the vector
bundle is holomorphic ({\it i.e.}, that its transition functions are
holomorphic functions over the base $X_3$). The consequences of the
condition $g^{a{\bar b}}F_{a{\bar b}}=0$ are not so easy to state,
however; solving this partial differential equation has
historically posed a significant challenge to the construction of
supersymmetric heterotic vacua, since the background Ricci-flat
Calabi-Yau metric $g^{a{\bar b}}$ and the field strength $F$
associated to the bundle $V$ are not known analytically except in
very special cases\footnote{For recent progress in solving these
  equations via numeric approximations, see for example
  \cite{Douglas:2006hz,Anderson:2010ke,Anderson:2011ed}.}. Thanks,
however, to the powerful Donaldson-Uhlenbeck-Yau theorem
\cite{duy1,duy2}, it is possible to translate this problem in
differential geometry into one in algebraic geometry. According to the
DUY theorem, a holomorphic bundle $V$ admits a connection $A$
that solves \eref{hym} if and only if $V$ is {\it slope poly-stable}. A
bundle $V$ is defined to be {\it slope stable} 
with respect to a given
K\"ahler form $\omega \in H^{1,1}(X_{3})$
if for all sub-sheaves
${\cal F} \subset V$, with $0<rk({\cal F}) < rk(V)$,
\begin{equation}
\mu({\cal F}) < \mu(V)\,,
\end{equation}
where for any sheaf, 
\begin{equation}\label{slope_def}
\mu({\cal F})=\frac{1}{rk({\cal F})}\int_{X_3} c_1({\cal F}) \wedge
\omega\wedge\omega \,.
\end{equation}
A bundle is called semi-stable\footnote{Note that all
  poly-stable bundles are automatically semi-stable, but the converse
  does not hold.} if $\mu({\cal F}) \leq \mu(V)$
for all sub-sheaves, and ``poly-stable'' if $V=\bigoplus_i V_i$ with
$V_i$ stable and $\mu(V)=\mu(V_i)$ $\forall i$. Regardless of the structure group $H$ of $V$,
vector bundles describing a good heterotic vacuum must be holomorphic,
slope poly-stable and satisfy
\begin{equation}\label{slopezero}
\mu(V)=0
\end{equation}
for the physical K\"ahler form $\omega$.

It is poly-stable bundles that we must consider in the context of
${\cal N}=1$ 4D heterotic Calabi-Yau vacua, and although the study of
such bundles and their moduli spaces is a rich and ongoing subject in
algebraic geometry, at present very little is known in general about
how to fully classify and enumerate the moduli space of stable bundles
(sheaves) on Calabi-Yau threefolds\footnote{See
  \cite{qin1,qin2,Anderson:2010ty} for some recent results in the
  math/physics literature on bundle moduli spaces on Calabi-Yau
%%CITATION = MATH/;%%
  threefolds in examples with special topology.}. One of
our goals in this work is to try to use heterotic/F-theory duality to
understand as much as possible about which stable bundles can exist on
Calabi-Yau threefolds and what properties characterize the associated
heterotic effective theories.

\subsection{The Bogomolov bound}
\label{sec:Bogomolov}
There are a number of constraints that slope-stability places on the
topology of a holomorphic vector bundle. One of the most important of
these is the so-called ``Bogomolov bound'' (see
\cite{kobayashi_hitchin} for a review), which states that if a rank
$N$ bundle $V$ is slope (poly-) stable with respect to a choice of
K\"ahler form $\omega=t^k \omega_k$ (with $k=1,\ldots h^{1,1}(X_3)$)
on the CY 3-fold $X_3$, then
\begin{equation}
\int_{X_3} \left(2N c_2(V)-(N-1)c_{1}^{2}(V) \right) \wedge \omega \geq 0
\end{equation}
For simplicity, let us consider first the case of vector bundles with $c_1(V)=0$, in which case the Bogomolov bound reduces to $\int_{X_{3}} c_2(V) \wedge \omega \geq 0$. 

Thus far our discussion of the consistency conditions on heterotic
vacua has been general. We restrict our attention now to those
threefolds that can give rise to F-theory duals, namely smooth,
elliptically fibered Calabi-Yau threefolds, $\pi: X_3 \to B_2$ (with
section). More specifically, we  restrict our consideration to the
case in which there is a single section ({\it i.e.}, the Mordell-Weil group
of sections is trivial and $h^{1,1}(X_3)=1+h^{1,1}(B_2)$) and the
manifold can be put in Weierstrass form as
\begin{equation}\label{het_weir}
{\hat Y}^2={\hat X}^3 +f(u){\hat X}{\hat Z}^4 + g(u){\hat Z}^6
\end{equation}
where $\{{\hat X}, {\hat Y}, {\hat Z} \}$ are coordinates on the
elliptic fiber (described as a degree six hypersurface in
$\mathbb{P}_{231}$) and $\{u\}$ are coordinates on the base $B_2$.

Let us consider the consequences of the Bogomolov bound for a bundle
over an $X_3$ defined as above. Recalling the geometric identities in
Appendix \S\ref{sec:app_identities},
as in \eq{c2special} we can expand the
second Chern class of the bundle as
\begin{equation}\label{c2V}
c_2(V) = \pi^*(\eta) \wedge \omega_{0} + \pi^*(\zeta)
\end{equation}
where $\pi^*(\eta)$ and $\pi^*(\zeta)$ are pullbacks of, respectively, $\{1,1\}$ and
$\{2,2\}$ forms on the base $B_2$. 
(Note that in other sections we often use the notation {\it e.g.}
$\eta$ both for the form on 
$B_2$ and for the pullback form -- technically $\pi^*(\eta)$ --
on ${\cal B}_3$.  Which form is used should be apparent in any given
equation from context.)
Expanding the K\"ahler form $\omega$ in the explicit basis of Appendix
\S\ref{sec:app_identities}, we have
\begin{equation}\label{gen_kahler}
\omega= a \omega_0 + \pi^*(\omega^{base})
\end{equation}
where $a$ is a constant and $\omega_{base}=b^{\alpha} \omega_{\alpha}$
($\alpha=1,\ldots h^{1,1}(B_2)$) is an ample divisor on $B_2$. Without
loss of generality, we can scale the K\"ahler form to set $a=1$
\begin{equation}\label{adiabatic_kahler}
\omega=  \omega_0 +M \pi^*(\omega^{base})
\end{equation}
for some constant $M$. Substituting this form for the K\"ahler moduli
into the Bogomolov bound, the constraint becomes
\begin{eqnarray}
\lefteqn{\int_{X} \left(\pi^*(\eta) \wedge \omega_0+ \pi^*(\zeta) \right)\wedge
(\omega_0 +M\pi^*(\omega_{base}))} \\
 & = & M \omega_0 \wedge \pi^*(\eta \wedge
\omega_{base}) + (\omega_0 \wedge \pi^*(\zeta) + \pi^*(\eta) \wedge
\omega_0 \wedge \omega_0) \geq 0\nonumber \,.
\end{eqnarray}
Moreover, we recall that by the triple intersection numbers in
\eref{x3intersec},
\begin{equation}\label{eq:intersection}
\int_{X_3} \omega_0 \wedge \pi^*(\eta \wedge 
\omega_{base})= \eta \cdot \omega_{base} \,,
\end{equation}
and $\pi^*(\eta) \wedge \omega_0 \wedge \omega_0=\eta \cdot K_{B_2}$. 
Thus, the Bogomolov bound becomes the condition
\begin{equation}\label{simple_condition}
\int_{X_{3}} c_2(V) \wedge \omega = M \eta \cdot \omega_{base}+\zeta+ \eta\cdot K_{B_2} \geq 0
\end{equation}
(with $\zeta$ viewed now as the coefficient of the $\{2,2\}$-form on
$B_2$).
Note that in (\ref{eq:intersection}) and (\ref{simple_condition})
while the LHS is computed by integrating over the threefold $X_3$, the
RHS is computed in terms of intersection products on the base $B_2$.

In general, to extract consistency conditions on $\eta$ from the Bogomolov bound, the condition in \eref{simple_condition} must be examined in a case-by-case
manner. That is, given a choice of $\zeta$ and a K\"ahler form in \eref{adiabatic_kahler}, it is
possible to derive consistency conditions on $\eta$ associated to the underlying bundle being slope-stable (though once again it is important to recall that the
Bogomolov bound is {\it necessary} but not sufficient for the
stability of $V$).

There is one limit, however, that is of particular interest in
heterotic/F-theory duality. In order to take the stable degeneration
limit of Section \S\ref{sec:duality}, it is necessary that we evaluate
this expression with not just any K\"ahler form, but one chosen in the
appropriate ``Adiabatic limit'' \cite{Friedman-mw} in which the volume
of the elliptic fiber is small compared to that of the base (and
the volume of $X_3$, as given in \eref{x3vol}, is large).  This limit is
achieved by taking $M\gg 1$ in \eref{simple_condition}. For $M$
sufficiently large, it is clear that the dominant constraint from the
Bogomolov bound in \eref{simple_condition} is that $\eta$ has positive
intersection with the K\"ahler form of the base $B_2$. In the
adiabatic limit, it is impossible for the Bogomolov bound to be
satisfied unless\begin{equation} \eta \cdot \omega_{base} \geq 0
\end{equation}
 If this is taken to hold for any ample K\"ahler form $\omega_{base}$
 of the base then, by definition, this is simply the condition that
 {\it $\eta$ is an effective divisor} in $B_2$.

\subsection{Matter spectra in heterotic theories}
\label{sec:hetmatter}

As described above, the presence of a vector bundle $V_i$ with
structure group $H_i$ on the Calabi-Yau threefold breaks $E_8$ to
$G_i$, the commutant of $H_i$ inside of $E_8$. All matter in the low-energy effective theory arises under dimensional reduction from
components of the $10$-dimensional gauge field, that is from the
decomposition of the $248$-dimensional representation of $E_8$ under
the direct product $G_i \times H_i$ (see \cite{GSW} for a standard
review):
\begin{equation}\label{248decomp}
{\bf 248} \to ({\rm Ad}(G),1)+\bigoplus_{A} (R_{A}, r_{A})
\end{equation}
where ${\rm Ad}(G)$ represents the adjoint representation of $G$ and
$\{(R_A,r_{A})\}$ denotes a set of representations of $G\times H$. For
example, the presence of a bundle with $H=SU(3)$ over $X_3$ breaks one
$E_8$ factor down to $G=E_6$ and the possible states in the theory are
determined by the decomposition 
\beq
{\bf 248} \to ({\bf 78},{\bf 1})+({\bf 27},{\bf 3})+({\bf \overline{27}},{\bf \bar{3}})+({\bf 1},{\bf 8})
\eeq
That is, in addition to the adjoint-valued $E_6$ gauge boson, the low energy theory can contain charged matter in the ${\bf 27}$ and ${\bf \overline{27}}$ representations, as well as $E_6$-singlet fields.

While the decomposition above is sufficient to determine the {\it type} of matter in the $4$-dimensional $E_6$ theory, to find the {\it multiplicity} of these massless scalar fields it is necessary to count the number of bundle-valued $1$-forms on the Calabi-Yau manifold. More precisely, under dimensional reduction, the zero-modes of the $4$-dimensional theory are determined by the dimensions of vector bundle-valued cohomology groups (in the representation $r_{A}$) over the Calabi-Yau threefold, such as
\begin{equation}
H^1(X_3, V), H^1(X_3, \wedge^2 V), H^1(X, {\rm End}_0(V)),~\text{etc.} \ldots
\end{equation}
For the $E_6$ example above this leads to the bundle-valued cohomology
groups shown in Table~\ref{t:multiplicity}
\begin{table}[h!]\begin{center}
\begin{tabular}{|c|c|}
\hline
\textnormal{Field} & \textnormal{Multiplicity} \\ \hline
${\bf 27}$ & $h^1(X, V)$ \\ \hline
$\overline{{\bf 27}}$ & $h^1(X, V^{\vee})$ \\ \hline
${\bf 1}$ & $h^1(X, {\rm End}_0(V))$ \\ \hline
\end{tabular}
\end{center}
\caption[x]{\footnotesize  
Matter multiplicities for the $E_6$ example}
\label{t:multiplicity}
\end{table}
where $V$ is the rank $3$ vector bundle valued in the fundamental of $SU(3)$ (and hence $V^{\vee}$ is associated to the ${\bf {\bar 3}}$ and ${\rm End}_0(V)$ to the ${\bf 8}$).

The chiral index\footnote{Physically, the chiral index in a heterotic
  compactification counts the number of generations minus the number
  of anti-generations of chiral particles. For example, the number of ${\bf
    27}$ multiplets minus the number of ${\bf \overline{27}}$'s in the
  $E_6$ theory given above, or the number of families in a heterotic
  Standard Model.} of the ${\cal N}=1$ theory is determined by the
Atiyah-Singer index theorem \cite{hartshorne1977algebraic} as the 
alternating sum  
\beq
{\rm Ind}(V)= h^0(X, V)- h^1(X,V) +h^2(X, V) -h^3(X,V)~~.
\eeq
But by slope stability of the bundle $V$
and the condition $c_1 (V)= 0$, 
\begin{equation}
H^0(X_3, V)=H^0(X_3, V^{\vee})=0
\end{equation}
(as well as the induced representations; {\it i.e.}, $H^0(X, \wedge^n V)=0$,
with $n<rk(V)$, etc). Finally, it should be recalled that by Serre duality \cite{hartshorne1977algebraic}, $h^{m}(X,V)=h^{3-m}(X, V^{\vee})$ and as a result, the chiral index can be expressed simply as the difference,
\begin{equation}\label{chiral_index}
{\rm Ind}(V)= -h^1(X, V)+h^1(X, V^{\vee})~~.
\end{equation}
In the case that $c_1(V)=0$ this is further given by
${\rm Ind}(V)={\rm Ch}_3(V)=\frac{1}{2}c_3(V)$, the third Chern
character. Finally, it should be noted that the index given in
\eref{chiral_index} is written in terms of the vector bundle associated to the
fundamental representation of the underlying principal $H$-bundle. The
chiral asymmetry in all other representations ({\it i.e.}, for the induced
vector bundles, $V^{\vee}, \wedge^m V, S^p V, \ldots$) is
fact determined by the index of the fundamental representation given above (see \cite{Anderson:2008ex} for
further details). In 
heterotic theories, the exact massless matter spectrum is frequently
easier to compute than in F-theory, and in this work we  use the
simple structure of heterotic matter to extract useful information
about the spectrum and chiral index arising from F-theory
compactifications on Calabi-Yau fourfolds.

For heterotic theories defined over elliptically fibered Calabi-Yau
threefolds, the bundle-valued cohomology groups defined above have a
simple decomposition in terms of the base/fiber geometry. Using the
techniques of Leray spectral sequences, there exists a decomposition 
\begin{equation}\label{leray1}
H^m(X,V)=\bigoplus_{p+r=m}H^r(B_2, R^p\pi_{*} (V))
\end{equation}
where $R^p\pi_{*} (V)$ is the $p$-th derived push forward
\cite{hartshorne1977algebraic} of $V$. On any open set ${\cal U}$,
$R^p\pi_{*} (V)$ can locally be represented  on $B_2$ by the pre-sheaf
\begin{equation}\label{leray2}
{\cal U} \to H^p(\pi^{-1}({\cal U}), V)
\end{equation}
This formalism allows for a very precise notion of ``localized''
matter in the heterotic theory (supported over loci in the base $B_2$)
which (in the case of simply-laced $G$) can be matched exactly to the
localized matter associated to $7$-brane intersections in the dual
F-theory geometry. We  explore this localized matter and the
chiral index further in Section \S\ref{chiralsec} and in the context of the
spectral cover construction of vector bundles below.

\subsection{The spectral cover construction}
\label{sec:spectral-cover}

Thus far our discussion of
heterotic compactifications and constraints has been completely
general, and throughout this work we attempt
as far as possible 
to keep our study of the holomorphic vector bundles $(V_1,V_2)$
independent of any particular method of bundle construction. 
It will, however, be useful in certain examples to appeal to one method of
constructing vector bundles on elliptically fibered manifolds in which
heterotic/F-theory duality is particularly well understood
\cite{Friedman-mw,Donagi:1998vw}. This is the well-known ``spectral
cover'' construction \cite{Friedman:1997ih,donagi}.

\subsubsection{Spectral covers}

The spectral cover construction can be used to build rank
$N$ bundles,
$V \rightarrow X_3$, with structure group $SU(N)$ or $Sp(2N)$. Moreover, for
some bundles with these structure groups, which are slope-stable in
the adiabatic region described in the previous section, the
Fourier-Mukai transform\footnote{The precise conditions for stability
  and consistency of spectral cover bundles will be discussed further
  in the next section.} \cite{Ron_Elliptic} provides a $1-1$ map (in
fact a full functor on the category of coherent sheaves) from $V$ to a
pair $({\cal S}, L_{{\cal S}})$ where ${\cal S}$ is a divisor in $X_3$
that is an $N$-fold cover of the base $B_2$ and $L_{{\cal S}}$ is a
rank-1 sheaf on ${\cal S}$. The class of ${\cal S}$ is given by
\begin{equation}\label{spec_class}
[{\cal S}]=N[\sigma]+\pi^*(\eta)
\end{equation}
where $\sigma$ is the zero section of $\pi: X_3 \to B_2$ and $\eta$ is
defined as in \eref{c2V}. As in \eref{het_weir}, let ${\hat X}, {\hat
  Y}, {\hat Z}$ be the coordinates of the elliptic fiber (where ${\hat
  Z}=0$ defines the section $\sigma$). Then in the case that the
structure group of $V$ is $SU(N)$, the 
{\it spectral cover} ${\cal S}$ can
be represented as the zero set of the polynomial
\begin{equation}\label{speccov}
s=a_0 {\hat Z}^N+a_2 {\hat X}{\hat Z}^{N-2}+a_3 {\hat Y}{\hat Z}^{N-3} +\ldots
\end{equation}
ending in $a_N {\hat X}^{\frac{N}{2}}$ for $N$ even and $a_N {\hat
  X}^{\frac{N-3}{2}}{\hat Y}$ 
for $N$ odd
\cite{Friedman-mw} (see
\cite{Friedman-mw} also for the analogous construction for $Sp(2N)$
structure group). The coefficients $a_j$ are sections of line bundles over the base
$B_2$ 
\begin{equation}
a_j \in H^0(B_2, K_{B_2}^{\otimes  j} \otimes {\cal O}(\eta)) 
= H^0 (B_2,{\cal O} (\eta  + jK_2))\,,
\end{equation}
which can locally be described as polynomial functions with
appropriate degrees.
Note that in the duality to F-theory, the coefficients $a_j$ play a
dual role as coefficients $f_k, g_k$ in the F-theory Weierstrass
model, providing a direct map between the moduli on the two sides of
the duality.
In order for the spectral cover to be an actual algebraic surface in
$X_3$ it is necessary that ${\cal S}$ be an effective class in
$H_4(X_3, \mathbb{Z})$. It is straightforward to show
\cite{FM,Friedman-mw,Friedman:1997ih} that this is true if and only if
$\eta$ is an effective class in $B_2$.
This can be seen by noting that $\eta$ must
be effective for $a_0$ to be nonvanishing, and
since $-K_2$ is effective no other coefficient $a_j$ can be
nonvanishing if $\eta$ is not effective.
In view of the Bogomolov
condition in the previous section, it is clear that spectral cover
bundles are built to be slope-stable in the adiabatic region of
K\"ahler moduli space\footnote{ More precisely, it is known \cite{FM}
  that for a spectral cover bundle there exists some value $M_0$ such
  that for $M \gg M_0$ in \eref{adiabatic_kahler}, the bundle
  associated to that spectral data is slope stable for the given
  region of K\"ahler moduli space. For K\"ahler twofolds the proof of
  stability is constructive and yields an explicit value of $M_0$
  defining the stable region of K\"ahler moduli space. For Calabi-Yau
  threefolds however, the arguments are not constructive and we are
  restricted to considering the limit $M \gg 1$ \cite{Friedman-mw,
    FM}.}.

There is a further condition that  must be imposed in order for the
spectral cover bundle $V$ to be slope stable. By construction, {\it
  irreducible} spectral covers are stable in the adiabatic region
given above \cite{Friedman-mw, Friedman:1997ih}. However, the
condition that the cover is irreducible places another condition on
$\eta$. It can be argued that ${\cal S}$ is irreducible\footnote{Note that if ${\cal S}$ is irreducible as an algebraic curve in $X_3$, the associated vector bundle under Fourier-Mukai transform will be indecomposable (i.e. not a direct sum $V_1 \oplus V_2 \oplus \ldots$). However, the converse does not hold. Some reducible spectral covers can still correspond to indecomposable vector bundles. See \cite{Aspinwall:1998he,Anderson:2013rka} for examples.} if $\eta$
is base-point free ({\it i.e.,} has no base locus in a decomposition
of the type (\ref{eq:Zariski})) and $\eta - Nc_1(B_2)$ is effective (see
\cite{Donagi:2004ia} for example). The condition of base point
free-ness will be explored in further detail for surfaces in Section \S
\ref{sec:bpf}, but for now it should simply be noted that this
condition guarantees that there exist irreducible curves in the class
$[\eta]$.

\vspace{10pt}

To fully determine the bundle $V$ and its topology after Fourier-Mukai
transform, it is necessary to specify not only the class $\eta$ in
\eref{spec_class}, but also the rank-$1$ sheaf $L_{\cal S}$
above. The condition that $c_1(V)=0$ fixes the first Chern
class of $L_{\cal S}$ to
be \cite{Friedman-mw}
\begin{equation}\label{l_top}
c_1(L_{\cal S})=N\left(\frac{1}{2}+\lambda \right)\sigma + \left(\frac{1}{2}- \lambda \right)\pi_{S}^{*}\eta + \left(\frac{1}{2}+N\lambda \right) \pi^{*}_{S} c_{1}(B_2)
\end{equation}
where $\pi_{{\cal S}}: {\cal S} \to B_2$, the
bundle $V$ has structure group $H=SU(N)$, and the parameter $\lambda$
is either integer or half-integer depending on $N$: 
\begin{equation}\label{lambda_n}
    \lambda= 
\begin{cases}
    m+\frac{1}{2},& \text{if }~N ~\text{is odd}\\
    m,              & \text{if}~ N~\text{is even}
\end{cases}
\end{equation}
where $m \in \mathbb{Z}$. This condition arises from the fact that
$c_1(L_{\cal S})$ must be an integral class in $H^{1,1}({\cal S},
\mathbb{Z})$. When $N$ is even it is clear that this integrality
condition imposes 
\begin{equation}\label{mod2_cond}
\eta  \equiv c_1(B_2)~\rm{mod}~2
\end{equation}
where ``mod $2$'' indicates that $\eta$ and $c_1(B_2)$ differ only by
an even element of $H^2(B_2, \mathbb{Z})$. 
The relation (\ref{l_top}) holds when the cohomology of ${\cal S}$
is spanned by the class $\sigma$ and the pullback of the cohomology in
the base.  While this is expected to be true generically, there can be
situations in which ${\cal S}$ has a larger Picard group (i.e. more independent divisor classes). Examples where this increase in $h^{1,1}({\cal S})$ may occur include Noether-Lefschetz loci in the complex structure of ${\cal S}$ and degenerate (singular) spectral covers (see \cite{Curio:2011eu} for some generalizations
  of \eref{l_top} for such examples).  In these more general situations, the constraint
(\ref{mod2_cond}) may not hold. We will see the need for such interesting possibilities (and their F-theory duals) in later sections.

Finally, with this data in hand it is possible to extract the full
topology of $V$, including the chiral index, ${\rm Ind}(V)=-h^1(X_3,
V)+h^1(X_3, V^{\vee})$. The Chern classes of a spectral cover bundle
$V$, specified by $\eta$ and the integers $N$ and $\lambda$, are
\cite{Friedman-mw,Friedman:1997ih,Curio:1998vu,Curio:2011eu} 
\begin{align}
& c_1(V)=0 \\
& c_2(V)=\eta \wedge \sigma -
  \frac{N^3-N}{24}c_1(B_2)^2+\frac{N}{2}\left(\lambda^2-\frac{1}{4}\right)\eta
  \wedge \left(\eta-Nc_1(B_2) \right) \\
& c_3(V)=2\lambda \sigma \wedge 
\eta \wedge \left(\eta-Nc_1(B_2) \right) =
2 \lambda
\eta \cdot \left(\eta-Nc_1(B_2) \right)
\label{c3spec}
\end{align}
Note that since $c_1(V)=0$, ${\rm Ind}(V)=ch_3(V)= \frac{1}{2}c_3(V)$.

The essential heterotic constraints on a bundle constructed via
spectral covers can be simply encapsulated by the $\{1,1\}$ form
$\eta$.  For a bundle $V_i$, if $\eta_i$ and $\eta_i-Nc_1(B_2)$ are effective,
$\eta_i$ is base-point free, and an $L_{{\cal S}}$ is chosen subject to \eref{l_top}, \eref{lambda_n} and \eref{mod2_cond}, a
spectral cover bundle is guaranteed to exist and to be slope stable
(for a region in K\"ahler moduli space in which heterotic/F-theory
duality is well understood). In Section \ref{sec:equivalence} we
explore the way that some of these same constraints appear in the dual
fourfold geometry, giving information about when these constraints
must be true based on topological data independent of a specific
method of bundle construction.

\subsubsection{Localized Matter and Spectral Covers}
\label{spec_cov_matter}

In $SU(N)$ spectral covers, at least some of the zero-modes of the
theory have a simple realization in terms of the geometry of the
spectral cover ${\cal S} \subset X_3$. By the Leray spectral sequence
arguments outlined above, it can be shown that the matter in the
theory determined by $H^1(X,V)$ is localized at the intersections
${\cal S} \cap
\sigma$ in $B_2$ (see Section (6.2) of \cite{Friedman-mw} for a
review). For example, for an $SU(2)$ spectral cover of the
form
\begin{equation}\label{su2speceg}
a_0{\hat Z}^2 +a_2 {\hat X}=0
\end{equation}
 ${\cal S}$ intersects the zero-section $\sigma$ at the zero-locus of
the section $a_2(u) \in H^0(B_2, K_2^{\otimes 2} \otimes
{\cal O}(\eta))$ over $B_2$. This localized matter appears as the ${\bf 56}$
multiplets of the $4$-dimensional $E_7$ theory. The exact multiplicity
of these fields can be found by a Leray calculation to determine the
exact zero-mode spectrum along the curve $a_2=0$ in
$B_2$. \footnote{Note that in the dual F-theory geometry matter curves
 appear in the shared base $B_2$ in exactly the same way. In the
  notation of \eref{fdef1}-\eref{gdef1}, the dual Weierstrass model to
  \eref{su2speceg} is $y^2=x^3+ (a_2 z^3+\ldots) x +( a_0 z^5+
  \ldots)$ and on the vanishing locus $a_2=0$ there is an enhancement
  of $E_7 \to E_8$ as expected.} 

More generally, given an $SU(N)$ spectral cover of the form shown in
\eref{speccov}, the localized matter counted by $H^1(X,V)$ ({\it
  i.e.}, the matter valued in the fundamental representation of $V$)
will be controlled by the zeros of $a_{N}(u) \in H^0(B_2, K_2^{\otimes
  N} \otimes {\cal O}(\eta))$.  Note that this matter need not be
chiral, and this is of course not the full matter spectrum of the
low-energy 4D theory -- for that other representations appearing in
the decomposition of the ${\bf 248}$ of $E_8$ in \eref{248decomp},
such as $H^1(X, \wedge^2 V), H^1(X, {\rm {\rm End}}_0(V))$ {\it etc.},
must be considered. For these, we must consider not just the
$N$-sheeted spectral cover associated to the fundamental
representation of the $SU(N)$ bundle, but other curves associated to
other induced vector bundles (such as $\wedge^k V, S^p V$, etc) as
well. At present, it is not known how to construct all such associated
spectral covers in full generality (see \cite{guerra,Hayashi:2008ba}
for some progress in this direction), though in the case of some of
these representations there will likewise be a notion of localized
matter \cite{Hayashi:2008ba}.

Finally, it is worth noting that although the presence of ``matter
curves'' in the class $[a_N=0]$ in $B_2$ indicates the presence of
charged matter in the 4D theory, the exact matter spectrum with
multiplicities cannot be determined without fully specifying {\it all}
the data of the bundle, including detailed properties of a particular
$V$ (not just its topology) and of course, the third Chern class
\eref{chiral_index}. With this data in hand, the restriction of $V$ to
both the fiber and base implicit in \eref{leray1} and \eref{leray2}
can be explicitly calculated. From the perspective of F-theory, we
expect this further data to be necessary, since it is known that the
exact (chiral) matter spectrum depends crucially not just on the
fourfold geometry, but also a choice of $G$-flux. We will return to
the issue of chiral matter and heterotic/F-theory duality again in
\sref{chiralsec}.

\subsubsection{Limitations of Spectral Covers}\label{spec_cov_limits}

Despite the fact that it constitutes one of the most studied and best
understood corners of the dual heterotic/F-theory landscape, the
spectral cover construction is far from general and care must be taken
in generalizing results derived in this context to the full vector
bundle moduli space or the generic dual fourfold geometries. As
pointed out above, the spectral cover construction is valid only for
special structure groups ({\it i.e.}, $H=SU(N)$ or $Sp(2 N)$) and perhaps more
importantly, its applicability is limited even in these settings. That
is, not all consistent $SU(N)$ or $Sp(2 N)$ bundles arising in heterotic compactifications can be represented by well-behaved spectral covers. 

Since this has an impact on the heterotic/F-theory comparisons
undertaken in this work, it is worth briefly reviewing some of these
constraints here. The Fourier-Mukai transform is a well-behaved
functor on the moduli space of sheaves, subject to the following
conditions: that the restriction $V|_{E_p}$ of the bundle to each
elliptic fiber $E_p$ is  

\begin{enumerate}
\item Semistable
\item Regular ({\it i.e.}, that the restricted rank $N$ bundle on the
  elliptic curve has an automorphism group of the minimum possible
  dimension: $\text{dim}(Aut(V|_{E_p}))=N$) 
\end{enumerate}

For the first of these, the semi-stability of $V$ is defined with respect to the restricted slope in \eref{slope_def}. For indecomposable stable bundles $V$ this will be true for generic fibers when the K\"ahler class is chosen to be 
\begin{equation}
\omega=\omega_0+M\omega_{\alpha}~~~,~~~M \gg 1
\end{equation}
where, as above, for  threefolds the stability proof is based on $M
>M_0$ for some unknown $M_0$ and is not constructive.  The
condition above is sufficient to guarantee that a stable region for
$V$ exists, compatible with the large volume, weakly-coupled limit that we
require for heterotic/F-theory dual pairs. It is important to note,
however, that this limitation of moduli space is certainly not
necessary for the consistency of heterotic theories
\cite{Anderson:2009sw,Anderson:2009nt}. This point is explored
further in Appendix \S\ref{sec:rigid}.

The second condition of ``regularity'' appearing above was introduced
by Atiyah in his classification of semi-stable sheaves on elliptic
curves \cite{Atiyah}. In particle, $V$ is called ``regular'' if when
it is restricted to every elliptic fiber it decomposes into a
poly-stable sum of line bundles ({\it i.e.}, by the divisor-line bundle correspondence, a set of points on $E_i$ summing to zero)
rather than a non-trivial extension (for example, Atiyah's $I_2$
bundle of the form $0 \to {\cal O}_{E_i} \to I_2 \to {\cal O}_{E_i}
\to 0$). Unlike the previous one, this condition cannot be stated as a
simple global restriction on the heterotic moduli space and is harder
to characterize for generic bundles. Indeed, it is significant that many
good heterotic bundles will fail the regularity criterion. For
instance, it is known that most bundles described via the monad
construction \cite{Distler:1987ee, OSS} are not regular
\cite{Bershadsky:1997zv}. In fact, there are indications that this
criteria can sometimes be consistently violated in the context of
perturbative heterotic/F-theory dual geometry (including in such
well-known examples \cite{Aspinwall:1998he} as the heterotic
``Standard Embedding'' in which $V=TX_3$). See \cite{Anderson:2013rka}
for some recent results on degenerate spectral covers and dual
F-theory geometry. The regularity condition will not be explored here
in detail, but it may be relevant in explaining some of the unsolved
questions regarding geometric constraints arising in dual
heterotic/F-theory pairs, including exotic G-flux (see Section \ref{sec:example-G-flux}) and the role of quantization conditions like \eref{lambda_n} and \eref{mod2_cond}.

\section{Equivalence of constraints}
\label{sec:equivalence}

We now consider the relation between constraints on 
F-theory geometry
and the constraints on bundle constructions on the heterotic side.  

%%%%%%%%%%%%%%%%%%%%%
\subsection{Effective condition on $\eta_i$}

We begin by showing that the F-theory
condition that $f, g$ do not vanish to degrees $4, 6$ 
on either $\Sigma_-$ or $\Sigma_+$
is equivalent to
the heterotic condition that $\eta_1$ and
$\eta_2$  are effective.
This follows directly from the analysis of Section
\ref{sec:constraints-global}, where it was shown that $-6K_2 \pm T$
must be effective in any F-theory geometry where $f, g$ are
well-behaved on $\Sigma_\pm$.  With the identification
\begin{equation}
\eta_\pm = \eta_{1, 2} =
-6K_2 \pm T
% \label{eq:}
\end{equation}
it follows that the condition that $\eta_\pm$  are effective is a
necessary condition for the existence of a good F-theory geometry.
This is thus a necessary condition for the existence of a good
heterotic dual.  This matches with what is known of the heterotic theory, where this
bound is necessary in order to satisfy the Bogomolov bound in the
adiabatic limit of the stable degeneration limit.  The fact that this
bound is necessary for any good F-theory geometry shows that this
bound on the heterotic side must be more general and applies to any
bundle construction, irrespective of of the stable degeneration limit.

In fact, from the F-theory side the constraint is significantly
stronger.  The constraint on the F-theory side states that either
$\eta_-= 0$, or $-4K_2 -T = \eta_--2c_1$ must be effective.  This
constraint, and the analogous constraint for $\eta_+$ must be
necessary conditions on the heterotic side for the existence of any
smooth bundle with the specified components of $c_2$.

\subsection{Effective constraint and gauge groups}

In \cite{Berglund-Mayr} the effective constraint on $\eta$ for a
bundle to exist on the heterotic side was generalized to situations
where the 4D gauge group can be seen to be restricted in specific ways
on the F-theory side.  These constraints can be readily attained by a
generalization of the analysis in the previous section.  For example,
if we consider a divisor $\Sigma_-$ that carries a gauge group no
larger than $E_7$, then $-4K_2-T$ must be effective for $f$ to have a
term of degree $\leq 3$ , so $\eta = 6 c_1-T \geq 2c_1$, where by $A
\geq B$ we mean that $A-B$ is effective.  This corresponds on the
heterotic side to a constraint on $\eta$ for bundles with structure
group $H = SU(2)$, so we conclude that $\eta \geq 2c_1$ is a necessary
condition for the existence of a bundle with structure algebra
$\gsu_2$ (or greater).  This matches with the result found in
\cite{Berglund-Mayr}.  Similarly, bundles with structure $\gsu_3$ or
$\ggg_2$ correspond to gauge algebras ${\cal G} =\ge_6,\gf_4$, which
have $g_4$ or a lower term in $g$ or $f$ vanishing, so $-6K_2-2t = 2
\eta -6c_1= 2 (\eta -3c_1)$ is effective.  The results for these and
the other minimal gauge group types are shown in
Table~\ref{t:bm-bounds}, again in agreement with \cite{Berglund-Mayr}.
Note that these conditions for bundles with structure group $SU(2),
SU(3)$ precisely agree with the condition that $\eta-N c_1 (B_2)$ is
effective that was needed in \S\ref{sec:spectral-cover}, showing that
this constraint is more general and independent of bundle
construction.  
In the case where $G$ is trivial and the structure group on the
heterotic side is $E_8$, it is only necessary that $-6K_2-5t = 5 (\eta
- 24 c_1/5)$ be effective (corresponding to a nontrivial $g_1$).  For
$SU(N)$ groups with $N > 3$, on the other hand, $f_0$ and $g_0$ must
be nonvanishing so $-6K_2-6t = 6 (\eta -5c_1)$ must be effective and
$\eta \geq 5c_1$.

\begin{table}
\begin{center}
\begin{tabular}{|ccc |}
\hline
 ${\cal G}$ & ${\cal H}$ & % $q \leq$ & 
bound\\
\hline
$ \ge_8$ & 1 & %5/6 &
 $\eta =0$\\
$\ge_7$ & $\gsu_2$ & %3/4 &
$\eta \geq 2c_1$\\
$\ge_6, \gf_4$ & $\gsu_3, \ggg_2$ & %2/3 &
 $\eta \geq 3c_1$\\
$\gso_8, \ggg_2$ & $\gso_8,\gf_4$ & %1/2 &
 $\eta \geq 4c_1$\\
$\gsu_3$ & $\ge_6$ & %1/3 &
 $\eta \geq \frac{9}{2}  c_1$\\
$\gsu_2$ & $\ge_7$ & %1/4 &
 $\eta \geq \frac{14}{3}c_1 $\\
$1$ & $\ge_8$ & %1/6 &
 $\eta \geq \frac{24}{5}c_1 $\\
\hline
\end{tabular}
\caption[x]{\footnotesize Constraints on $\eta$ for certain structure
  groups $H$ of heterotic bundles,
identified from F-theory bounds on
  $\eta$ for a given 4D gauge group $G$; $\eta \geq ac_1$ means that
  $\eta -ac_1 (B_2)$ is an effective divisor on $B_2$.  For example, a
  heterotic bundle with structure group $SU(2)$ is only possible when
  $\eta -2c_1$ is effective.
These results match those found in \cite{Berglund-Mayr}.}
\label{t:bm-bounds}
\end{center}
\end{table}

\subsection{F-theory constraints and $SO(32)$ models}
\label{sec:constraint.32}

As discussed in Section \ref{sec:duality}, for any given base $B_2$ there is
a unique twist $T = 2 c_1 (B_2)$ so that F-theory on the resulting
${\cal B}_3$ is dual to the $SO(32)$ heterotic theory on a generic
elliptically fibered Calabi-Yau threefold over $B_2$.  In this case
the F-theory conditions are that $f_2$ is a section of ${\cal O}
(-4K_2-2T)={\cal O}_{{\cal B}_3}$ and $g_3$ is a section of ${\cal O}
(-6K_2-3T) ={\cal O}_{{\cal B}_3}$.  Both of these are therefore simply complex
numbers so that the cubic $x^3 + f_2x + g_3$ has three complex roots
and the resulting gauge algebra is $\gso_8$.  This matches the dual
heterotic theory where a generic choice of bundle will break the full
$SO(32)$ (really Spin (32)$/Z_2$)
down to $SO(8)$, just as in the 6D case where the F-theory dual of
heterotic on K3 is F-theory on an elliptically fibered Calabi-Yau
threefold over the base $F_4$.

%%%%%%%%%%%%%%%%%%%%%%%%%%%

%\subsection{High degree of vanishing matter curves or Other Conditions of Interest???}
%{\bf Add discussion here}

\subsection{Base-point free condition}
\label{sec:bpf}

We now consider the conditions that the heterotic base-point free
condition imposes on F-theory geometry.    On the heterotic side, the
constraint is that $\eta_i$ is base-point free on $B_2$.  
As discussed in \S\ref{sec:geometry-constraints},
on a surface
this simply means that there does not exist any effective divisor
(curve) $D$ of negative self-intersection such that $\eta_i \cdot D <
0$.  In the F-theory picture this means that
\begin{equation}
(-6K_2 \pm T) \cdot D \geq 0
\label{eq:free-f}
\end{equation}
for all effective divisors $D$ in $B_2$.  Let us examine the
consequences of this condition for curves satisfying $D \cdot D = -n$
where $n = 1, 2$.

For a curve $D$ with
$n = 1$, we have $-K_2 \cdot  D = 1$, since $(K_2 + D) \cdot D =
2g-2 = -2$ where $D$ is a rational curve, so \eq{eq:free-f} becomes
\begin{equation}
6 \pm T \cdot D \geq 0 \,.
\label{eq:k}
\end{equation}
This is automatically satisfied from \eq{eq:t-1}.  Thus, the
base-point free condition imposes no additional conditions for
F-theory bases associated with twists over curves of self-intersection
$-1$.  In particular, this means that any F-theory base ${\cal B}_3$ formed
as a $\P^1$ bundle over a del Pezzo base $B_2$ automatically
gives rise to $\eta_i$ that satisfy the base-point free condition on
the heterotic side, since $B_2$ contains no curves of
self-intersection lower than $-1$.

Now consider the case of a curve $D$ with self-intersection $ -n
=-2$.  In this case, we have $-K_2 \cdot D = 0$, so \eq{eq:free-f}
becomes
\begin{equation}
\pm  T \cdot D \geq 0 \,.
\label{eq:l}
\end{equation}
This relation is only satisfied with both signs when $T \cdot D =
0$.  Thus, the base-point free condition will be violated whenever the
twist over a $-2$ curve has $T \cdot D = \pm 1$.  From \eq{eq:t-2},
these are the only nonzero possibilities.

We can now analyze the consequences for the gauge group on the
divisors $\Sigma_\pm$ when the base-point free condition is violated.
This corresponds to the condition (C) analyzed in Section
\ref{sec:constraints-global}.  Assume that $D$ is a curve of
self-intersection $-2$ with an associated twist having $T \cdot D =
+1$.  In this situation, $(-nK_2-(n-k)T) \cdot D = -(n-k)$.  
As described in Section \ref{sec:F-theory-twists},
this means
that $g_5, g_4, f_3, f_2$ vanish on $D$ to degree at least 1, $g_3,
g_2, f_1, f_0$ vanish on $D$ to degree at least 2, etc.  An immediate
consequence is that $f, g$ cannot vanish on $\Sigma_-$ to degrees $3,
5$ or higher, or they would vanish on $\Sigma_-\cap D$ to degrees $4,
6$.  This means that there cannot be a generic $\ge_7$ or $\ge_8$
gauge group on $\Sigma_-$.  Furthermore, if the degrees of vanishing
are $3, 4$ then $g_4$ cannot be a perfect square, since if it was then
it would vanish to degree two on $D$ and again $f, g$ would vanish to
degrees $4, 6$ on $\Sigma_-\cap D$.  
As reviewed in \S\ref{sec:duality-geometry},
this condition means that in the
$3, 4$ vanishing case the generic gauge group must be $\gf_4$ and not
$\ge_6$.  Thus, in the non-base-point free cases where an F-theory
construction is possible, the generic gauge group cannot be
$\ge_6,\ge_7,$ or $\ge_8$.  A similar consideration holds for the
generic gauge group on $\Sigma_+$ when $T \cdot D = -1$.

These conditions can be made more explicit in the toric context.
Given a divisor $D_i$ with $\tilde{t}_i = T \cdot D_i = 1$,
we have a local set of rays in the
toric fan as in \eq{eq:transformation}
\begin{eqnarray}
s_- & = & (0, 0, 1) \label{eq:canonical-s}\\
\tilde{w}_{i -1} & = & (1, 0, 0)\label{eq:canonical-1}\\
\tilde{w}_{i} & = & (0, -1, 0)\label{eq:canonical-2}\\
\tilde{w}_{i +1} & = & (-1, -2, 1) \,.\label{eq:canonical-3}
\end{eqnarray}
All monomials $m = (a, b, c) \in N^*$ in ${\cal F},{\cal G}$ then
satisfy the inequalities
(\ref{eq:inequality-1}-\ref{eq:inequality-3}),  as well as
\eq{eq:inequality-5}, which becomes
$c \geq a + 2b-B$ for $B = 4, 6$.
In addition, there must be at least one monomial $m$ that satisfies
\eq{eq:inequality-4}, $b-c > B$, for $B = 4$ or 6.

First, we can ask if it is possible to have a gauge group factor on
$\Sigma_-$ associated with a summand $\ge_7$ or $\ge_8$.  From
Table~\ref{t:Kodaira}, this would mean that every monomial $m = (a, b,
c)$ would have $c \geq -1$ for both  $B = 4$ and $B = 6$.
From Figure~\ref{f:twist-constraint}, it is clear that this is not
possible, however.  The only simultaneous solution
in $(b, c)$ to $c < b-B, b
\leq B, c \geq -1$  is $b = B, c = -1$, and this is ruled out for both
values of $B$ by $c \geq a + 2b-B$ since $a \geq -B$.  Thus, when the
base-point free condition is violated through $\tilde{t}_i= +1$ over a
$-2$ curve,
the gauge algebra summand
associated with $\Sigma_-$ cannot be $\ge_7$ or $\ge_8$.
A similar result follows for $\Sigma_+$ when $\tilde{t}_i= -1$.

Now let us consider the possible summands $\ge_6$ and $\gf_4$, both
associated with vanishing degrees of $3, 4$ for $f, g$ on $\Sigma_-$.
In this case the constraint for $B = 6$ is $c \geq -2$.  There is a
simultaneous solution to the inequalities for this value of $c$, given
by $m =( a, b, c) = (-6, 5, -2)$, again as depicted in
Figure~\ref{f:twist-constraint}.  So this combination of vanishing
degrees can be realized.  The distinction between $\ge_6$ and $\gf_4$
can be seen most easily from the leading term in the Weierstrass
coefficient $g = g_4 z^4 + g_5z^5 + \cdots$, in a local expansion in
coordinates around the relevant divisor, which in this case is
$\Sigma_-$.  If $g_4$ is a perfect square, the gauge algebra is $\ge_6$,
and otherwise it is $\gf_4$.  Since the only allowed
monomial at order $z^4$ is $g_4 = y$, which is not a perfect square,
the gauge algebra must always be $\gf_4$ when the base-point free
condition fails on $\eta_1 = 6c_1 + T$ and the degree of vanishing of
$f, g$ is $3, 4$ on $\Sigma_-$.  The same result follows for
$\Sigma_+$ when $\tilde{t}_i = -1$.

The upshot of this analysis is that the only gauge algebras that are
possible for the structure group of the bundle in the dual heterotic
model when the base-point free condition is violated are the
commutants of the possible gauge algebras of the 4D theory, namely
$\ge_8,\ge_7,\ge_6,\gf_4,\gso_8,$ and $\ggg_2$.  This is in good
agreement with what is known of heterotic/F-theory duality in these
cases, since the spectral cover construction for $SU(N)$
and $Sp(N)$ structure group bundles is not possible when the base
point free condition is violated. Furthermore, these results demonstrate that the base-point-free condition must be necessary for {\it any} heterotic bundle with structure group $SU(N)$ or $Sp(N$), independent of the spectral cover construction.

For the gauge groups associated with the
more general exceptional algebras, it is
expected that other bundle constructions such as the
more general ``cameral cover'' construction
\cite{Ron_Elliptic,donagi,Donagi:1998vw} will exist for the heterotic
bundles (though explicit conditions on bundle topology are at present
not as well understood in this context as they are in the case of spectral covers) and as a result, the base-point free condition 
is not necessarily a requirement for the construction of a sensible bundle.  
The analysis of this section suggests that all F-theory models on
$\P^1$ bundles over $B_2$'s that are generalized del Pezzo surfaces
have well-defined heterotic duals, even when the base-point free
condition is violated, though new tools may be needed for explicit
construction of the appropriate bundles on the heterotic side.

\subsection{A note of caution: G-flux}\label{gflux_caution}
To close this section, we
return briefly to a caveat mentioned in Section
\S\ref{sec:het_f_motivation} regarding the results presented here that
are based purely on F-theory geometry.  In deriving the bounds on
structure groups and $\eta$ in Sections
\ref{sec:duality}-\ref{sec:equivalence} we have {\it ignored G-flux}
which must be taken into account for a full description of the F-theory
physics and dual heterotic bundle moduli space.
Some general aspects of G-flux in 4D F-theory models, and relevant
references are given in \S\ref{chiralsec}, \S\ref{sec:example-G-flux}.

In some cases it may be possible for non-trivial G-flux (in the
singular limit of the fourfold geometry) to change the apparent
symmetry group that would be inferred from the Weierstrass
equation. Although counterintuitive from the perspective of Abelian
G-flux in a smooth M-theory limit, such symmetry-breaking by flux can
be generic in the singular limit and is expected to occur in a wide
range of 4D F-theory models. This has recently been explored in the
context of local F-theory models as ``T-branes''
\cite{Donagi:2003hh,Cecotti:2010bp} (or equivalently ``gluing data''
\cite{Donagi:2011jy,Donagi:2011dv}) and in the global context in both
$4$- \cite{Marsano:2012bf,Braun:2013cb} and $6$-dimensional
compactifications \cite{Anderson:2013rka}.

The basic mechanism by which G-flux can break an apparent symmetry
appearing from the geometric F-theory analysis in terms of a
Weierstrass model is most clear in the dual heterotic picture.  As
discussed in \S\ref{sec:duality-geometry}, the geometry of the
F-theory base ${\cal B}_3$ determines almost all of the topology of
the corresponding bundle on the heterotic side.  The components
$\zeta_i$ in \eref{eq:f2-decomposition}, however, are not determined by the $4$-fold geometry.  On the
heterotic side, non-trivial bundles with a second Chern class entirely in
$\zeta_i$ can break the gauge group just as effectively as bundles
with non-vanishing $\eta_i$ that have a clear dual in F-theory geometry.
The symmetry breaking bundles corresponding to the topology $c_2(V_i) \sim \zeta_i$ will not be visible in the F-theory geometry, and can only be seen in F-theory when G-flux is
correctly incorporated.

In the context of the present investigation, we hope to explore the
full moduli/vacuum space of the dual theories including G-flux in
future work. For now, however, we simply provide an
illustration of where the purely geometric criteria may miss solutions
involving exotic G-flux, in an example appearing in \S\ref{sec:example-G-flux}.

\section{Examples}
\label{sec:examples}

We examine some specific examples of F-theory models on bases ${\cal B}_3$
that are $\P^1$ bundles over various bases $B_2$ and which
illustrate various features discussed in the main part of the paper.

\subsection{$\P^1$ bundles over $\P^2$}

We begin with the simplest example, taking ${\cal B}_3$ to be
a $\P^1$ fibration over $B_2 = \P^2$.  
The effective divisors on $\P^2$ are multiples $nH$ of the hyperplane
class $H$ with $H \cdot H = 1$.  The (anti-)canonical class is
$-K_2 = 3H$.  The general constraint on
the twist $T =tH$ is that $| t | \leq 18$; this is the analogue of the
constraints
(\ref{eq:t-0}-\ref{eq:t-big}) for a curve of self-intersection $+1$.
This class of F-theory models and their heterotic duals was described
in \cite{Klemm:1996ts, Berglund-Mayr}.

From the general classification of allowed F-theory models it follows
that there is a valid model with $t =  18$ ($-6K_2-tH = 0$),  and that
there are valid models for $0 \leq t \leq 12$ ($-4K_2-tH$ effective).
Models with negative $t$ are equivalent under reflection to those with
positive $t$.  The twists $13 \leq t \leq 17$ correspond to bases
${\cal B}_3$ in which $g_5$ vanishes on some curves on the $\ge_8$
locus $\Sigma_-$, which must be blown up for a smooth threefold base,
analogous to the Hirzebruch surfaces $F_{9, 10, 11}$ in the 6D construction.
Since the base $B_2$ has no curves of self-intersection $-2$ or below,
the base-point free condition is never violated.  The resulting 4D
supergravity models have a range of gauge group factors according to
the value of $t$.  We thus have a total of 14 distinct bases ${\cal
  B}_3$ corresponding to different twists in a $\P^1$ bundle over
$\P^2$ giving F-theory models with distinct smooth heterotic duals.

These features can be seen explicitly through a toric
construction, where 
the parts of the
fan from $B_2$ are
\begin{equation}
w_0 = (0, 1, 0), w_1 = (1, 0, 0), w_2 = (-1, -1, t) \,.
\label{eq:o}
\end{equation}
Here we have used linear transformations to set the component of $T$
to 0 for the first two vectors.
The F-theory condition that  $f, g$ do not automatically
vanish to degrees $4, 6$ on $\Sigma_-$ corresponds to the condition that
$t \leq 18$ in the toric picture since
the plane spanned by $w_1, w_2, w_3$ intersects the third
axis at $(0, 0, t/3)$, so $t/3 \leq 6$.
(In general, the condition that $g$ not vanish
on a divisor such as $\Sigma_-$ at degree $6$
can be described in any toric case as the condition that the plane
spanning the $w_i$'s intersect the $z$ axis at a value less or equal
to 6, as can be verified geometrically.\footnote{Thanks to L.\ Swanson
for discussions on this point.})
The corresponding condition for $\Sigma_+$ gives $t \geq -18$.  

On the heterotic side, the choice of $t$ corresponds to the bundle
decomposition where $\eta_{1, 2} = 18 \pm t \geq 0$ is the number of
instanton factors in each component of the gauge group, which must be
nonnegative associated with the condition that $\eta_i$ is effective.

Specific examples with generic gauge algebras $\gsu (2)$ and $\ggg_2$ arise in the cases $t = 4, 5$.  For $t = 4$, $f$ and $g$ vanish
to degrees $1, 2$ on $\Sigma_-$, as can be seen from the fact that
$f_0, g_0$, and $g_1$ must all vanish as the corresponding divisors
$-nK_2-(n-k)tH = -4H, -6H, -2H$ are all non-effective.  
In this case the gauge algebra $\gsu(2)$  cannot be enhanced to an $\gsu(5)$, though the algebra can
be enhanced to the exceptional series ${\mathfrak g}_2, \gf_4,\ge_6, \ge_7,
\ge_8$.
In the case
$t = 5$, $f, g$ vanish to degrees (2, 3), and the gauge algebra is the
generic $\ggg_2$.

The unique model with an $SO(32)$ heterotic dual is the model with $t
= 6$ ($tH = -2K_2$), with gauge algebra $\gso (8)$, parallel to the 6D
model on $\F_4$.

\subsection{$\P^1$ bundles over $\F_m$}
\label{sec:examples-fm}

Now consider F-theory models where the base $B_2$ is a Hirzebruch
surface.  Some of these models were also discussed in
\cite{Klemm:1996ts}.  The cases $\F_0$ and $\F_1$ are qualitatively
similar to the models described in the previous section,
and are always base-point free as there are no $-2$ curves in the
base.  The cases $\F_m$ with $m > 2$ correspond to singular geometries
on the heterotic side.  We briefly describe the cases $\F_0$ and
$\F_1$ and then focus on the case $\F_2$.  For any Hirzebruch surface
$\F_m$, the cone of effective divisors is generated by divisors $S, F$
with $S\cdot S= -m, F \cdot F = 0, S\cdot F = 1$, and $-K_2 = 2S+ (2 +
m)F$.

For $\F_0$, we have $-K_2 = 2S+ 2F$, and we can parameterize $T = a S+
bF$.  There are symmetries under $T \rightarrow -T$ and $a
\leftrightarrow b$.  The general constraints on twists over curves of
self-intersection 0 give $-12 \leq a, b \leq 12$.  There is a
single $\ge_8$ model with $a = b = 12$ ($T = -6K_2$).  For all other
good bases $-4K_2\pm T$ are effective, so $-8 \leq a, b \leq 8$.  Up
to symmetries this gives $81 + 1 = 82$ distinct twists associated with
valid F-theory models.
As in the models over $B_2 = \P^2$, there are a variety of gauge
groups associated with the different twists.

For $\F_1$, we have $-K_2 = 2S+ 3F$, and we can again parameterize $T
= a S+ bF$.  There is a symmetry under $T \rightarrow -T$.  The
general constraints on twists over curves of self-intersection 0 and
$-1$ give $| a |= T \cdot F \leq 12, | b-a |=T \cdot S \leq 6$.
There is a single $\ge_8$ model with $a = 12,b = 18$ ($T = -6K_2$).
For all other good bases $-4K_2\pm T$ are effective, so $| a |\leq 8,
| b |\leq 12$.  Up to the sign symmetry of $T$ this gives $108 + 1 =
109$ distinct twists associated with valid F-theory models.  
Again, there are a variety of gauge groups
associated with the different twists, and all models have $\eta_\pm =
-6K_2 \pm T$ base-point free.

A situation mentioned in \S\ref{sec:constraints-toric} occurs for
several $\P^1$ bundles ${\cal B}_3$ over $B_2 = \F_1$, where a
codimension two singularity arises on a curve despite the absence of
gauge groups from codimension one singularities.
A sample example of this occurs for the twist $T = 2F$.  In this case,
$-nK-(n-k) T = 2nS+ (n + 2k)F$ has a negative intersection with $S$
when $n > 2k$, so $f_0, f_1, g_0, g_1, g_2$ all vanish on $S$, giving
a $(2, 3)$ codimension two singularity type over the curve $S\cap
\Sigma_-$ although there is no gauge group on $\Sigma_-$ as
$-nK-(n-k) T$ are all effective for $n = 4, 6$, $0 \leq k < n$.

Now we consider models with $B_2 =\F_2$.
We have $-K_2 = 2S+ 4F$, and we parameterize $T
= a S+ bF$, with a symmetry under $T \rightarrow -T$
exchanging $\Sigma_\pm$.  The
general constraints on twists over curves of self-intersection 0 and
$-2$ constrain $| a |= T \cdot F \leq 12, | b-2a |=T \cdot S \leq 1$.
There is a single $\ge_8$ model with $a = 12,b = 24$ ($T = -6K_2$).
The models with $b = 2a$ are base-point free; from the constraint that
$-4K_2-T$ be effective the base-point free models have $0 \leq a \leq
8$ up to symmetry, so along with the $\ge_8$ model there are 10
base-point free configurations.  For the non-base-point free
configurations, up to the sign symmetry on $T$ we can choose $b = 2a
+1$, from which
\begin{equation}
-3K_2-T = ( 6-a) S+ ( 12-2a-1) F
% \label{eq:}
\end{equation}
must be effective, and similar for $-3K_2 + T$.  This constrains
\begin{equation}
-8 \leq a \leq 5
\end{equation}
so there are 14 non-base-point free configurations, with gauge groups
up to $\gf_4$ on $\Sigma_-$.

We can describe these cases explicitly in toric language.
The toric fan for ${\cal B}_3$
contains the rays
\begin{eqnarray}
s_{\pm} & = &  (0, 0, \pm 1)\\
w_0 & = &  (0, 1, 0)\\
w_1 & = &  (1, 0, 0)\\
w_2 & = &  (0, -1, a)\\
w_3 & = &  (0, -1, b)\,.
\end{eqnarray}
The twist $T$
is parameterized by the integers $a, b$. 
An explicit computation of the monomials  in the dual lattice
that satisfy
$\langle m, w_\alpha \rangle \geq -4, -6$
confirms that the cases described above are the only ones for which
the F-theory model is acceptable and that in all these cases $f, g$
have acceptable degrees of vanishing on all divisors and curves.

We consider explicitly the cases where $\eta_-= -6K_2-T$ fails the
base-point free condition.
When $a < 2$, $b = 2a +1$, there is no vanishing of $f, g$ on
$\Sigma_-$.  
For the twist combination $(a, b) = (2, 5)$, the vanishing degrees 
are
1, 2, so the gauge algebra contribution from $\Sigma_-$ is $\gsu_2$.
This should correspond on the heterotic side to an $E_7$
structure bundle on the
Calabi-Yau described by the generic elliptic fibration over $\F_2$
that violates the base-point free condition, with
$\eta = 6c_1 -T = 10S+ 19F$, which is not
base-point free since $\eta \cdot S= -1$.

A similar analysis for the twist combination $(a, b) = (3, 7)$ gives
vanishing degrees of $f, g$
on $\Sigma_-$ 2, 3, for a gauge algebra factor of ${\mathfrak g}_2$ and a
dual heterotic bundle structure group of $\gf_4$.  In this case the
bundle has $\eta = 9S+ 17F$.  For the twists $(a, b) = (4, 9)$ and
$(5, 11)$ the vanishing degrees are 3, 4, so the gauge algebra
contribution on $\Sigma_-$ is $\gf_4$, and the heterotic structure
group is ${\mathfrak g}_2$.

This gives a number of explicit examples of F-theory constructions
that violate the base-point free condition, where the dual heterotic
model should nonetheless exist with a bundle having an exceptional
structure group.

\subsection{An F-theory model over  $B_2 =\F_3$ with 
a codimension two singularity but
no gauge
  group}
\label{sec:examples.3-no-group}

As discussed in \S\ref{sec:duality-bases}, all elliptically fibered
Calabi-Yau geometries over the base $B_2 =\F_3$ are singular  due to
the Kodaira singularity over the $-3$ curve $D$ in  the base with vanishing
of $f, g$ to degrees (2, 2) or greater.  This means that no F-theory
models on any base ${\cal B}_3$ over $B_2 =\F_3$ can have a smooth
heterotic dual.  In most cases, the absence of the heterotic dual is
made particularly clear by the appearance of an additional gauge group
factor in the 4D F-theory model over the divisor in ${\cal B}_3$
associated with $D$.  In the singular heterotic dual theory this would
correspond to an additional gauge factor arising at the singularity in
the Calabi-Yau geometry.  It is interesting to note, however, that for
certain values of the  twist $T$, a $\P^1$ bundle ${\cal B}_3$ over
$B_2 = \F_3$ can be constructed so that there is no extra nonabelian
gauge group factor.  For example, with the twist $T = F$, there is no
divisor that must carry a gauge group factor, though there is a
codimension two singularity where $(f, g)$ vanish to degrees $(3, 4)$
on the curve $S\cap \Sigma_-$.  These assertions can easily be checked
explicitly using the monomials computed in the toric description.

\subsection{dP$_2$}

The second del Pezzo surface, dP$_2$, is constructed by blowing up
$\P^2$ at two points, giving a pair of exceptional divisors $E_1,
E_2$ with $E_1 \cdot E_1 = E_2 \cdot E_2 = -1, E_1 \cdot E_2 = 0$.
The proper transform of the line passing through the two points is a
third -1 curve $F = H - E_1 - E_2$, with $F \cdot E_1 = F \cdot E_2 =
1$.  The cone of effective divisors is spanned by $F, E_1, E_2$, which
we can write as $(1, -1, -1), (0,  1, 0), (0, 0, 1)$ in a basis where
the intersection product is diag$(1, -1, -1)$.
dP$_2$ also has a simple toric presentation, but we use this more
abstract formulation for the del Pezzo examples
to illustrate how the methods of this paper can
be implemented outside the toric context.
The (anti-)canonical class of dP$_2$ is the proper transform of $- K =
3H$ on $\P^2$,
\begin{equation}
 - K = 3F + 2E_1 + 2E_2 \,.
% \label{eq:}
\end{equation}
If we parameterize the twist as
\begin{equation}
T = a F + b E_1 + c E_2 \,,
% \label{eq:}
\end{equation}
there are symmetries under $b\leftrightarrow c$ and
$(a, b, c) \leftrightarrow (- a, - b, - c)$.

We can now count the set of allowed twists $T$ using the conditions
described in \S\ref{sec:F-theory-twists} and
\S\ref{sec:constraints-global}.  From the analysis in
\S\ref{sec:constraints-curves}, we know this gives a necessary and
sufficient set of conditions for the set of allowed ${\cal  B}_3$'s.
There is a single twist $T = -6K$ of class (A).
There are no -2 curves in the base, so all other ${\cal  B}_3$'s are
of type (B).  The constraints that
\begin{equation}
 -4K \pm T = (12 \pm a) F + (8 \pm b) E_1 + (8 \pm c) E_2
% \label{eq:}
\end{equation}
are effective
constrain $| a | \leq 12, | b | \leq 8, | c | \leq 8$.  The
constraints that $| T \cdot E_1 |, | T \cdot E_2 |, | T \cdot F | \leq
6$ imply $| a - b | \leq 6, | a - c | \leq 6, | a - b - c |\leq 6$.
Up to the symmetries listed above,
there are 471 distinct $T$'s that satisfy these conditions, so the
number of distinct $\P^1$ bundles ${\cal  B}_3$ over $B_2 =$ dP$_2$
that give good F-theory models is 472.
This agrees with a direct analysis using toric methods, as described
in Section \ref{sec:enumeration}.

\subsection{dP$_3$}

The story for dP$_3$ is similar to that for dP$_2$.   Blowing up
$\P^2$ at 3 generic points gives 3 exceptional divisors $E_1, E_2,
E_3,$ and  three  $-1$ curves $X_1 = H - E_2 - E_3, X_2 = H -  E_1 - E_3, X_3
= H - E_1 - E_2$ from the proper transforms of the lines connecting
each pair of points.  In a basis with intersection form diag$(1, -1,
-1, -1)$ we have
\begin{eqnarray}
X_1 = (1, 0, -1, -1) & & E_1 = (0, 1, 0, 0) \\
X_2 = (1, -1, 0, -1) & & E_2 = (0, 0, 1, 0) \\
X_3 = (1, -1, -1, 0) & & E_3 = (0, 0, 0, 1) 
\end{eqnarray}
There are symmetries under the 6 permutations of the indices $i = 1,
2, 3,$ and under $E_i \leftrightarrow X_i$, which maps
\begin{equation}
 (a, -b_1, -b_2, -b_3) \leftrightarrow
(2a -b_1 -b_2  -b_3, a -b_2 -b_3, a -b_1 -b_3, a -b_1 -b_2)
% \label{eq:}
\end{equation}
In the toric picture this can be seen as the 12-fold dihedral symmetry
group $D_6$ of the regular hexagon.  A divisor $D = (a, - b_1, - b_2,
- b_3)$ is effective if $a \geq 0, b_1 + b_2 + b_3 \leq 2a$.  The
 (anti-) canonical class of $B_2 =$ dP$_3$ is again the proper
transform of $H$
\begin{equation}
 - K = (3, -1, -1, -1) =\sum_{i} X_i+\sum_{i} E_i.
% \label{eq:}
\end{equation}
There is one base ${\cal  B}_3$ of type (A), with $T = -6K$.  To
enumerate bases ${\cal  B}_3$ of type (B), following the analysis of
\S\ref{sec:constraints-curves}, it is sufficient to identify all
twists $T = (a, - b_1, - b_2, - b_3)$ so that the local twist
conditions
\begin{eqnarray}
| T \cdot E_i | & = &  | b_i | \leq 6 \\
| T \cdot X_i | & = &   | a - b_j - b_k | \leq 6,\;\;\;\;
i, j, k\ {\rm  distinct}\ \in \{1, 2, 3\}
\end{eqnarray}
are satisfied and $-4K \pm T = (12 \pm a, -4 \mp b_1, -4 \mp b_2, -4
\mp b_3)$ is effective, which implies
\begin{equation}
 | a | \leq 12,\;\;\;\;\;\;| 2a - b_1 - b_2 - b_3 | \leq 12 \,.
% \label{eq:}
\end{equation}
A simple enumeration shows that up to the $D_6$ symmetry group there
are 775 solutions of all these conditions, so a total of 776 distinct
possible bases ${\cal B}_3$ that are $\P^1$ bundles over dP$_3$.
As for dP$_2$, this result agrees with the explicit enumeration done
using toric methods described in Section \ref{sec:enumeration}.

\subsection{dP$_4$}
\label{sec:dp4}

For dP$_n$ the analysis is again similar to dP$_3$ and dP$_2$, though
there is no toric construction, there are more symmetries, and the
effectiveness condition is increasingly complicated as the number of
-1 curves on the wall of the cone of effective divisors increases.

For dP$_4$ there are 4 exceptional divisors $E_i,$ $E_1 = (0, 1, 0, 0,
0), \ldots E_4 = (0,  0, 0, 0, 1)$, and 6 proper transforms of lines
$X_{i j}$, $X_{12}= (1,-1, -1, 0, 0), \ldots$.  The intersection form
is diag (1, -1, -1, -1, -1), with (anti-)canonical  class $- K = (3,
-1, -1, -1, -1)$.
There are symmetries under arbitrary permutations of the
$i$'s, as well as additional symmetries of the form
\begin{equation}
E_1 \leftrightarrow X_{23}, \;\;\;\;\;
E_2 \leftrightarrow X_{13}, \;\;\;\;\;
E_3 \leftrightarrow X_{12}, \;\;\;\;\;
 X_{i4}, E_{4} \; {\rm fixed}
% \label{eq:}
\end{equation}
The full symmetry group is of order $5!  = 120$, and can be seen most
clearly by redefining $\tilde{X}_{0i}= E_i, \tilde{X}_{i
  j}=\frac{1}{2}  |\epsilon_{i j k l}| X_{k l}$, for which the nonzero
intersection products are  $\tilde{X}_{\mu \nu} \cdot \tilde{X}_{\mu
  \nu} = -1,
\tilde{X}_{\mu \nu}\cdot
\tilde{X}_{\lambda \sigma}= 1$ when $\mu, \nu, \lambda, \sigma\in
\{0, 1, 2, 3, 4\}$ are distinct; in terms of the $\tilde{X}$'s, the symmetry group is
simply the set of permutations on all 5 possible index values.

In dP$_4$, a divisor $D = (a, - b_1, - b_2, - b_3, - b_4)$ is
effective iff $b_i \leq a \,\forall i, \sum_i  b_i \leq 2a,$ which can
be seen from the conditions that  $D$ must be formed from a positive
integral linear combination of $X_{i j}$'s and $E_i$'s.
The twist $T = (s, t_1, t_2, t_3, t_4)$ must satisfy the conditions
\begin{equation}
 | T \cdot E_i | = | t_i | \leq 6, \;\;\;\;\;
| T \cdot X_{i j} | = | s - t_i - t_j | \leq 6 \,.
% \label{eq:}
\end{equation}
As in the previous del Pezzo examples, there is one solution with $T =
-6K$, and we can enumerate all solutions with $-4K \pm T$ effective,
which along with the twist conditions (\ref{eq:t-0}-\ref{eq:t-big})
give necessary and sufficient conditions for a good base ${\cal
  B}_3$.  An explicit enumeration shows that after taking account of
symmetry there are 6976 distinct bases ${\cal  B}_3$ of type (B), for
a total of 6977 ${\cal  B}_3$ that are $\P^{1}$ fibrations over
dP$_4$.

For higher dP$_n$, and for generalized del Pezzo surfaces, the
analysis can be carried out in a similar fashion.  As $n$ increases,
however, the details of the calculation become more complicated.  For
dP$_5$, for example, there is an additional -1 curve from a conic
passing through all 5 blown up points, $C = 2H - E_1 - E_2 - E_3 - E_4
- E_5$, which complicates the effectiveness condition on divisors.  In
principle, however, for any base $B_2$, the number of twists
satisfying the local twist conditions is finite, and the determination
of the full set of ${\cal B}_3$'s over $B_2$ can be done efficiently
and explicitly.

\subsection{An example of  an upper and lower bound on $\eta$}\label{H_stuck_eg}
One of the more novel observations of this study is the fact that for
certain fourfold geometries there exist generic symmetries that {\it
  can be neither broken (Higgsed) or enhanced at any  points in
  the complex structure moduli space of $Y_4$}. These restrictions
arise because of a variety of features, however all the failures of
``enhancement'' occur because of too-high a degree of vanishing of
$(f,g)$ on divisors and curves as described in Section
\ref{sec:equivalence}.

The consequences of having a twist $T$ of the
$\mathbb{P}^1$-fibered base ${\cal B}_3$ on the F-theory side that
gives rise to such a restrictive condition on the gauge group
corresponds in the heterotic
geometry to a choice of partial bundle topology $\eta$ for which only one
structure group $H$ is possible (subject once again to the caveats
arising from ignoring G-flux, see Section
\ref{gflux_caution})

Let us consider here an example of this type for which only one
symmetry is possible and all Higgsing/enhancing is forbidden. 
This is the case for the base ${\cal B}_3$ defined by the $\P^1$
bundle with twist $T=5S+11F$ on $\mathbb{F}_2$. Constructing the
generic Weierstrass model over this base ${\cal B}_3$, it is
straightforward to verify that this $Y_4$ manifests a generic $F_4$
symmetry.

In the dual heterotic theory, this corresponds to a $G_2$ bundle over
the threefold $X_3$ with $\eta=7S+13F$. More precisely, the 
heterotic bundle $V_1$ has $\eta_{-}=7S+13F$ and is a $G_2$
bundle. The second bundle, associated to $\eta_{+}$, has generic $E_8$
structure group and hence one $E_8$ factor is generically
completely broken and
will not concern us further. More explicitly, we have the following
Weierstrass equation for $Y_4$:

\begin{equation}\label{e6f4}
Y^2=X^3+(f_3z^3+f_4z^4 +...)X+(g_4z^4+g_5z^5+g_6z^6 + ...)
\end{equation}
(Here $(X,Y, Z = 1)$ are the coordinates on the elliptic fiber of the
$CY_4$, while $z=0$ defines the 7-brane locus ({\it i.e.}, the section
$\Sigma_-$) inside of ${\cal B}_3$ and is of $E_6/F_4$ type according to
Kodaira-Tate.) As mentioned in previous sections, if $g_4$ is a perfect
square then the symmetry is $E_6$, and for more general polynomials
it is $F_4$.

For the twist $T=5S+11F$, \eref{e6f4} is the generic form of the
Weierstrass model for arbitrary complex structure. The fact that the
$F_4$ symmetry is generic ({\it i.e.}, cannot be Higgsed) from the
point of view of the F-theory Weierstrass model has the same natural
low-energy 4D
interpretation on both sides of the duality -- there is simply no
charged matter available to get a vev in vacuum.

For the given twist, all additional tunings of the complex structure
that might increase the gauge group on $\Sigma_-$
induce non-CY singularities. As an example, consider the
specialization of $g_4=\alpha^2$ for some polynomial $\alpha$ of the
appropriate degree. Here the vanishing degree of $(f,g, \Delta)$
increases from the generic values of $(4,5,10)$ on the curve
$\Sigma_-\cap S$ to
$(4,6,12)$ on the same curve, and hence the
singularity cannot be resolved without going to a different F-theory
base ${\cal B}_3$ by blowing up the curve.

In this case, the restriction on enhancement has a clear
interpretation in terms of the heterotic bundle geometry. An
enhancement of the symmetry of $F_4 \to E_6$ for example, corresponds
in the heterotic theory to a {\it reduction in rank} of the associated
bundle from $G_2 \to SU(3)$. For the case at hand this would indicate
that by tuning the complex structure of $Y_4$ we were inducing a
``splitting'' of the vector bundle. In terms of the associated vector
bundles the following reduction of representations in $G_2 \to SU(3)$,
\begin{equation}
{\bf 7}\to {\bf 3} + {\bf {\bar 3}}+ {\bf 1}
\end{equation}
would lead to
\begin{equation}\label{vector_decompG2}
V_{7}=V_{3} \oplus {V_{3}}^{\vee} \oplus {\cal O}_{X_3}
\end{equation}
the fact, however, that this tuning leads to a badly singular $Y_4$ indicates
that a generic $G_2$ bundle with $\eta=7S+13F$ cannot be decomposed
as in \eref{vector_decompG2} for {\it smooth $SU(3)$ bundles
  $V_3$}. Instead, any such decomposition must lead to non-locally
free sheaves ({\it i.e.}, heterotic ``small instantons''
\cite{Donagi:1999jp,Ovrut:2000qi,Buchbinder:2002ji}) and a degenerate
limit of the theory.

For this choice of twist, we have an additional confirmation of this
heterotic result in the fact that $\eta$ is not base-point-free. Thus
by the arguments of Section \ref{sec:equivalence}, we cannot define
any smooth $SU(3)$ spectral cover bundle to play the role of $V_3$ in
\eref{vector_decompG2}. Although the spectral cover construction is
not guaranteed to be representative for the bundle moduli space in
general, the consistency conditions on $Y_4$ applied to these
``non-enhanceable'' geometries indicate that if a generic symmetry $F_4$
is not base-point free, it will be impossible to enhance the
symmetry for special values of the complex structure (compatible with
a CY resolution). This provides an interesting window into the moduli
space of all such $G_2$-bundles by providing general restrictions on possible
decompositions like the one given above.

\subsection{Examples with non-trivial chiral matter}\label{chiralsec}
In the previous examples we have seen that the F-theory fourfold
geometry frequently encodes otherwise hard to obtain information about
the moduli space of vector bundles on heterotic CY threefolds. In this
section, we  use the heterotic theory to obtain new information
about the matter spectrum of a $4D$ effective F-theory. To accomplish
this, we return to the formulas for chiral matter given in Section
\ref{sec:hetmatter}.

As an example, let us consider $E_6$ theories in the dual
heterotic/F-theory geometry. As discussed in Section
\ref{spec_cov_matter}, for an $SU(N)$ bundle described via a spectral
cover the chiral index is \cite{Curio:1998vu,Curio:2011eu} 
\begin{equation}\label{intersec_example} 
{\rm Ind}(V)=-h^1(X,V)+h^1(X, V^{\vee}) = \lambda [\eta] \cdot [\eta +NK_2]
\end{equation}
where $\lambda$ is defined by \eref{l_top} and \eref{lambda_n}. Thus,
the chiral index is proportional to a simple geometric intersection of
the curve $[\eta]$ in the base with the matter curve $[\eta +
  NK_2]$.

To understand the significance of this geometry, the case of $E_6$ theories is particularly interesting because for $SU(3)$ bundles described as spectral covers we can guarantee that the constant $\lambda$ is non-vanishing. Recall from \eref{lambda_n} that for some integer $m$
\begin{equation}\label{lambda_again}
    \lambda= 
\begin{cases}
    m+\frac{1}{2},& \text{if }~N ~\text{is odd}\\
    m,              & \text{if}~ N~\text{is even}
\end{cases}
\end{equation}
Thus, for $SU(3)$ bundles/$E_6$ theories it is required that $\lambda \neq 0$ and the question of whether or not the theory has chiral matter can be reduced to a of question of
intersection theory for the matter curve $[\eta + 3K_2]$ in
the $2$-fold base.  We will be interested in whether or not this curve
is reducible and whether or not it has non-trivial intersection with
$[\eta]$.

To illustrate the possibilities, we can consider the four generic $E_6$ theories over the base $B_2=\mathbb{F}_1$. There the twists
\begin{equation}
T=nS+9F~~~,~~~3 \leq n \leq 6
\end{equation}
all give rise to $E_6$ symmetries on $\Sigma_{-}$. This is easy to
check since the coefficient $g_4$ in the Weierstrass equation
\begin{equation}\label{e6f4_again}
Y^2=X^3+(f_3z^3+f_4z^4 +...)X+(g_4z^4+g_5z^5+g_6z^6 + ...)
\end{equation}
satisfies
$g_4 \in H^0(B_2,{\cal O} (\eta)^{\otimes 2} \otimes K_2^{\otimes 6})
=H^0 (B_2, {\cal O} (-6K_2-2T))
=H^0(\mathbb{F}_1,2(6-n)S)$, which
indicates that $g_4$ is a perfect square in these cases\footnote{Note
  that the line bundle cohomology over $\mathbb{F}_n$ can be shown to
  satisfy: $h^0(\mathbb{F}_n, {\cal O}(S))=1$ for the divisor
  $S^2=-n$. Hence if $s$ is the toric coordinate associated to the
  divisor $S$, the generic (only) element of $H^0(\mathbb{F}_n, nS)$
  is $s^n$. } and hence (as described in the previous section) the
fiber type is split to $E_6$ (rather than the generic, non-split
$F_4$). Moreover, for $3 \leq n \leq 6$ we have the heterotic
topology:
\begin{equation}
\eta=(12-n)S+9F~~~~\text{and}~~~~\eta + 3 K_2= (6-n)S
\end{equation}
For these $SU(3)$ bundles, $\eta$ is effective and base-point-free,
$\eta -3c_1$ is effective, and it is straightforward to verify that
simple line bundles $L_{\cal S}$ of the form \eref{l_top} can be found
with $\lambda \neq 0$. Thus, we are guaranteed that smooth spectral
cover bundles exist 
(stable in the appropriate adiabatic region in K\"ahler moduli space).  

In the case that $g_4=\alpha^2$ the polynomials in \eref{e6f4_again} appear as a $SU(3)$ spectral cover inside the heterotic $CY_3$ of the form
\begin{equation}\label{su3eg}
g_5 {\hat Z}^3+ f_3 {\hat X}{\hat Z}+ \alpha {\hat Y}=0
\end{equation}
where $({\hat X},{\hat Y},{\hat Z})$ are coordinates on the $CY_3$ elliptic fiber (the above equation gives three points on the elliptic fiber for each point on the $\mathbb{F}_2$ base, as expected for an $SU(3)$ spectral cover). As usual, $f_4,g_6$ in the fourfold Weierstrass equation above appear as the coefficients in the $CY_3$ Weierstrass
\begin{equation}
{\hat Y}^2={\hat X}^3 +f_4 {\hat X} + g_6
\end{equation}
Recall from the arguments of \sref{spec_cov_matter} that the coefficient $\alpha \in H^0({\cal O}(\eta) \otimes K_2^{\otimes 3})=H^0(\mathbb{F}_2,{\cal O}((6-n)S)$ in \eref{su3eg} defines the ``matter curve'', $\alpha=0$. This is where the ${\bf 27}$-type matter is localized in both the heterotic/F-theory geometries.

Thus, for this class of bundles \eref{lambda_n} and \eref{c3spec} can
be used to straightforwardly compute 
\begin{equation}
{\rm Ind}(V)=\text{no. of}~ {\bf \overline{27}}'s-\text{no. of}~ {\bf 27}'s = (6-n)(n-3)
\end{equation}
From this we see that the cases $n=3$ and $n=6$ have chiral index
zero, but for $n=4,5$ the theory {\it must} have chiral matter. Given
the full defining data of the bundle, the exact multiplicity of the
${\bf 27}$'s and ${\bf \overline{27}}$s could be computed using Leray
spectral sequences \cite{hartshorne1977algebraic}, but even at this
preliminary level, the results of the chiral index are intriguing. For
these dual geometries, given a value of $\eta$, it can immediately be
determined whether or not the theory contains chiral matter.

Of course, we derived the necessity of a non-vanishing chiral index for an $SU(3)$ bundle described as a spectral cover, but the third Chern class (and hence the Chiral index) is a topological invariant in the bundle moduli space. As a result, so long as a good $SU(3)$ spectral cover bundle exists, we can use it as a probe to extract the structure of the full moduli space of hermitian bundles, all of which must have non-vanishing index! As  discussed in \S\ref{sec:bpf}, all CY fourfold geometries with generic $E_6$ symmetry and smooth heterotic duals satisfy the base-point-free condition and can be described by well-behaved spectral cover bundles as in Section \ref{sec:spectral-cover}. 

It would be interesting to investigate the question of chiral matter
in this context more directly on the F-theory side in the future, in
particular
by including G-flux.
Some general aspects of how G-flux can be incorporated in 4D F-theory
models are described in
 \cite{Grimm:2009sy,Grimm:2009ef,Bies:2014sra, Bizet:2014uua}.
Progress has been made in understanding how chiral matter in
F-theory models can be determined in the presence of G-flux based on
aspects of the spectral cover construction \cite{Andreas:1999ng,Candelas:2000nc,Donagi:2008ca,
  Beasley:2008dc, Hayashi:2008ba, Braun:2011zm, Marsano:2011hv,
Krause:2011xj} and more directly from the M-theory description
\cite{Grimm:2011fx}, but a more complete and directly computable
formulation is desirable.
An inspection of \eref{e6f4_again}
shows that the points determined by the intersection
\begin{equation}
[\eta] \cdot [\eta +NK_2]
\end{equation}
in the F-theory geometry corresponds exactly to the simultaneous vanishing of $g_4$ and $g_5$ at points in the base -- that is, the chiral index in these cases is counted by co-dimension $3$ singular loci in the $4$-fold geometry. The observation that chiral matter and co-dimension $3$ singularities (and associated $G$-flux) could be linked (at least in $K3$-fibered $4$-folds) has been observed for some time \cite{Andreas:1999ng,Candelas:2000nc} and has been used more recently in F-theory model building \cite{Hayashi:2008ba,Hayashi:2009ge}.

It would be interesting to study more generally whether simple
correlations such as those between \eref{chiral_index} and
\eref{lambda_n} and \eref{c3spec} exist between $\eta$ and the chiral
index, independent of the existence of a heterotic dual. In fact, it
is possible that the F-theory $4$-fold could explicitly give
indications of such correlations through its topology. For example, it
is well known that some $4$-folds cannot be good F-theory vacua
without including non-trivial G-flux. In these cases their second
Chern class (or more generally Wu class) is incompatible with trivial
G-flux in the presence of quantization conditions. A study of the
topology of $Y_4$ and its links to intersection structure such as that
in \eref{intersec_example} could yield important information along
these lines (for similar investigations see
\cite{Collinucci:2010gz,Collinucci:2012as} which explore Chern and Wu
classes of $Y_4$ with simple singularities). We hope to explore this
in future work.

\subsection{Generic G-flux that breaks gauge symmetry}
\label{sec:example-G-flux}

In \S\ref{gflux_caution}, we discussed how G-flux can break the gauge
group associated with a purely geometric construction, through the
structure of the second Chern class of the dual heterotic bundle.  As
an example of this mechanism, consider the case of base
$B_2=\mathbb{P}^1 \times \mathbb{P}^1$ with twist $T=6c_1(B_2)$;
{\it i.e.}, $\eta_{-}=0$. According to the arguments of Section
\ref{sec:duality}, we would naturally have determined from the
F-theory geometry $Y_4$ that $\Sigma_{-}$ carried an $E_8$ symmetry
({\it i.e.}, the fiber degeneration is type II*) and hence that there were
{\it no smooth bundles} $V_1$ on $X_3$ with $\eta_{-}=0$ ({\it i.e.}, that
$V_1$ is trivial and $V_2$ satisfies $\eta_{+}=12c_1(B_2)$). However,
this is too quick since this argument ignores the fact that $\eta$ alone is not
enough to determine even $c_2(V_1)$. To see this, consider the
following smooth heterotic geometry.

On the base $B_2$, consider the poly-stable rank 2 vector bundle
defined as a kernel (of the map $m$) via the following short exact
sequence ({\it i.e.}, a ``monad'' bundle
\cite{Distler:1987ee}):
\begin{equation}
0 \to {\cal V}_{1} \to {\cal O}(0,1)^{\oplus 2} \oplus {\cal O}(1,0)^{\oplus 2} \stackrel{m}{\rightarrow} {\cal O}(1,1)^{\oplus 2} \to 0 \label{weird_eg}
\end{equation}
(where ${\cal O}(a,b)={\cal O}(aS + b F)$ and $S,F$ are the hyperplanes in each $\mathbb{P}^1$ factor). For an appropriately block-diagonal choice of the map $m$, ${\cal
  V}_1$ is a simple twist of the poly-stable tangent bundle of
$\mathbb{F}_0$ with vanishing slope;
{\it i.e.}, ${\cal V}_{1}=T\mathbb{P}^1
\otimes {\cal O}(1,1)$. For generic choices of $m$ this bundle is
slope-stable for all of the K\"ahler cone of $\mathbb{P}^1 \times
\mathbb{P}^1$. Over the entire elliptically fibered threefold, $\pi:
X_3 \to B_2$, we can likewise define the stable, slope-zero pull-back
bundle $\pi^{*}({\cal V}_{1})=V_1$ with $c_1(V_1)=0$ and $c_2(V_1)$
non-trivial solely {\it from the pull-back of the $(2,2)$-form on the
  base, $\pi^{*}(\zeta)$}
\begin{equation}
c_2(V_1)=\pi^{*}(\zeta)=\pi^*\left(4 \omega_1 \omega_2 \right)~~\Rightarrow \eta_{-}=0
\end{equation}
where $\omega_i$ are the $(1,1)$-forms dual to the divisors $S,F$ in $\mathbb{F}_0$.

The pull-back bundle defined by \eref{weird_eg} is an example of a
smooth, everywhere stable $SU(2)$ bundle which breaks $E_8 \to E_7$ in
the heterotic effective theory. Thus, in contradiction to the
conventional indication of the F-theory $E_8$-type Weierstrass
equation, it is clear that this is an everywhere well-defined $E_7$
theory. In the other $E_8$ factor a generic bundle with
$c_2(V_2)=12c_1(\mathbb{F}_0)+88\pi^*(\omega_1 \omega_2)$ (as required
by anomaly-cancellation, see \eref{anom_canc} and \eref{c2tx}) breaks
all the symmetry. At first pass it would seem that the bundle in
\eref{weird_eg} cannot be naively be described by a smooth spectral
cover and that as a result the Heterotic/F-theory dictionary is
unclear. For spectral covers $\eta=0$ indicates that $[{\cal S}]=2
[\sigma]$ from \eref{spec_class} and we would be tempted to conclude
here that the spectral cover \eref{speccov} with $\eta=0$ described
only the Fourier-Mukai transform of the trivial rank $2$ bundle ${\cal
  O}^{\oplus 2}$. However, this is forgetting half of the data of the
Fourier-Mukai transform: in particular the rank 1 sheaf $L_{{\cal S}}$
over ${\cal S}$ (see Section \ref{sec:spectral-cover} and
\eref{l_top}). Taking into account the possibility of rank $1$ sheaves
on the non-reduced scheme ${\cal S}$ which arise from higher rank
sheaves (in this case rank $2$) on $\sigma=0$
\cite{Donagi:1995am,ein_donagi} it is clear that more general bundles
$V_1$ are possible after FM transform\footnote{As mentioned in Section
  \S\ref{spec_cov_limits}, the holomorphic tangent bundle to an
  elliptically fibered threefold, $TX_3$, is frequently found to have
  a degenerate spectral cover description of this type ({\it i.e.}, reducible
  or non-reduced ${\cal S}$) \cite{Bershadsky:1997zv,
    Berglund:1998ej}.}. In the standard heterotic/F-theory dictionary,
the data of these rank 1 sheaves (whether ordinary line bundles or
higher rank sheaves in the non-reduced or reducible case as above) is
mapped into G-flux \cite{Curio:1998bva,Donagi:1998vw,
  Hayashi:2008ba}. 

As this example illustrates, such possibilities must be taken into
account if one hopes to fully determine the properties of heterotic
vector bundle moduli space from its F-theory dual. For now, we 
consider only the data of $Y_4$ itself, focusing on purely geometric
structure and properties, and leave an investigation of
the intriguing possibilities of G-flux for future work.

\section{Consequences for heterotic bundles}\label{sec:het_cons}

Many of the new results in this paper are conclusions/constraints
regarding properties of the moduli space of bundles (more precisely, the moduli space of semi-stable sheaves) arising in heterotic theories and links between bundle topology and
structure group. In this section we provide a brief summary of these results.

Unlike in six dimensions, four-dimensional heterotic/F-theory duality
provides new and non-trivial insight into the structure of the
heterotic moduli space ${\cal M}_{\omega}(c(V))$ of semi-stable sheaves with fixed
topology on $X_3$. At present, very few techniques are known for
determining the dimension and structure of ${\cal M}_{\omega}$ on
Calabi-Yau threefolds and there are many open questions which are of
interest to both physics and mathematics. These include applications
to string phenomenology (for example the large scale scans
for ``Standard Model'' bundles undertaken in
\cite{Anderson:2011ns,Anderson:2012yf,Anderson:2013xka}) 
as well as more mathematical questions such as the possible existence
of
new mathematical rules for linking topology $(c(V))$ to conditions for
vanishing/triviality of ${\cal M}_{\omega}(c(V))$ and the computation
of higher rank Donaldson-Thomas invariants. 
For elliptically fibered
Calabi-Yau threefolds, heterotic/F-theory duality provides a rich set
of new computational tools and we view this work as a preliminary
step in using these tools to determine the full structure of ${\cal
  M}_{\omega}$.  We briefly summarize here the main new results:\\

\noindent {\bf  Effectiveness conditions on $\eta$}:
\begin{itemize}
\item
As discussed in \S\ref{sec:Bogomolov} and \S\ref{sec:spectral-cover},
the parameter $\eta$ determining part of the second Chern class of the
heterotic bundles must obey several effectiveness constraints in
different contexts.  $\eta$ must be effective in the stable
degeneration limit, and $\eta-N c_1 (B_2)$ must be effective for a
spectral cover construction of an $SU(N)$ bundle.  We have found that
these constraints are more general.  For any F-theory construction
with a smooth heterotic dual, $\eta$ must be effective, and $\eta -N
c_1$ must be effective for $N = 2, 3$ for gauge groups $E_7, E_6$ (or
smaller) corresponding to heterotic theories with structure bundles
$SU(2), SU(3)$, independent of the stable degeneration limit or method
of bundle construction.
\end{itemize}

\noindent {\bf {Base-point-freeness and bundles with exceptional structure group}}:
\begin{itemize}
\item
Previous work  aimed at describing vector bundles over Calabi-Yau
threefolds in the context of heterotic/F-theory duality, such as
\cite{Friedman-mw}, has focused on bundles with $SU(N)$ and $Sp(N)$
structure groups, constructed using spectral covers in the stable
degeneration limit.  Here we have considered consistency conditions on topology in a construction-independent way and demonstrated that the base-point-freeness condition on $\eta$ is necessary for $SU(N)$ and $Sp(N)$ structure groups, independent of the the method of bundle construction.

Moreover, our study has shown that these constraints on the topology
of the vector bundle do not seem to be universal.  We have considered
a broader class of heterotic/F-theory dual models and identified a
large range of models in which the base-point free condition on the
components $\eta$ of the second Chern class need not be satisfied in
the dual F-theory model.  In all these models the structure group on
the heterotic side is an exceptional group or $SO(8)$.  Thus, F-theory
allows us to identify the conditions on the second Chern class that
are necessary, and apparently sufficient, for vector bundles with
exceptional and $SO(8)$ structure group to be constructed over a broad
class of Calabi-Yau threefolds.  These results could be mathematically
useful in explicitly constructing or characterizing such bundles. In
particular, it would be intriguing to utilize these conditions in
formulating topological consistency conditions for bundles constructed
through the cameral cover construction (which are at present not as
explicitly described as those for the spectral cover construction).

\item
One interesting feature of the heterotic models with F-theory duals where the base-point
free condition is violated is that they all involve 
elliptic fibrations over
generalized del Pezzo surfaces  that contain  curves of self-intersection $-2$.  These
surfaces are limits of usual del Pezzo surfaces where the points where
$\P^2$ is blown up are brought together in specific ways.  In
principle, the Calabi-Yau threefolds formed over generalized del Pezzo
surfaces should simply be special limits in the moduli space of
the generic elliptically fibered Calabi-Yau threefold over the
corresponding del Pezzo. This limit can be controlled precisely and
may provide an avenue for the explicit construction of the vector
bundles with exceptional structure groups that arise in these
cases. More precisely, in these limits as the complex structure of
$B_2$ is tuned to produce the generalized del Pezzo surfaces, the
Mori cone of effective divisors in $B_2$ jumps discontinuously (though
$h^{1, 1}(X_3)$ remains unchanged), and the
K\"ahler cone 
of the Calabi-Yau threefold decreases correspondingly.
This change in the K\"ahler cone impacts the properties of the moduli space of stable bundles that can arise, and also seems to restrict the existing
bundles to have exceptional structure groups in many
cases. \footnote{See
  \cite{Donagi:2009ra,Anderson:2010mh,Anderson:2011ty,Anderson:2013qca,Curio:2011eu,Curio:2012iv}
  for similar ``Noether-Lefschetz'' type-problems and ``jumping'' in
  complex structure/bundle moduli space.}

\end{itemize}

\noindent {\bf {$SO(32)$ heterotic/F-theory duality
and the connectivity of string moduli space}}:

\begin{itemize}
\item
By using topological terms in the 4D effective supergravity action to
characterize heterotic F-theory duality \cite{Grimm-Taylor}, we can
identify topologically which F-theory models are dual to heterotic
models for $SO(32)$ as well as $E_8 \times E_8$ models without
requiring a stable degeneration limit.  We have explicitly identified
those F-theory models that are dual to $SO(32)$ heterotic string
theory over a general smooth elliptically fibered Calabi-Yau threefold
base, and shown that in all such cases the generic model has a gauge
group of $SO(8)$, which cannot be broken further by Higgsing.

\item
On the F-theory side, all the models we have considered are connected
in a smooth geometric moduli space.  Over each base ${\cal B}_3$ there
is a moduli space of Weierstrass models that provides a
nonperturbative completion of the perturbative heterotic moduli space
of bundles (sheaves) over the dual heterotic elliptically fibered
Calabi-Yau threefold.  Furthermore, the distinct bases are connected
by tensionless string transitions that correspond to small instanton
transitions on the heterotic side.  For those F-theory bases with
$SO(32)$ heterotic duals, there are also $E_8 \times E_8$ duals;
F-theory/heterotic duality may illuminate the connection between these
two distinct heterotic perturbative limits and the resulting
relationship on Calabi-Yau threefolds between the moduli spaces of
$SO(8)$ structure group bundles and other bundle structures associated
with the $E_8 \times E_8$ theory. This extends to four dimensions
results on the geometry of $E_8 \times
E_8/SO(32)$ dual heterotic pairs that were previously understood in
higher-dimensional contexts
\cite{Narain:1985jj,Ginsparg:1986bx,Morrison-Vafa-I,
  Morrison-Vafa-II,Bershadsky:1996nh,Witten:1995gx,Aspinwall:1996nk}.
\end{itemize}

\noindent {\bf {Upper bounds on $H$}}:
\begin{itemize}
\item 
As first explored in \cite{Rajesh:1998ik,Berglund-Mayr}, the presence
of generic, non-Higgsable symmetries for singular $Y_4$ geometries
indicates that for a given $\eta$ there is an upper bound on the size
of the structure group $H$ for any bundle in the moduli space. As
explained in \S\ref{sec:equivalence}, if a generic symmetry $G$ cannot
be Higgsed in the $4$-dimensional effective theory, this implies that
for the given topology ($\eta$) there exist {\it no bundles with
  structure group larger than $H$}, the commutant of $G$ in
$E_8$. Phrased differently, in order to define a bundle with structure
group $H$ over the elliptically fibered CY threefold, there is a
minimum ``size'' for $\eta$. These conditions (given in the absence of
$G$-flux) are listed in Table \ref{t:bm-bounds}.
\item 
We further observe that these rules are at present only a first step
in determining ${\cal M}_{\omega}(c(V))$ and its constraints. As
described in \S\ref{gflux_caution}, for some choices of
bundle/$4$-fold topology it may be that generic $G$-flux breaks the
apparent symmetry $G$ indicated by the Weierstrass equation of
$Y_4$. In these cases, the bundle structure group may be bigger than
indicated by the bounds on $\eta$ in Table
\ref{t:bm-bounds}. 
%We hope to incorporate the further constraints
%arising from $G$-flux in future work.% \cite{us_to_appear}.
%\item 
%It is important to note that $G$-flux can only further break the
%generic symmetry, never enhance it. Thus the $4$-dimensional gauge
%symmetry $G$ can at most be smaller than that assumed in
%\S\ref{sec:equivalence} (see \S\ref{sec:example-G-flux} for
%example). 
%As a result, the structure group could in some cases be
%generically larger than that indicated by Table \ref{t:bm-bounds},
When this occurs, it must involve non-trivial values for $\zeta$ in
\eref{eq:f2-decomposition}; this indicates a new layer of structure
linking not only $\eta$ with $H$, but also with $(\zeta, c_3(V))$ --
the two integer values specifying the remaining bundle topology not
studied in this work. Such a correlation would involve a finer level
of structure linking $H$ and $c(V)$ than has been so far explored in
the literature.
We hope to explore these issues in further work.
\end{itemize}

\noindent {\bf {Lower bounds on structure group, $H$}:} \\
In addition to the non-Higgsable symmetries described above and their
heterotic consequences, in the F-theory geometry we have seen many
examples of $Y_4$ with a generic symmetry $G$ which {\it cannot be
  consistently enhanced} 
%for many special values of the complex structure of $Y_4$ 
% WT:?
(see \S\ref{sec:equivalence}). In the dual
heterotic geometry, these geometric observations provide constraints
on when a given vector bundle can be consistently decomposed into a
reducible sum: 
\beq
V \to {\cal V}_1 \oplus {\cal V}_2 \oplus {\cal O} \oplus \ldots
\eeq
with a smaller (reduced) structure group.
\begin{itemize}
\item 
The results of \S\ref{sec:bpf} indicate that the base-point-free
condition described in \S\ref{sec:spectral-cover} and \S\ref{sec:bpf}
cannot be consistently violated in the case of $H=SU(N)$ ({\it i.e.},
for a hermitian bundle, regardless of the method of construction). As a
result, for any technology applied to a bundle
associated with a
consistent {\it non-base-point free choice of $\eta$},
it is clear that there is no way to reduce the
structure group to $SU(N)$. We find that all the
non-base-point free examples of $\eta$ correspond to structure groups
$H=SO(8)$, $G_2$, $F_4$, $E_6$, $E_7$ or $E_8$. In these cases, the
bundle structure group can never be consistently decomposed $H \to H_1
\times H_2 \ldots$ with $H_i \subset H$ hermitian; in these cases the
gauge group $G$ thus cannot be enhanced to $E_6$ or $E_7$.
\item 
In a similar spirit, for many examples (see \S\ref{su2su3_sec}) there
are generic $SU(3)$ or $SU(2)$ symmetries on $Y_4$, with 
$(f,g)$ vanishing to degree $(2,2)$ or $(1,2)$
and no
possible enhancements to higher $SU(N)$ gauge symmetries
are possible (this would require $(f,g)$
non-vanishing and $\Delta$ vanishing to degree $N$). As
in the case of the non-base-point free examples above, this constrains
the ways in which bundles with exceptional structure groups can be
decomposed into hermitian factors.
\item 
We find examples for which there appear to be both upper and lower
bounds on $H$; in these cases for the given value of $\eta$ the
moduli space of semi-stable sheaves can contain bundles with exactly
one allowed structure group only. See for example \S\ref{H_stuck_eg}, where
an example of a topology is given for which
$F_4$ is the only consistent structure group.
\item 
The lower bounds on $H$ constrain the structure of possible
sub-sheaves ${\cal F} \subset V$, and determine a bound below which $V$
cannot decompose as $V= {\cal F} \oplus V/{\cal F}$, etc. The
determination of such substructure has important consequences for the
Harder-Narasimhan filtration of $V$, and the group quotient structure
of ${\cal M}$ itself \cite{huybrechts_lehn}.
\item 
Finally,  these ``lower'' bounds on $H$ may
be strengthened by the presence of $G$-flux. 
While $G$-flux cannot enhance the gauge group in a way that violates
these lower bounds on $H$, it could act to reduce the generic gauge symmetry
arising purely from geometry, which could lead to even stronger lower
bounds on $H$ in some cases.
\end{itemize}

%%%%%%%%%%%%%%%%%%%%%%%%%%%%%%%%%%%
\section{Enumeration of heterotic/F-theory dual pairs with toric bases $B_2$}
\label{sec:enumeration}

We have systematically analyzed all toric F-theory bases ${\cal B}_3$ (constructed as $\P^1$ bundles) that have smooth heterotic duals on Calabi-Yau threefolds that are
elliptically fibered with section.  The toric bases $B_2$ that
support such models are the generalized del Pezzo (gdP)
surfaces, a subset of 16 of the complete set of 61,539 toric surfaces
enumerated in \cite{mt-toric} that can act as bases for elliptically
fibered Calabi-Yau manifolds with section.  Over the 16 toric gdP
bases we find 4962 distinct ${\cal B}_3$'s that have smooth heterotic
duals.  Each choice of ${\cal B}_3$ corresponds to a specific
Calabi-Yau threefold $X_3$ giving the generic elliptic fibration over
$B_2$, with a particular choice in the part of the
bundle topology characterized by $\eta_i$ in (\ref{eq:f2-decomposition}).  The
analysis was performed by considering all possible twists $T$
compatible with the bounds (\ref{eq:t-0}-\ref{eq:t-2}), and explicitly
analyzing the monomial and singularity structure of the resulting
Weierstrass model in the toric description.  For each model that does
not have a (4, 6) singularity on a divisor or curve, we determine the
gauge group content and whether $\eta_1, \eta_2$ are base-point free.
The resulting enumeration of ${\cal B}_3$'s for some specific $B_2$'s matches that described in the previous section using the more general constraints described in
\S\ref{sec:F-theory-constraints}.  The results of the toric analysis
are listed in Tables~\ref{t:table},~\ref{t:group-matrix}.  For each
base $B_2$ we have indicated the number of distinct ${\cal
  B}_3$'s over that base (the number of possible ``twists'' $T$ in the
$\P^1$ bundle giving acceptable ${\cal B}_3$'s), and the number of
these ${\cal B}_3$'s that violate the base-point free condition for
one or both gauge group factors.
A subset of the F-theory bases ${\cal B}_3$ tabulated in
Table~\ref{t:table} have been explored in previous work; in
particular, the $\P^1$ bundles over $\P^2$ and $\F_m$ described in the
first four lines of the table, and the 18 toric Fano varieties were
also described in \cite{Klemm:1996ts, Mohri, Berglund:1998ej}.

\begin{table}
\begin{center}
\begin{tabular}{|  l | c  |c | c | c | c |% c | c | c | c |
}
\hline
base  $B_2$&$h_{1, 1}$ & \# ${\cal B}_3$'s & NB (1) & NB (2) 
& \# cod 3
%&$\gf_4$ &$\gso_8$ &$\gsu_3$ &$\gsu_2$\\
\\
\hline
(1, 1, 1)\hfill  ($\P^2$) &1& 14 & 0 & 0& 0\\
\hline
(0, 0, 0, 0)\hfill($\F_0$)& 2 &82 & 0 & 0 & 0\\
(1, 0, -1, 0)\hfill($\F_1$) & 2&109 & 0 & 0 & 0\\
(2, 0, -2, 0)\hfill($\F_2$)& 2 &24 & 14 & 0 & 0\\
\hline
(0, 0, -1, -1, -1) \hfill ($dP_2$) & 3 &472 & 0 & 0 & 0\\
(1, -1, -1, -2, 0) &  3 &173 & 100 & 0 & 0\\
\hline
(-1, -1, -1, -1, -1, -1) \hfill  ($dP_3$) & 4 & 776 & 0 & 0 & 0\\
(0, -1, -1, -2, -1, -1) &  4 &729 &  396 & 0 & 0\\
(0, 0, -2, -1, -2, -1) &  4 &312 & 213 & 42 & 0\\
(1, 0, -2, -2,  -1, -2)  & 4 &62 & 31 & 25 & 32\\
\hline
(-1, -1, -2, -1, -2, -1, -1) & 5 &1119 & 755 & 140 & 0\\
(0, -1, -1, -2, -2, -1, -2)   & 5 &406 & 219 & 150 & 217\\
\hline
(-1, -1, -2, -1, -2, -2, -1, -2) &  6 &351 & 149 & 185 & 173\\
(-1, -2, -1, -2, -1, -2, -1, -2) &   6 &214 & 119 & 69 & 0\\
(0, -2, -1, -2, -2, -2, -1, -2) &  6 &83 & 18 & 59 & 45\\
\hline
(-1, -2, -2, -1, -2, -2, -1, -2, -2)  & 7 &36 & 8 & 26 & 29\\
\hline
\hline
 total & &
4962 & 2022 & 696& 496\\
\hline
\end{tabular}
\end{center}
\caption[x]{\footnotesize Table of all smooth F-theory bases with
  smooth heterotic duals that are $\P^1$ bundles over toric bases
  $B_2$.  The
base $B_2$ is characterized by the sequence of
  self-intersections of toric divisors.  
NB (Non-Base point free) indicates the number of
  bases ${\cal B}_3$ that violate the
base-point free condition on one (1)
or both (2)
  sides $\Sigma_\pm$.
The final column is the number of models that have a toric codimension
3 locus where  
$f, g$ vanish to degrees 4, 6.}
\label{t:table}
\end{table}

One interesting observation from the data in Table~\ref{t:table} is
that more than half of the possible toric F-theory geometries violate the base-point free condition on at least one
of the gauge factors, so that more than half of the corresponding
heterotic models have generic bundles with exceptional or $SO(8)$ structure group.
The only models that can violate the base-point free condition
are those with generalized del Pezzo bases having $-2$ curves, and for
such bases a very high fraction of models violate the base-point free
condition on at least one side.  Since the vast majority of the
hundreds of
possible non-toric bases $B_2$ compatible with a smooth heterotic dual
are generalized del Pezzo's, we expect that the fraction of all models
with smooth heterotic duals that have exceptional or $SO(8)$ structure
group is quite high.

In Table~\ref{t:group-matrix} we tabulate the number of models in the
full set that have each of the possible distinct gauge algebras ${\cal
  G}_1 \oplus {\cal G}_2$.  Note that this gauge algebra represents
the minimal (most generic) gauge algebra for each base.  For each
${\cal B}_3$, tuning Weierstrass monomials can lead to enhanced gauge
groups through ``unHiggsing,'' corresponding on the heterotic side to
special loci in bundle moduli space where the structure group $H$
becomes smaller and $G$ correspondingly larger.  The minimal gauge
algebra summands ${\cal G}_i$ for each base ${\cal B}_3$ are
determined from the dual monomials in $f, g$ in the toric picture
using Table~\ref{t:Kodaira}.  In places where the degrees of vanishing
of $f, g$ do not uniquely determine the gauge algebra type, the gauge
algebra is fixed by the monodromy around the codimension one divisor,
which can be read off from the structure of the monomials following
the discussion in \S\ref{sec:duality-geometry}.  When the vanishing
degrees of $f, g, \Delta$ are $2, 2, 4$, the gauge algebra is $\gsu_2$
unless $g = g_2 (u, v)z^2 +{\cal O} (z^3)$ with $g_2 (u, v)$ a perfect
square, where $z = 0$ on the divisor locus in question and $u, v$ are
coordinates on the divisor.  When there is no restriction other than
the vanishing of certain monomials, $g_2 (u, v)$ is only guaranteed to
be a perfect square if it contains only a single even monomial $u^{2n}
v^{2m}$.  Similarly, when $f, g, \Delta$ vanish to degrees 3, 4, 8,
the gauge algebra factor is only $\ge_6$ when the leading part of $g$
is a perfect square, which again is only possible when it is a single
even monomial.  For vanishing degrees $2, 3, 6$ the story is slightly
more subtle, but again easy to analyze in terms of the monomials.  The
generic gauge algebra factor is $\ggg_2$.  The algebra becomes
$\gso_8$ when $f_2^3 = c g_3^2$, with $c$ an overall constant
(complex) coefficient, which is only possible when each contains only
a single monomial $f_2 = au^{3n}v^{3m}, g_3 = bu^{2n}v^{2m}$.  
%In
%principle, 
The gauge algebra can also be $\gso_7$, which occurs when $g_3 = 0$
and $f_2 (u, v)$ is not a perfect square; this only occurs for two of
the 3D bases ${\cal B}_3$ considered here.\footnote{These $\gso_7$
  examples were missed in earlier versions of this paper due to a
  coding error.  These examples are described explicitly, along with a
  more general discussion of isolated non-Higgsable $\gso_7$
  components, in \cite{mt-NHC-4D}.  Thanks to D.\ R.\ Morrison for
  discussions on issues related to these cases.}  (Note, however, that
for more general bases that do not have smooth heterotic duals we do
expect $\gso_7$ to arise more frequently as a generic gauge algebra
component, in conjunction with $\gsu_2$ algebra components.  This
occurs, for example, in 6D models when the base $B_2$ contains
intersecting curves of self-intersection $-2, -3, -2$, which support
the gauge algebra $\gsu_2 \oplus\gso_7 \oplus\gsu_2$
\cite{mt-clusters}.)

\begin{table}
\begin{center}
\begin{tabular}{| r | c | c | c | c | c | c | c | c | c |c |}
\hline
$\times$ & $ \cdot $ & $\gsu_2 $ & $\gsu_3 $ & $\ggg_2$ & $\gso_7$ &
$\gso_8 $ & $\gf_4 $ &
  $\ge_6 $ & $\ge_7 $ & $\ge_8 $\\
\hline
$\cdot  $
& 712 %640
 &   &   &   &   &   &   &   &  & \\
$\gsu_2  $
& 499 %415
 & 47 %33
 &   &   &   &   &   &   &   &\\
$\gsu_3  $
& 121 %112
 & 11 %8
 & 2 %2
 &   &   &   &   &   &   &\\
$\ggg_2 $
& 590 %589 %492
 & 62 %41
 & 7 %3
 & 34 %25
 &   & &   &   &   &  \\
$\gso_7 $
& 2
 & 0
 & 0
 & 0 & 0
    & &   &   &   &  \\
$\gso_8  $
& 275 %276 %232
 & 14 %9
 & 1 %1
 & 12 %8
& 0 & 3 %2
 &   &   &   &  \\
$\gf_4  $
& 1243 %1245 %1039
 & 74 %53
 & 6 %2
 & 54 %43
& 0 & 9 %9
 & 32 %29
 &   &   &  \\
$\ge_6  $
& 184 %165
 & 2 %2
 & 0 %0
 & 2 %2
& 0 & 0 %0
 & 2 %2
 & 0 &   &  \\
$\ge_7  $
& 890 %726
 & 24 %14
 & 0 %0
 & 14 %10
& 0 & 2 %2
 & 13 %12
 & 0 %0
 & 4 %4
 &  \\
$\ge_8 $
& 15 & 0 & 0 & 0 & 0 & 0 & 0 & 0 & 0 & 0\\
\hline
\end{tabular}
\end{center}
\caption[x]{\footnotesize Gauge algebras ${\cal G}_1 \oplus
{\cal G}_2$ arising in  generic models for
the 4962 F-theory bases ${\cal B}_3$ with toric $B_2$
and smooth heterotic duals.
Note that for many bases ${\cal B}_3$ a variety of
distinct models with enhanced gauge groups can be realized when
moduli are tuned to specific loci, corresponding to distinct
elliptically fibered fourfolds in the F-theory picture.}
\label{t:group-matrix}
\end{table}

In Table~\ref{t:group-matrix} we see that the great majority of
models (85\%) have some gauge group automatically imposed from the
geometry, which cannot be removed by Higgsing charged moduli fields.
Furthermore, most of the models either have gauge factors that are not
subgroups of $SU(5)$, or contain $SU(2)$ or $SU(3)$ factors that
cannot be enhanced to $SU(5)$ as discussed in the previous sections.
Note that the gauge group described here is purely that determined by
the geometry.  As discussed in \S\ref{gflux_caution}, in some situations
the gauge group may be modified when G-flux is taken into account.  We
leave further investigation of this effect to future work.

Note that there are many toric ${\cal B}_3$'s that are $\P^1$ bundles over
toric $B_2$ bases for which gauge algebra summands $\ge_8$ arise
that are not included in this tabulation because there are
codimension two curves in ${\cal B}_3$ living in the $\ge_8$ locus where the
degree of vanishing of $f, g$ reaches 4, 6.  These are a special class
of examples of situations where a curve in the base ${\cal B}_3$ must be
blown up to have a base ${\cal B}_3'$ that can act as the base of an
elliptically fibered Calabi-Yau fourfold.  After this blow-up, the
F-theory model no longer has a smooth heterotic dual and is not included in
this analysis.  In this situation the blown up base is also
generically non-toric.  These cases are closely analogous to base
surfaces $B_2$ for 6D F-theory models that contain $-9, -10,$ and
$-11$ curves; along such curves there is an $\ge_8$ gauge algebra
summand and 3, 2, or 1 points where $f, g$ vanish to degrees $4, 6$
and the base must be blown up for a smooth F-theory model
\cite{mt-clusters, mt-toric, Martini-WT}.

Several unusual features arise in many of the 4962 models we have
constructed.  As mentioned in \S\ref{sec:geometry-constraints}, in
some models that are otherwise well-behaved there are codimension
three singularities of order (4, 6).  It is not known whether these
singularities herald a sickness of the associated 4D supergravity
theories \cite{Morrison-codimension-3}; as discussed in
\S\ref{chiralsec}, such codimension three singularities may also be
associated with chiral matter, G-flux, or abelian gauge symmetries.  
In some cases, a codimension three (4, 6) singularity arises at a
toric point given by the intersection of three toric divisors.
There are a total of 496
models with this feature (or bug) in the toric set; we have tabulated
the number of threefolds ${\cal B}_3$ where this occurs for each base
surface $B_2$ in the last column of Table~\ref{t:table}.
One of the simplest examples of a threefold with this property is the
$\P^1$ bundle over the 10th base $B_2$ in Table~\ref{t:table},
characterized by divisors in the base with self-intersections
(1, 0, -2, -2, -1, -2) and a twist divisor $T = D_5$, where $D_5$ is
the divisor in $B_2$ with self-intersection -1.
This model has no codimension one singularities associated with
nonabelian gauge groups, but $f$ and $g$ vanish to degrees (4, 6) at
the point $\Sigma_-\cap D_3 \cap D_4$.
Note that, as mentioned in \S\ref{sec:constraints-curves} there can
also be codimension three (4, 6) singularities arising at non-toric
points, such as generically occurs on curves where $f, g$ vanish to
degrees (4, 5).  We have not attempted to classify the models with
such singularities here.

Another feature that can arise is a codimension two singularity on a
curve that does not lie on any divisor carrying a gauge group.  While
in general codimension two singularities indicate matter charged under
the nonabelian gauge groups of the corresponding divisors, in this
situation this interpretation is not possible.  It is possible that
these singularities are simply cusps in the discriminant locus with no
physical meaning, or they may herald the presence of abelian $U(1)$
factors.  This occurs in roughly half (2495 of the 4962 total) of
the threefold bases ${\cal B}_3$.  The simplest example is the $\P^1$
bundle over $\F_1$ with twist $T = 2F$; this base gives no codimension
one singularity associated with nonabelian gauge groups, but has a (2, 3)
vanishing of $f, g$ on $S\cap \Sigma_-$.
We leave further investigation of the models with these features to
future work.

\section{Conclusions and open questions}
\label{sec:conclusions}

In this paper we have given a global characterization of a broad class
of 4D ${\cal N} = 1$ string vacua that admit both a heterotic
description and a dual F-theory description.  The class of vacua we
have considered are described in the heterotic theory through
compactification on a smooth Calabi-Yau threefold that is elliptically
fibered with section over a base $B_2$ and carries a smooth vector
bundle (and in some cases, $5$-branes wrapping the elliptic fiber),
and in F-theory through compactification on an elliptic fibration over
a base threefold that is itself a $\P^1$ bundle with section over the
same base $B_2$.  We have shown that the number of topologically
distinct vacua in this class is finite, and we have explicitly
enumerated all models where the base $B_2$ is toric.

By focusing on the underlying geometrical and topological structure of
the theories, we have developed tools and identified features of these
models that do not depend on specific limits or bundle constructions
on the heterotic side of the duality.  We have identified from the
F-theory side a simple set of constraints that are necessary and
sufficient for the existence of a Calabi-Yau compactification
geometry; these constraints are expressed in terms of the ``twist''
defining the $\P^1$ bundle on the F-theory side and related components
of the second Chern class of the bundles on the heterotic side.  These
constraints give a detailed characterization of the
circumstances under which slope-stable bundles with general structure
groups should exist both for heterotic $E_8 \times E_8$ and $SO(32)$
theories on smooth Calabi-Yau threefolds.  The structure of chiral
matter on the heterotic side has implications for the interplay
between chiral matter and G-flux on the F-theory side.

The results described in this work represent a small step towards a
systematic characterization of the broader class of ${\cal N} = 1$ 4D
supergravity theories that can be realized in string theory.  There
are clearly many  directions in which this work could be expanded
further.  Many more detailed aspects of the physics of the large class of models
described here can be explored further using the tools described
here.  This work also provides a basis for a further systematic
expansion of our understanding of heterotic/F-theory duality, as well
as tools for expanding the range of applicability of both heterotic
and F-theory approaches to string compactification.  We conclude with
a brief summary of some of these possible future directions.

\subsection{Detailed physics of smooth heterotic/F-theory dual pairs}

In this work, following \cite{Grimm-Taylor}, we used topological
structure, in the form of axion-curvature squared terms in the 4D
supergravity theory, to identify dual heterotic and F-theory
geometries.  This gives an association between F-theory constructions
and heterotic bundles that is independent of the stable degeneration
limit \cite{Friedman-mw} in which the duality has been most thoroughly
studied; the approach taken here has enabled a systematic classification of all smooth
dual geometries where the F-theory model is described in terms of a
threefold base ${\cal B}_3$ that is a $\P^1$ bundle over a base
$B_2$.  Within this class of smooth dual geometries, there are many
questions that could be explored further.

The roughly 5000 models where $B_2$ is toric provide an
extensive dataset of dual heterotic/F-theory constructions that may be
useful in a variety of contexts.  To aid further development in this
direction, we have provided the details of this set of F-theory
compactification geometries in a file that can be downloaded from
\cite{files}.  This file contains a listing for each of the bases
$B_2$ of the complete set of allowed twists $T$ describing a $\P^1$
bundle over that $B_2$ that does not have $(4, 6)$ codimension one or
two singularities, as well as the generic gauge algebras ${\cal G}_1
\oplus{\cal G}_2$ generated by for
codimension one singularities over that base
($SO(8)$ in those cases that have an $SO(32)$ heterotic dual).
The methods of this paper can be applied more generally for any of the
several hundred generalized del Pezzo bases that support a smooth
elliptically fibered Calabi-Yau threefold.  One natural extension of
the work here would be the explicit construction and classification of
this broader class of (generically non-toric) bases, along the lines
of the example ${\rm dP}_4$ worked out in \S\ref{sec:dp4}
(which by itself already gives rise to roughly 7000 additional ${\cal
  B}_3's$; other non-toric generalized del Pezzo surfaces are expected
to similarly generate large numbers of additional examples).

For the models considered and enumerated here, many more detailed
questions remain to be addressed.  For each of the $\sim 5000$ toric
${\cal B}_3$'s, there are many branches of the moduli
space in which the generic gauge group is enhanced by ``unHiggsing'',
corresponding to a tuning of Weierstrass moduli in the F-theory
picture and special loci in bundle moduli space on the heterotic
side.  Many general aspects of the branching structure of these moduli
spaces remain to be investigated.  We have identified from the
F-theory side specific conditions under which the dual heterotic model
should admit a bundle with exceptional structure group; in many of
these cases there is no explicit mathematical construction known for
such models, finding such constructions represents another class of
open problems related to this work.  And, as mentioned throughout the
text, we have not incorporated the effects of G-flux on the F-theory
side; this mechanism will in general lift many geometric F-theory
moduli and produce chiral matter.  We hope that the explicit
correspondence we have developed here will help in elucidating these
issues further.

Although we have focused in this paper on general aspects of heterotic
and F-theory constructions that are independent of specific models,
some lessons have emerged that may be relevant for more
phenomenological ``model building''.  One general lesson from the
systematic study of F-theory models both in 6D and in 4D, illustrated
particularly clearly in six dimensions \cite{mt-toric, Martini-WT}, is
that a large fraction of the elliptically fibered Calabi-Yau manifolds
that can be used to compactify F-theory give rise to large
``non-Higgsable'' gauge groups.  While a clear understanding of the
connection between geometrically non-Higgsable gauge groups and 4D
physics requires a better incorporation of the effects of G-flux, the
models we have studied here are among those in which the minimal
geometric gauge groups are smallest, and may provide the most
promising candidates for realistic models of physics.  In terms of
potentially phenomenologically relevant gauge groups, we have found
that many F-theory geometries contain geometric $SU(2)$ and/or $SU(3)$
factors that cannot be enhanced to $SU(5)$, but that can for example
be enhanced to $SO(10)$, $E_6$, or $E_7$.  Much work has been done in
constructing phenomenologically oriented F-theory models based on an
$SU(5)$ unification structure (see \cite{Heckman-review,
  Weigand-review, Maharana:2012tu} for a review of some of this work, and
\cite{Marsano-F-theory,Blumenhagen-F-theory,Cvetic-gh,Knapp-10,
  Knapp:2011wk, Marsano:2012yc} for some specific global GUT models).  It would be
interesting to study more broadly how the generic gauge group
structures that we have explored here might play into more general
model building approaches, perhaps in the context of GUT groups other
than $SU(5)$.
As we have focused on generic geometric structure in the models
studied here, we have also not investigated the tuning of abelian
gauge group factors.  Much recent work \cite{abelian,
  Morrison:2014era} has focused on the role 
of global $U(1)$ factors in F-theory models.  In most cases such
$U(1)$ factors arise only at very special tuned loci in the
Weierstrass moduli space over any given F-theory base; it was recently
found, however, that $U(1)$ factors can be generic over certain
special F-theory bases in 6D \cite{Martini-WT}, including bases
related to non-toric generalized del Pezzo bases considered here.
Further investigation of $U(1)$ factors in the class of dual
heterotic/F-theory models provides another interesting direction for
further work.

At present, the explicit
heterotic/F-theory dual ``dictionary'' has been most fully determined
in a corner of moduli space in which the heterotic bundles can be
described via spectral (more generally cameral) covers
\cite{Curio:1998bva,Donagi:1998vw,donagi, Friedman-mw}. 
However, the
results of this work shows that many consistent, perturbative
heterotic theories cannot be described by a naive application of these
constructions. For example, the bundle with $\eta=0$ of
\eref{weird_eg} can clearly not be described as an ordinary, smooth
spectral cover (see \S\ref{sec:example-G-flux} for a discussion).
Other specific examples arise in the class of models we
have analyzed here. For instance, considering the set of $4962$ toric
dual pairs described in Section \ref{sec:enumeration}, there appear to
be many good F-theory geometries for which we cannot directly
construct the heterotic dual bundles with standard tools, even when
the heterotic structure bundle is $SU(N)$.  For example, of the $947$
$4$-folds with a generic $E_7$ symmetry (on at least one patch),
$897$ of these fail to satisfy the parity
condition \eref{mod2_cond} for $SU(2)$ bundles constructed as generic,
irreducible spectral covers. This indicates that if good heterotic
duals exist they must either a) not be constructible as ordinary
$SU(2)$ spectral covers ({\it i.e.}, they possess one of the
limitations described in Section \ref{spec_cov_limits}) or b) the
bundle moduli space contains no irreducible spectral covers at all
(see \cite{Curio:2011eu} for example, for more exotic
possibilities). 
It is also possible that this parity condition  may be a more general
constraint and may indicate some problem with the associated F-theory
models, for example that may indicate a conflict with the existence of
a consistent
choice of G-flux.
In any case, there remains something to be
understood in the explicit moduli mapping of the heterotic/F-theory
dual pair which could yield important new insights into both theories.

As mentioned above, we have not considered here the effects of
G-flux in modifying the underlying Calabi-Yau geometry of F-theory.
One of the most interesting  aspects of
$4$-dimensional heterotic/F-theory duality is the fact that
deformations that change the gauge symmetry ({\it i.e.}, deformations
of the complex structure of $Y_4$ or of the bundles $V_i$ on $X_3$)
can be obstructed. These obstructions can appear both through D- and
F-term contributions to the potential in the low energy theory. A
better understanding of this potential would have impact not only on
the problem of moduli stabilization in heterotic/F-theory effective
theories but could also lead to novel dynamical effects in the
$4$-dimensional theories -- including, for example, the obstruction of 
tensionless string/small-instanton transitions and possible duality to
non-commutative $D3$ branes.
We hope to explore these topics in future
work.

\subsection{Expanding heterotic/F-theory duality}

In this work we have focused on the simplest class of
heterotic/F-theory dualities, where  both sides have elliptically
fibered Calabi-Yau geometries with a single section, and the F-theory
base ${\cal B}_3$ is a $\P^1$ bundle that also has a single section.
Since we expect that the heterotic and F-theory constructions are
simply different mathematical approaches to describing the same
physical theory in distinct limits, we expect that it should be
possible to extend this duality to a much broader class of vacua,
possibly at the expense of needing to introduce more complicated
mathematical objects in the theory on one or both sides.

One clear question is the extent to which heterotic/F-theory duality
can be systematically described when the heterotic Calabi-Yau geometry
and/or bundle structure
becomes singular, leading mathematically to a description in terms of
more singular objects such as sheaves.  While many examples of this
have been studied in the literature, the appearance of structure such
as enhanced gauge groups and additional geometric moduli  arising
through tensionless string transitions is more transparent
geometrically from the point of view of Weierstrass models on the
F-theory side.  The framework developed here may provide a useful
context in which to systematically extend the duality in these
directions.  The simple geometric framework of
F-theory has the potential to clarify some of the mathematical
questions that are rather subtle in this context on the heterotic side.

More generally, there are classes of geometries that are slightly more
general than those considered here in which heterotic/F-theory duality
is  not understood.  These include bases ${\cal B}_3$ built as more
general $\P^1$ fibrations (rather than $\P^1$ bundles) and situations
where
either the heterotic or F-theory elliptic fibrations have
multiple sections (higher rank Mordell-Weil group) or no section at all.  In particular, for Calabi-Yau geometries
that are elliptically fibered but have more than one section, or a
multi-section, the story is not yet completely clear.  F-theory models with
multiple sections (higher rank Mordell-Weil group) are understood
simply as models with additional $U(1)$ factors, which generally
should have natural heterotic duals.  F-theory models with a
multi-section but no global section have recently been incorporated
into the global moduli space of Weierstrass models
\cite{Braun:2014oya,Morrison:2014era}.  It is less clear, however, how
to construct an F-theory dual for a heterotic model on a Calabi-Yau
threefold with multiple sections or a multi-section.  While in
principle such threefolds can be realized as special limits in the
Weierstrass moduli space of elliptic fibrations (using the Jacobian
fibration associated with threefolds having a multi-section, as in
\cite{Morrison:2014era}), which
should give a corresponding construction on the F-theory side, the
details of the physics of this correspondence have not been worked
out.
In this paper we considered only cases where the F-theory threefold
base ${\cal B}_3$ is itself a $P^1$ bundle with section.
It is also possible to consider situations where ${\cal B}_3$ is a
$P^1$ fibration without a section or indeed even more general geometries \cite{Donagi:2012ts,Heckman:2013sfa}.

Another example of a situation where heterotic/F-theory duality is not
well understood comes from the fact that for Calabi-Yau threefolds,
the moduli space of bundles ${\cal M}$ can have multiple components
(see Appendix \ref{sec:rigid}).  If is not known how F-theory duals to
such situations can be understood.  A natural hypothesis is that for
each component of the heterotic moduli space there would exist
topologically identical, non-diffeomorphic Calabi-Yau $4$-folds on the
F-theory side.  This is another interesting avenue for further
investigation.

Finally, on the heterotic side there are compactifications on
Calabi-Yau threefolds that are not elliptically fibered.  Such
geometries can be reached by nonperturbative transitions from
elliptically fibered Calabi-Yau threefolds, so should in principle be
connected to the underlying geometric moduli space of F-theory
compactifications.  At present, there is no known mechanism by which
F-theory can include such vacua.  F-theory is at present still an
incomplete physical theory, however; there is no direct action
principle for the theory that incorporates all degrees of freedom.  In
an optimistic scenario, further development of heterotic/F-theory
duality may provide some insight into a more complete formulation of
the theory and a broader and more unified characterization of the full
space of ${\cal N} = 1$ 4D supersymmetric string theory vacua.

\vspace{.5in}

{\bf Acknowledgements}: We would like to thank Ron Donagi, Antonella
Grassi, James Gray, Thomas Grimm, Jonathan Heckman, Samuel Johnson,
Denis Klevers, Gabriella Martini, David Morrison, Daniel Park, and
Lucia Swanson for helpful discussions.  This research was supported by
the DOE under contract \#DE-FC02-94ER40818, and was also supported in
part by the National Science Foundation under Grant No. PHY-1066293.
We would like to thank the Simons Center for Geometry and Physics (LA
and WT), the Aspen Center for Physics (WT), and the Center for the
Theoretical Physics at MIT (LA) for hospitality during part of this
work.

\appendix

\section{Properties of elliptically fibered Calabi-Yau three- and fourfolds}
\label{sec:app_identities}
In this Appendix we briefly review a collection of useful results
regarding the geometry and topology of elliptically fibered Calabi-Yau
manifolds (see \cite{Friedman-mw,Klemm:1996ts} for a more complete
treatment). 
We focus on smooth elliptically fibered Calabi-Yau
threefolds, $\pi: X_3 \to B_2$ with a single section (which defines
$B_2$ as an algebraic sub-manifold within $X_3$). If $X_3$ is in
Weierstrass form, a minimal set of divisors\footnote{We focus here on
  a minimal form of elliptically fibered threefold in which all
  exceptional curves in the fiber have been blown down.} that span
the Picard group of $X_3$ is given by the zero section, $D_0$ and
divisors pulled back from the base of the form
$D_{\alpha}=\pi^{*}(D^{base}_{\alpha})$, where $D^{base}_{\alpha}$,
$\alpha=1, \ldots h^{1,1}(B_2)$ is an ample divisor on $B_2$. For such
smooth, minimal elliptic fibrations, $h^{1,1}(X_3)=h^{1,1}(B_2)+1$. We
will denote the basis of $\{1,1\}$-forms dual to the divisors above as
$\{\omega_0, \omega_{\alpha}\}$.

By virtue of this simple fibration structure, the triple intersection numbers of these divisors exhibit a universal behavior. First, since the base is a $2$-fold it is clear that
\begin{equation}
D_{\alpha} \cap D_{\beta} \cap D_{\gamma}=0
\end{equation}
Moreover, from the very definition of what it means for $D_0$ to be
section (and not a multisection) it is guaranteed that for any
two-form $\zeta$ on $B_2$ (dual to a single point), $D_0 \cap \zeta=1$
(that is, the zero section intersects each elliptic fiber precisely
once). It follows from this fact that
\begin{equation}
D_0 \cap D_{\alpha} \cap D_{\beta} = m_{\alpha \beta}
\end{equation}
where $m_{\alpha\beta}=D^{base}_{\alpha} \cap D^{base}_{\beta}$. These facts are enough to derive the following important cohomological identity on $\{2,2\}$ forms,
\begin{equation}\label{22-identity}
\omega_0 \wedge \omega_0 = K \wedge \omega_0
\end{equation}
where $K$ is the canonical class of the base, $K=-c_1(B_2)=K^{\alpha}\omega_{\alpha}$. With these results, the triple intersection numbers of $X_3$ 
\begin{equation}
d_{ABC}=\int_{X_3} \omega_{A} \wedge \omega_{B} \wedge \omega_{C}
\end{equation}
where $\omega_{A}=\{\omega_0, \omega_{\alpha} \}$, are given by
\begin{align}\label{x3intersec}
&d_{000}=m_{\alpha\beta}K^{\alpha}K^{\beta}& & d_{00\alpha}=m_{\alpha\beta}K^{\beta} \\
&d_{0\alpha\beta}=m_{\alpha\beta}& & d_{\alpha\beta\gamma}=0 \nonumber
\end{align}

With these intersection numbers and a chosen K\"ahler form $\omega=t^0 \omega_0 + t^{\alpha}\omega_{\alpha}$, the volume of $X_3$ takes the form
\begin{equation}\label{x3vol}
{\rm Vol}(X_3)=\frac{1}{3!}\int_{X} \omega\wedge \omega\wedge
\omega=\frac{1}{3!}\left(d_{000}(t^0)^3+3d_{00\alpha}(t^0)^2t^{\alpha}+3d_{0\alpha\beta}t^{0}t^{\alpha}t^{\beta}
\right) 
\end{equation}

The fibration structure guarantees that the second Chern class of $X_3$ can be written as \cite{Friedman-mw}
\begin{equation}\label{c2tx}
c_2(TX_3)=12c_1(B_2)\wedge \omega_0 + c_2(B_2)+11c_1(B_2)^2
\end{equation}
where in addition the topology of $B_2$ satisfies
\begin{equation}
\chi(B_2)=\int_{B_2} c_2(B_2)=2+h^{1,1}(B_2)~~~,~~~\int_{B_2} c^{2}_{1}(B_2)=K^{\alpha}K^{\beta}m_{\alpha \beta}=10-h^{1,1}(B_2)
\end{equation}

Finally, using the redundancy relation on $\{2,2\}$ forms in \eref{22-identity} it is possible to write the second Chern class of any bundle, $V$, on $X_3$
\begin{equation}\label{c2special}
c_2(V)=\eta \wedge \omega_0 + \zeta
\end{equation}
where $\eta$ and $\zeta$  are pullbacks through $\pi$ of $\{1,1\}$
 $\{2,2\}$ forms
from $B_2$.

\section{A brief exploration of rigid bundles}\label{sec:rigid}

A novel feature of four-dimensional compactifications of heterotic
string theory/F-theory is the possibility of multiple components in the dual
(vector bundle/fourfold) moduli spaces. In the case of
heterotic/F-theory duality, such multiple components to the moduli
space have not yet been studied in detail. Indeed, thus far in the
literature the correspondence between vector bundle moduli spaces in
heterotic theories and the complex structure moduli space of
Calabi-Yau fourfolds has only been studied in the case of single,
connected components \cite{Friedman-mw}.

For the moduli space of stable sheaves on Calabi-Yau threefolds a
natural case of interest is given when the local moduli space in fact
contains an isolated, rigid component. In general, given two bundles with the same topology, it is difficult to decide whether or not they reside in the same component of a global bundle moduli space. However, if one bundle is rigid (and the other has local deformation moduli), it is clear that the rigid bundle consists of its own distinct component to bundle moduli space. As a result, a search for rigid bundles is one of the simplest probes for multiple components of bundle moduli spaces.

For elliptically fibered
Calabi-Yau threefolds, it is straightforward to come by examples of
such rigid vector bundles. For instance, over the Calabi-Yau threefold
defined as a single degree $\{3,3\}$ hypersurface in the product,
$\mathbb{P}^2 \times \mathbb{P}^2$, of two projective spaces, consider
the following poly-stable $SU(2)$ bundle
\begin{equation}
V_2=L+L^{\vee}={\cal O}(-H_1+H_2)+{\cal O}(H_1-H_2)
\end{equation}
where $H_1, H_2$ are the restrictions of the hyperplanes of each
ambient $\mathbb{P}^2$ factor to $X_3$. On this space, $V_2$ is rigid,
since the dimension of the space of bundle-valued singlets is given by
\begin{equation}
h^1(X_3, L^{\otimes 2})=h^1(X_3, {L^{\vee}}^{\otimes 2})=0
\end{equation}
for generic values of the complex structure of $X_3$ \cite{Anderson:2008ex}. Geometrically the spaces $H^1(X_3, L^{\otimes 2})$, $H^1(X_3, {L^{\vee}}^{\otimes 2})$ constitute the space of non-trivial extensions (for example $H^1(X_3, L^{\otimes 2})$ parameterizes the space of non-trivial extensions $0 \to L \to V_2 \to L^{\vee} \to 0$) which parameterize how $L, L^{\vee}$ may be non-trivially ``glued'' back into an indecomposable $SU(2)$ bundle. Since the space of such extensions vanishes in these cases, the split bundle $L+L^{\vee}$ has no infinitesimal deformations, that is, it is rigid. 

Although the example above is interesting from the point of view of vector bundle moduli spaces, it is not clear what the impact of such examples will be in the heterotic-F-theory pairs constructed in this work. Although the threefold above is elliptically fibered, it has no section and cannot be written in Weierstrass form. At present its F-theory dual (if any) is unknown.

In this Appendix, we make a tentative exploration of whether it is possible to obtain isolated components to the moduli space of $SU(2)$ bundles, such as the one described above, over the class of Calabi-Yau threefolds considered here -- that is, elliptically fibered threefolds with a single section, obeying the topological identities listed in Appendix \ref{sec:app_identities}. Once again, we can search for bundles of the form
\begin{equation}\label{splitsu2}
L+L^{\vee}
\end{equation}
where $L$ is a holomorphic line bundle on $X_3$ satisfying $\mu(L)=0$ in the K\"ahler cone, as required for supersymmetry by \eref{slopezero}. This reducible bundle in \eref{splitsu2} will be rigid if $h^1(X,L^{\otimes 2})=h^1(X,(L^{\vee})^{\otimes 2})=0$

As we will argue below, however, for the geometry in consideration in
this work, such examples appear to be rare and we have obtained no
explicit examples. This result is fully consistent with the fact that
in the dual F-theory geometry we find a single Calabi-Yau fourfold
(with a single connected (and non-trivial) component to its complex
structure moduli space) for each choice of vector bundle
topology/twisting parameter, $T$. 

To see this, we must consider the line bundle cohomology of $L^{\otimes 2}$ on $\pi: X_3 \to B_2$. Here we are aided by the formalism of Leray Spectral sequences \cite{hartshorne1977algebraic}. According to this spectral sequence for the fibration $\pi: X_3 \to B_2$, we have a natural bi-grading such that for any bundle $V$ on $X_3$,
\beq\label{spec_seq2}
H^p(X, V) = \sum_{p=l+m}E^{l,m}_{\infty} 
\eeq
where 
\beq\label{Edef}
E_{1}^{l,m}=H^l(B_2, R^{m}\pi_{*} (V))
\eeq
and
and $R^{m}\pi_{*}(V)$ is the $m$-th direct image sheaf of the bundle $V$  (pushed forward under the fibration $\pi$). We need not concern ourselves with the iteration of the sequence via the maps $d_r: E_{r}^{p,q} \rightarrow {E_{r}}^{p+r,q-r+1}$, since the spectral sequence terminates at $E_1$. To see this, note that on any open set $\mathcal{U}$ on $\mathbb{P}^1$, the $m$-th direct image sheaf, $R^{m}\pi_{*} (V)$ can be locally represented by the pre-sheaf
\beq
\mathcal{U} \rightarrow H^m(f^{-1}(\mathcal{U}), R^{m}\pi_{*} (V))
\eeq
For elliptic fibrations, however, the fiber (locally isomorphic to $f^{-1}(\mathcal{U})$) is one dimensional. As a result, $R^{m}\pi_{*} (V)$ is non-vanishing only for $m=0,1$ and the spectral sequence terminates at
\beq
E_{\infty}=E_1
\eeq
To analyze the cohomology, we further need to observe how a line bundle of the form
\beq\label{l_eg_def}
L={\cal O}(a\sigma + b^\alpha \pi^*(D_\alpha))
\eeq
behaves under the push-forward functor. The first useful useful
observation is that line bundles of the form ${\cal O}(b^\alpha
D_\alpha)$ built from divisors pulled back from the base $B_2$
satisfies $\pi^{*}{\cal O}_{B_2}(b^\alpha D_\alpha)={\cal
  O}_{X_3}(b^\alpha \pi^*(D_\alpha))$. This implies that under
push-forwards we have the so-called ``projection formula''. For a
fibration $\pi: X_3 \to B_2$, and any bundles $V$ on $X$ and $U$ on
$B$,
\begin{equation}
R^{q}\pi_{*}(V\otimes \pi^{*}U)=R^{q}\pi_{*}(V)\otimes U
\end{equation}

To determine the cohomology of $L={\cal O}_{X_3}(a\sigma +b^\alpha \pi^*(D_\alpha))$ then, we need only consider $R^i\pi_*({\cal O}_{X_3}(a\sigma +b^\alpha \pi^*(D_\alpha))=R^i\pi_*({\cal O}_{X_3}(a\sigma))\otimes {\cal O}_{B_2}(b^\alpha D_\alpha)$. Moreover, the structure of $R^i\pi_*({\cal O}_{X_3}(a\sigma))$ can straightforwardly be determined by considering the Koszul sequence:
\beq
0 \to {\cal O}_{X_3}(-\sigma) \to {\cal O}_{X_3} \to {\cal O}|_{\sigma=B_2} \to 0
\eeq
Twisting this by ${\cal O}_{X_3}(\sigma)$ yields
\beq
0 \to {\cal O}_{X_3} \to {\cal O}_{X_3}(\sigma) \to {\cal O}(\sigma \cdot \sigma)|_{\sigma=B_2} \to 0
\eeq
But by \eref{22-identity} and the definition of $\sigma$ as the zero-section this is simply
\beq
0 \to {\cal O}_{X_3} \to {\cal O}_{X_3}(\sigma) \to {\cal O}(K_2 \cdot \sigma)|_{\sigma=B_2} \to 0
\eeq
and pushing forward to $B_2$ gives the short exact sequence
\beq
0 \to {\cal O}_{B_2} \to R^0\pi_{*}({\cal O}_{X_3}(\sigma)) \to K_2 \to 0
\eeq
For $B_2$ the base of a CY threefold, the above sequence splits and we have determined the direct image sheaf:
\beq
R^0\pi_{*}({\cal O}_{X_3}(\sigma))={\cal O}_{B_2} \oplus K_2 
\eeq
The calculation outlined above can be iterated inductively to find the higher direct image sheaves $R^i\pi_{*}(\cO_{X_3}(a\sigma))$ for $a >1$ in a similar manner. It is straightforward to demonstrate that
\begin{align}\label{known_stuff}
R^{0}\pi_{*}(\cO_{X_3}(2\sigma))=&\cO_{B_2} \oplus K_2 \oplus K_2^{\otimes 2} & R^{1}\pi_{*}(\cO_{X_3}(2\sigma))=0 \\
R^{0}\pi_{*}(\cO_{X_3}(3\sigma))=&\cO_{B_2} \oplus K_2 \oplus K_2^{\otimes 2} \oplus K_2^{\otimes 3} & R^{1}\pi_{*}(\cO_{X_3}(3\sigma))=0 \\
\vdots &  &  \vdots \\
R^{0}\pi_{*}(\cO_{X_3}(a\sigma))= &
{\rm Sym}^{a}(\cO_{B_2} \oplus  K_2) & R^{1}\pi_{*}(\cO_{X_3}(a\sigma))=0
\end{align}
Similar results for $a<0$ can be found by using Grothendieck duality:
For any sheaf $F$ on $X_3$, the push-forward functors obey the following relation:
\beq
R^{1-i}\pi_{*}(F^{\vee}\otimes \omega_{\mathbb{X}_3|B_2})=(R^i\pi_{*}F)^{\vee}~~~~,~i=0,1
\eeq
where $\omega_{X_3|B_2}=K_{X_3}\otimes \pi^{*}(K_2^{\vee})=\pi^{*}(K_2^{\vee})$ is the ``dualyzing sheaf'' \cite{hartshorne1977algebraic}. With these results in hand, we can now in principle calculate all line bundle cohomology on $X_3$. In order to build reducible, rigid $SU(2)$ bundles like those above we must note that for the line bundles of interest, $\mu(L)=0$, and hence \cite{Anderson:2008ex}
\beq\label{h0h3}
H^0(X,L)=H^3(X,L)=0
\eeq
Thus, to construct a rigid $SU(2)$ bundle we must use the results above for line bundle cohomology and further ask, for what values of $a,b$ in \eref{l_eg_def} can we have $H^1(X,L)=H^1(X,L^{\vee})=0$? By \eref{spec_seq2} and \eref{Edef} it is clear that we require
\beq
H^1(B_2, R^0\pi_* L)=H^0(B_2, R^1\pi_* L)=0
\eeq
in order to satisfy $H^1(X,L)=0$ and
\beq
H^2(B_2, R^0\pi_* L)=H^1(B_2, R^1\pi_* L)=0
\eeq
for $H^2(X,L)=H^1(X,L^{\vee})=0$ (by Serre duality). This, coupled with \eref{h0h3} means that the direct image sheaves $R^0\pi_* L$ and $R^1 \pi_* L$ must have entirely vanishing cohomology on $B_2$. However, as we will see below, this does not occur for simple threefolds of the type we are considering here. 

To see that it is rare for $H^i(B_2, R^0\pi_* L)=0$ $\forall i$, it is useful to consider the index of $R^0\pi_* L$ using the Riemann-Roch theorem. With $L$ as in \eref{l_eg_def}
\beq
R^0\pi_* L=(\cO_{B_2} \oplus K_2 \oplus \ldots K_2^{\otimes a}) \otimes \cO_{B_2}(b^\alpha D_{\alpha})
\eeq
we have that for each term in the sum, the index is additive. Thus
\beq
\chi(R^0\pi_* L)=\chi(\cO_{B_2}(b^\alpha D_{\alpha})+\chi(K_2 \otimes \cO_{B_2}(b^\alpha D_{\alpha})) \ldots 
\eeq
Letting $D=b^{\alpha}D_{\alpha}$ and using the fact that for any divisor, $A \subset B_2$,
\beq
\chi(A)=1+\frac{1}{2}A\cdot (A-K_2)
\eeq
we have
\beq
\chi(R^0\pi_* L)=(a+1)+D\cdot D
\eeq
If we demand that the index vanishes as a necessary condition for
entirely vanishing cohomology, we require $D \cdot D= -(a+1)$ for some
curve $D \subset B_2$. Putting this together with other geometric
constraints in the problem, we see that the Bogomolov bound of
\S\ref{sec:Bogomolov} places a positivity condition on $c_2(V)$. For
$V=L \oplus L^{\vee}$ this enforces that $a \geq 0$. Finally, it can
be noted that for the case $a=0$, the line bundles can be verified to
have non-vanishing cohomology. Thus, here we will consider $a > 0$. 

For the geometries considered in this work, we have at most $-2$
curves, thus without loss of generality we can restrict ourselves to
line bundles of the form $\cO(\sigma + b^{\alpha} D_{\alpha})$ where
$D=b^{\alpha}D_{\alpha}$ is a $-2$ curve. Although there do exist
curves of this type (for example the divisor $S$ in $\mathbb{F}_2$),
it can be verified on a case-by-case basis that here the necessary
condition is not in fact sufficient and $H^1(B_2, R^0\pi_* L) \neq
0$. Although we have not rigorously ruled out all possible $-2$ curves
in our set of base manifolds $B_2$, systematic searches have found no
examples with entirely vanishing cohomology. Thus, we expect that for the
simple geometries outlined in Appendix \S\ref{sec:app_identities}, no
reducible $SU(2)$ bundles of the form shown in \eref{splitsu2} exist
as rigid components in the moduli space.  
It would be nice, however, to have a more general abstract proof of
this result.
If rigid bundles can be found within the context of heterotic/F-theory
duality, it would be interesting to investigate the dual F-theory
constructions.  Some possibly related F-theory models may
exist; for 6D compactifications over $B_2=\P^2$, some models with
exotic matter were identified in \cite{kpt} that are similarly
rigid in the
sense that they have no moduli that
preserve the gauge group and matter content.

\end{document}